\documentclass{aa}  

\usepackage{graphicx}
\usepackage{txfonts}
\usepackage{gensymb}
\usepackage{lscape}
\usepackage{xcolor}
\usepackage[version=3]{mhchem}
\usepackage{caption,subcaption}

\usepackage[colorlinks = true,
            linkcolor = blue,
            urlcolor  = blue,
            citecolor = blue,
            anchorcolor = blue]{hyperref}

\begin{document} 
\title{Investigating the asymmetric chemistry in the disk around the young star HD~142527}
\author{M. Temmink\inst{1} \and
       A. S. Booth\inst{1} \and
       N. van der Marel\inst{1} \and
       E. F. van Dishoeck\inst{1,2}}
\institute{Leiden Observatory, Leiden University, 2300 RA Leiden, the Netherlands \\
          \email{temmink@strw.leidenuniv.nl} \and
          Max-Planck-Institut f\"ur Extraterrestrische Physik, Giessenbachstraße 1, D-85748 Garching, Germany}
\date{Received xx, 2023; accepted xx, 202x}

\abstract
{The atmospheric composition of planets is determined by the chemistry of the disks in which they form. Studying the gas-phase molecular composition of disks thus allows us to infer what the atmospheric composition of forming planets might be. Recent observations of the IRS~48 disk have shown that (asymmetric) dust traps can directly impact the observable chemistry, through radial and vertical transport, and the sublimation of ices. The asymmetric HD~142527 disk provides another good opportunity to investigate the role of dust traps in setting the disk's chemical composition. In this work, we use archival ALMA observations of the HD~142527 disk to obtain an as large as possible molecular inventory, which allows us to investigate the possible influence of the asymmetric dust trap on the disk's chemistry. We present the first ALMA detections of [\ce{C} I], \ce{^{13}C^{18}O}, \ce{DCO^+}, \ce{H_2CO} and additional transition of \ce{HCO^+} and \ce{CS} in this disk. In addition, we have acquired upper limits for non-detected species such as \ce{SO} and \ce{CH_3OH}. For the majority of the observed molecules, a decrement in the emission at the location of the dust trap is found. For the main \ce{CO} isotopologues continuum over-subtraction likely causes the observed asymmetry, while for \ce{CS} and \ce{HCN} we propose that the observed asymmetries are likely due to shadows cast by the misaligned inner disk. As the emission of the observed molecules is not co-spatial with the dust trap and no \ce{SO} or \ce{CH_3OH} are found, thermal sublimation of icy mantles does not appear to play a major role in changing the gas-phase composition of the outer disk in HD~142527 disk. Using our observations of \ce{^{13}C^{18}O} and \ce{DCO^+} and a \textit{RADMC-3D} model, we determine the \ce{CO} snowline to be located beyond the dust traps, favouring cold gas-phase formation of \ce{H_2CO}, rather than the hydrogenation of \ce{CO}-ice and subsequent sublimation.}


\maketitle

\section{Introduction}
Planets form in disks consisting of both gas, dust and ice surrounding newly formed stars. The composition of these planets is set by the composition of the disk \citep{OB21,Eistrup22}. Unravelling the physical and chemical processes that shape disk composition are thus fundamental in understanding both a planet's composition and how this can be linked to its formation history. Elemental abundances and ratios, such as the \ce{C}/\ce{H}-, \ce{O}/\ce{H}-, \ce{C}/\ce{O}-, \ce{C}/\ce{N}- and \ce{N}/\ce{O}-ratios, are the most important quantities that allow to dissect the linkage between the planetary compositions and the disk's chemistry \citep{ObergEA11,RBoothEA17,EistrupEA18,PacettiEA22}. \\
\indent The chemistry of the disk itself, however, is closely linked to the physics and dynamics of the disk. As an example, the location of molecular snowlines, which depend on the dust temperature structure (see, for example, \citealt{MileyEA21}), are of great importance in setting both the available gas- and ice-content in the disk. The snowlines lead to radial variations in the elemental gas-phase abundances of C, O and N \citep{RBoothEA17,KrijtEA20,vdMarelEA21b}. In addition, the location of substructures, or dust traps, with respect to these snowlines plays another important role in setting these elemental abundances. In a disk with no substructures or a dust trap inside the snowline (e.g. the \ce{CO} snowline), dust settling and subsequent radial drift will cause the dust particles to cross the snowline and their icy mantles to sublimate, yielding higher gas-phase \ce{C}, \ce{O} and \ce{N} abundances (see, for example, \citealt{RBoothEA17}). If a dust trap is, however, located beyond the snowline(s), radial drift will prevent the dust particles from crossing the snowlines and the molecules remain frozen-out on the grains, yielding gas in the inner disk depleted of \ce{C}, \ce{O} and/or \ce{N} \citep{BB19,BanzattiEA20,McClureEA20,SturmEA22}. \\
\indent In recent years, the Atacama Large Millimeter/submillimeter Array (ALMA) has provided fantastic opportunities to study planet-forming disks, as it has given access to both observations of the mm-sized dust and gas at high angular resolution and sensitivity. Studies with ALMA have revealed a large amount of substructures, present not only in the dust, but also in the gas. Three of those studies are the Disk Substructures at High Angular Resolution Project (DSHARP; \citet{DSHARPI}), the survey of gaps and rings in Taurus disks \citep{LongEA18} and the Molecules with ALMA at Planet-forming Scales (MAPS; \citealt{MAPSI}) ALMA Large Programs. All three programs have identified a large variety of substructures, mostly a combination of gaps and rings, in different types of disks and have shown that the dust properties and chemical composition can vary over very small scales. Especially the MAPS program shows the importance of observing a large variety of molecules for investigating the variations in the chemical composition. Single disk studies, such as PDS 70 (e.g. \citealt{FacchiniEA21}) or TW Hya \citep{QiEA13a,QiEA13,ObergEA21TWH,TvSEA21,CalahanEA21TWH,CleevesEA21TWH}, provide additional insights in chemical variations. \\
\indent There is, however, so far only one disk in which the observed chemistry can be directly linked to the observed dust substructure, the Oph-IRS~48 disk. The disk has a large asymmetric dust concentration on the southern side of the disk at a radius of $\sim$60 au \citep{vdMarelEA13}. Recent ALMA observations \citep{vdMarelEA21,ABoothEA21,BrunkenEA22,LeemkerEA23} have shown that the emission of a variety of both simple (e.g. \ce{SO}, \ce{SO_2}, \ce{NO}, \ce{H_2CO}) and complex organic molecules (e.g. \ce{CH_3OH}, \ce{CH_3OCH_3}) is co-located with the dust trap. The observations led to the conclusion that the dust trap is an ice trap, where co-spatial molecules are frozen out on the dust grains. Sublimation of these ices into the gas-phase, following the radial and vertical transport of the dust grains, allows the molecules to be detected and provides the current explanation for the co-spatial molecular emission. Especially the low \ce{CS}/\ce{SO}- and \ce{CN}/\ce{NO}-ratios, suggesting a low gas-phase \ce{C}/\ce{O}-ratio, and the high excitation temperatures found for \ce{H_2CO} and \ce{CH_3OH} ($T_\textnormal{ex}\sim$100-200 K) support this hypothesis. \\
\indent To further investigate the role of dust substructures in setting the disk's chemical compositions, to investigate whether the IRS~48 disk is a special case and to obtain a unique peek into the ice composition of disks, we use ALMA archival data to investigate the chemistry of the HD~142527 disk. The HD~142527 disk is, besides the IRS~48 disk, one of the few other disks that has easily resolvable, large-scale dust substructure and happens to be the second most asymmetric dust disk presently known \citep{CasassusEA13,vdMarelEA21c}. 

\section{Observations and methods}
\subsection{The source: HD~142527}
The HD~142527 system is located at a distance of 157 pc \citep{GC18} and consists of a 1.69 $M_\odot$, F6-type star \citep{FairlambEA15,FvdM20} and a M-dwarf companion on an eccentric orbit ($e\gtrsim$0.2), at separations of $\sim$44-90 mas ($\sim$7-14 au) \citep{BillerEA12,LacourEA16,BalmerEA22}. The stars are encompassed by a circumbinary planet-forming disk, which is known to have a large, horseshoe-shaped asymmetry in the millimetre-sized dust, peaking at $\sim$175 au in radius \citep{CasassusEA13}. This outer-disk has an inclination of 27\degree (\citealt{vdMarelEA21c} and references therein) and a position angle (PA) of 160\degree \citep{vdPlasEA14}. Furthermore, the system is known to have a significantly misaligned inner disk (inclination of 23\degree and PA of 14\degree, \citealt{MarinoEA15,BohnEA22}), which causes shadows to be visible in both the scattered light images (e.g., \citealt{FukagawaEA06,CanovasEA13,AvenhausEA17}) and the emission of various \ce{CO} isotopologues \citep{BoehlerEA17}. \citet{YoungEA21} have shown that even a small misalignment of the inner disk can lead to the azimuthal variations in the line emission and/or column densities (up to two orders of magnitude) of various molecules as the result of azimuthal temperature variations due to the shadowing from the inner disk. \\
\indent Previous ALMA studies have shown that emission of various molecules, for example \ce{CS} and \ce{HCN}, is also asymmetric in the HD~142527 system, albeit not coincidental with the dust trap \citep{vdPlasEA14}, as opposed to the IRS 48 disk. \citet{vdPlasEA14} have shown that the \ce{CS} $J$=7-6 and \ce{HCN} $J$=4-3 emission appears to be suppressed in the region of the continuum emission. They provide two possible explanations for the molecular asymmetries: (1) a lower dust temperature causing the freeze-out of \ce{CS} and \ce{HCN}. The lower temperature follows from the lowered ratio of the mass opacity in the optical and far-IR due to the increased averaged grain size \citep{MN93,vdPlasEA14}, as larger grains are more subject to trapping at pressure maxima. (2) the quenching of line emission following the continuum emission having a higher optical depth. This latter option can emerges as continuum over-subtraction in the case that both the dust and line emission are optically thick \citep{BoehlerEA17}. Although some molecules show spatially asymmetric integrated line emission, the overall gas distribution, as for IRS 48, is symmetric throughout the disk, based on the observations of CO and its various isotopologues (e.g., \citealt{CasassusEA13,BoehlerEA17}). 

\subsection{Data}
To further investigate the role of the asymmetric dust trap and the disk ice composition, if any, in setting the chemistry of the HD 142527 disk, we have hunted for observable line features in various ALMA archival datasets. The analysed datasets were selected based on the highest chance of observing transitions of \ce{CS}, \ce{SO}, \ce{SO_2}, \ce{H_2CO} and \ce{CH_3OH}, of which the latter four have been detected in the IRS 48 dust trap, allowing for proper comparisons, with upper level energies of $E_{up}<$150 K and Einstein A coefficients of $\log{A_{ij}}>$-5. The datasets used throughout this work are summarised in Table \ref{table:Datasets} and have the spatial resolution that is sufficiently high to resolve both the cavity and the asymmetric dust trap. All datasets were calibrated using the provided pipeline scripts and the specified \textit{Common Astronomy Software Applications (CASA)} \citep{CASA} version therein. Subsequent phase self-calibration and imaging were carried out using \textit{CASA} version 6.5.1.

\subsubsection{Imaging process}
The imaging of the continuum and molecular lines were carried out using the \textsc{tclean}-task, where we have made use of the multiscale algorithm. Before imaging the molecular lines, we have conducted a continuum subtraction using the task \textsc{uvcontsub} with a fit-order of 1. All images have been made using the Briggs weighting scheme, with, depending on the line strength, a robust parameter of +0.5 or +2.0, and velocity channel width of 0.5 km s$^{-1}$. The chosen channel width ensures consistency between the different datasets and provides sufficient velocity resolution for the aims of this work, considering that we do not investigate line kinematics. To ensure all emission was captured during the cleaning process, we have made use of a Keplerian mask and we cleaned down to a threshold of $\sim$4$\times$ the RMS in the initial, uncleaned image. All targeted molecular lines, including the line properties, the adopted robust parameter and found emission properties, can be found in Table \ref{table:Molecules}.

\subsubsection{Imaging the 2015.1.01137.S dataset} \label{sec:2015.1.01137.S}
The dataset 2015.1.01137.S has been treated slightly different compared to the other datasets. This ALMA Band 8 dataset has been used to image the [\ce{C} I] $^3$P$_1$-$^3$P$_0$ and \ce{CS} $J$=10-9 transitions. During the imaging process, it was noted that one of the three execution blocks (observing date: 19 May 2016) had higher angular resolution for the [\ce{C} I] transition and all datapoints for the \ce{CS} transition were flagged in this execution. Due to the differences across the execution blocks, the choice was made to not self-calibrate this dataset. The [\ce{C} I] transition has been imaged in three different ways: (1) using all three execution blocks with a robust parameter of 0.5, (2) using the higher angular resolution 

\begin{landscape}
\begin{table}
\caption{Table showing the detected and non-detected molecules in the various archival datasets.}             
\label{table:Molecules}     
    \begin{tabular}{c c c c c c c c c c c c}
    \hline\hline       
    Molecule & Transition & Frequency & E$_\textnormal{up}$ & $\log\left(A_{ij}\right)$ & g$_\textnormal{up}$ & Robust & Beam & Peak flux & RMS & Int. Flux & N$_\textnormal{avg}^{(a)}$\\
    & & [GHz] & [K] & & & & & \multicolumn{2}{c}{[mJy beam$^{-1}$]} & [mJy km s$^{-1}$] & [cm$^{-2}$] \\
    \hline
    \ce{C}$^{(b)}$: & $^3$P$_{1}$-$^3$P$_0$ & 492.16065100 & 23.6 & -7.09724 & 3 & 0.5 & 0.58"$\times$0.49 (-85\degree) & 372.34 & 18.36 & (6.1$\pm$0.6)$\times$10$^3$ & $>$(4.2$\pm$0.4)$\times$10$^{17}$ \\
    \hline
    \ce{^{12}CO} & 2-1 & 230.538000 & 16.6 & -6.16050 & 5 & 0.5 & 0.76"$\times$0.60" (-62\degree) & 619.40 & 12.64 & (7.1$\pm$0.7)$\times10^3$ & $>$(3.8$\pm$0.4)$\times10^{15}$ \\
     & 3-2 & 345.795990 & 33.2 & -5.60266 & 7 & 0.5 & 0.57"$\times$0.35" (65\degree) & 877.74 & 9.21 & (25.1$\pm$0.3)$\times10^3$ & $>$(5.2$\pm$0.5)$\times10^{15}$ \\
     & 6-5 & 691.473076 & 116.2 & -4.67011 & 13 & 0.5 & 0.48"$\times$0.33" (-82\degree) & 3018.36 & 92.31 & (95.7$\pm$9.6)$\times10^3$ & $>$(2.2$\pm$0.2)$\times10^{16}$ \\
    \ce{^{13}CO} & 2-1 & 220.398684 & 15.9 & -6.51752 & 10 & 0.5 & 0.78"$\times$0.64" (-64\degree) & 498.32 & 11.22 & (6.8$\pm$0.7)$\times10^3$ & $>$(1.0$\pm$0.1)$\times10^{16}$ \\
     & 3-2 & 330.587965 & 31.7 & -5.95976 & 14 & 0.5 & 0.51"$\times$0.42" (78\degree) & 520.40 & 3.91 & (16.9$\pm$1.7)$\times10^{3}$ & $>$(8.0$\pm$0.8)$\times10^{15}$ \\
    \ce{C^{18}O} & 2-1 & 219.560354 & 15.8 & -6.22103 & 5 & 0.5 & 0.79"$\times$0.64" (-64\degree) & 247.67 & 7.89 & (2.7$\pm$0.3)$\times10^3$ & $>$(2.3$\pm$0.2)$\times10^{15}$ \\
     & 3-2 & 329.330553 & 31.6 & -5.66324 & 7 & 0.5 & 0.52"$\times$0.42" (80\degree) & 309.11 & 5.04 & (6.7$\pm$0.7)$\times10^{3}$ & $>$(2.2$\pm$0.2)$\times10^{15}$ \\
    \ce{^{13}C^{18}O} & 3-2$^{(c)}$ & 314.119675 & 30.2 & -5.72343 & 8 & 2.0 & 0.32"$\times$0.24" (82\degree) & 13.66 & 2.00 & 297.0$\pm$29.7 & (1.9$\pm$0.2)$\times10^{15}$ \\
    \hline
    \ce{HCO^+} & 4-3 & 356.734223  & 42.8 & -2.44709 & 9 & 0.5 & 0.55"$\times$0.34" (66\degree) & 489.37 & 9.24 & (11.4$\pm$1.1)$\times10^3$ & $>$(7.8$\pm$0.8)$\times10^{13}$ \\
     & 8-7 & 713.341228 & 154.1 & -1.51943 & 17 & 2.0 & 0.51"$\times$0.34" (-88\degree) & 837.33 & 159.32 & (9.7$\pm$1.0)$\times10^3$ & (1.3$\pm$9.1)$\times10^{14}$ \\
    \ce{DCO^+} & 4-3 & 288.143858 & 34.6 & -2.3224 & 9 & 2.0 & 0.36"$\times$0.25" (89\degree) & 18.36 & 4.05 & 98.0$\pm$9.8 & (1.4$\pm$0.1)$\times10^{11}$\\
    \ce{HCN} & 4-3 & 354.505478 & 42.5 & -2.68602 & 27 & 2.0 & 0.60"$\times$0.39" (72\degree) & 211.57 & 7.69 & (1.6$\pm$0.2)$\times10^3$ & (5.7$\pm$0.6)$\times10^{12}$ \\
     & 8-7 & 708.877005 & 154.1 & -1.75839 & 51 & 2.0 & 0.41"$\times$0.32" (-79\degree) & 335.82 & 73.60 & $<$(1.5$\pm$0.2)$\times10^3$ & $<$(6.5$\pm$0.6)$\times10^{11}$ \\
    \ce{CS} & 7-6 & 342.882850 & 65.8 & -3.07737 & 15 & 0.5 & 0.55"$\times$0.45" (76\degree) & 129.41 & 3.44 & (1.1$\pm$0.1)$\times10^3$ & (1.6$\pm$0.2)$\times10^{13}$ \\
     & 10-9 & 489.750921 & 129.3 & -2.60415 & 21 & 2.0 & 0.89"$\times$0.79" (82\degree) & 301.91 & 42.00 & 693.65$\pm$69.37 & (2.0$\pm$0.2)$\times10^{13}$ \\
    \ce{SO} & 1$_2$-0$_1$ & 329.385477 & 15.8 & -4.84664 & 3 & 2.0 & 0.62"$\times$0.47" (86\degree) & 19.26 & 4.76 & $<$65.3$\pm$0.6 & $<$(1.9$\pm$0.2)$\times10^{13}$ \\
    \ce{SO_2} & 10$_{4,6}$-10$_{3,7}$ & 356.755190 & 89.8 & -3.48406 & 21 & 2.0 & 0.60"$\times$0.39" (73\degree) & 41.92 & 11.73 & $<$175.1$\pm$17.5 & $<$(7.4$\pm$0.7)$\times10^{12}$ \\
    \ce{^{34}SO_2} & 5$_{3,3}$-4$_{2,2}$ & 342.208857 & 35.1 & -3.50911 & 11 & 2.0 & 0.60"$\times$0.48" (70\degree) & 13.29 & 3.28 & $<$44.5$\pm$0.4 & $<$(8.2$\pm$0.8)$\times10^{11}$ \\
    \hline
     \ce{H_2CO} & 2$_{0,2}$-1$_{0,1}$ & 145.602949 & 10.5 & -4.10719 & 5 & 2.0 & 0.67"$\times$0.38" (-90\degree) & 19.76 & 3.53 & 677.1$\pm$67.7 & (2.9$\pm$0.3)$\times10^{14}$ \\
     & 4$_{0,3}$-3$_{0,3}$ & 290.623405 & 34.9 & -3.16102 & 9 & 2.0 & 0.36"$\times$0.25" (89\degree) & 44.77 & 3.97 & (1.2$\pm$0.1)$\times10^3$ & (1.1$\pm$0.1)$\times10^{14}$ \\
     & 4$_{2,3}$-3$_{2,2}$ & 291.237766 & 82.1 & -3.28314 & 9 & 2.0 & 0.35"$\times$0.25" (89\degree) & 18.37 & 3.73 & 215.3$\pm$21.5 & (7.8$\pm$0.8)$\times10^{13}$ \\
     & 4$_{3,2}$-3$_{3,1}$ & 291.380442 & 140.9 & -3.51654 & 27 & 2.0 & 0.32"$\times$0.26" (70\degree) & 25.19 & 4.83 & $<$127.8$\pm$12.8 & $<$(9.6$\pm$1.0)$\times10^{12}$ \\
     & 4$_{3,1}$-4$_{3,0}$ & 291.384362 & 140.9 & -3.51653 & 27 & 2.0 & 0.32"$\times$0.26" (70\degree) & 24.55 & 4.97 & $<$130.5$\pm$13.0 & $<$(9.8$\pm$1.0)$\times10^{12}$ \\
     & 4$_{2,2}$-3$_{2,1}$ & 291.948067 & 82.1 & -3.27994 & 9 & 2.0 & 0.36"$\times$0.24 (90\degree) & 29.91 & 5.68 & 316.0$\pm$31.6 & (1.1$\pm$0.1)$\times10^{14}$ \\
     & 3$_{2,1}$-4$_{0,4}$ & 691.921008 & 68.1 & -5.96670 & 7 & 2.0 & 0.54"$\times$0.37" (-81\degree) & 353.97 & 78.70 & $<$(1.3$\pm$0.1)$\times10^3$ & $<$(1.4$\pm$0.1)$\times10^{16}$ \\
    \ce{CH_3OH} & 7$_{1,6}$-7$_{0,7}$ & 314.859528 & 80.1 & -3.4562 & 60 & 2.0 & 0.31"0.24" (81\degree) & 7.03 & 1.41 & $<$38.6$\pm$3.9 & $<$(1.4$\pm$0.1)$\times10^{12}$ \\
     & 6$_{-2,5}$-5$_{-1,4}$ & 315.266861 & 71.0 & -3.92716 & 52 & 2.0 & 0.31"$\times$0.24" (81\degree) & 6.83 & 1.37 & $<$37.5$\pm$3.7 & $<$(3.6$\pm$0.4)$\times10^{12}$ \\
     & 8$_{3,5}$-9$_{2,7}$ & 330.793887 & 146.3 & -4.26857 & 68 & 2.0 & 0.61"$\times$0.48" (84\degree) & 16.67 & 3.46 & $<$48.4$\pm$4.8 & $<$(6.7$\pm$0.7)$\times10^{13}$ \\
     & 6$_{2,4}$-5$_{1,5}$ & 713.982466 & 74.7 & -2.839 & 52 & 2.0 & 0.55"$\times$0.35" (89\degree) & 724.81 & 163.53 & $<$(2.8$\pm$0.3)$\times10^3$ & $<$(2.4$\pm$0.2)$\times10^{13}$ \\
    \hline
    \end{tabular}
\tablefoot{The line properties (frequency, upper level energy, Einstein $A$ coefficient and upper state degeneracy) have been acquired from \textit{CDMS}. The integrated flux and disk-integrated column densities for the non-detected molecules are shown as upper limits. \\
$^{(a)}$: The displayed column densities have been calculated taking a rotational temperature of $T_\textnormal{rot}$=35 K. \\
$^{(b)}$: The listed fluxes and average column densities for the [\ce{C} I] line have been obtained from the image in which all executions blocks have been used.
$^{(c)}$: The $^{13}$C$^{18}$O transition is a blending of the $J$=3-2 F=5/2-5/2, F=5/2-3/2 and F=7/2-5/2. The shown line properties correspond to those of the F=7/2-5/2 line.}
\end{table}
\end{landscape}

\noindent execution block with a robust parameter of 2.0 and (3) using the other two executions blocks (both observed at the 23$^\textnormal{rd}$ of May, 2016) with a robust parameter of 0.5. The \ce{CS} $J$=10-9 transition has been imaged using the remaining two execution blocks.

\subsection{\textit{GoFish}: Searching for weak line emission}
When no clear emission signature could be visually distinguished in the image cubes, we have used the \textit{GoFish} package \citep{GoFish} to search for weak Keplerian emission signatures. \textit{GoFish} uses a spectral stacking technique, first introduced by \citet{YenEA16}, which accounts for the Keplerian rotation of the disk and aims to improve the signal-to-noise ratio of weak line emission by aligning and stacking the spectra taken from different (e.g. blue- and redshifted) sides of the disk. 

\subsection{Analysis methods}
\subsubsection{Integrated fluxes \& upper limits}
Disk integrated fluxes, displayed in Table \ref{table:Molecules}, have been obtained from the image cubes using the \textsc{specflux}-task. For the \ce{CO}-isotopologues, except \ce{^{13}C^{18}O}, we have extracted the flux density using circular apertures with radii between 2.5" and 3.0". As these transitions have the highest signal-to-noise and we expect the emission to be optically thick, using a circular aperture instead of a Keplerian mask should not yield any significant differences for the integrated fluxes or the disk-averaged column densities. For all other molecules we have used Keplerian masks, which encircles all the emission. The uncertainties displayed in Table \ref{table:Molecules} are the 10\%-errors, accounting for the ALMA flux calibration errors. \\
\indent For the non-detected molecules, we have acquired 3$\sigma$ upper limits using a circular area with a radius of 3.0", which covers the entire disk. The upper limits have been calculated based on the method described in \citet{CarneyEA19}, using
\begin{align}
    \sigma = \delta v \sqrt{N}\sigma_\textnormal{rms}.
\end{align}
Here, $\delta v$ is the channel velocity width and $\sigma_\textnormal{rms}$ is the channel noise in mJy~beam$^{-1}$. $N$ is the number of independent measurements, obtained through division of the amount of pixels in the circular aperture by the number of pixels in the beam. 

\subsubsection{Column densities}
Column densities for the upper level ($N_u^\textnormal{thin}$) have been acquired from the relation between the integrated line flux ($I_\nu$) (e.g., \citealt{GL99,TvSEA21}), under the assumption that the emission is optically thin,
\begin{align}
    I_\nu = \frac{A_{ul}N_u^\textnormal{thin}hc}{4\pi\Delta v}.
\end{align}
$A_{ul}$ denotes the Einstein $A$ coefficient and $\Delta v$ is the velocity width of the emission line. For each emission line, the velocity width has been taken to be the FWHM of a Gaussian profile fitted to the integrated line spectrum, which was acquired with \textit{GoFish}. By rewriting this equation and taking $I_\nu=S_\nu/\Omega$, where $S_\nu$ is the flux density in Jy and $\Omega$ is the solid angle subtended by the emission, the optically thin approximation of the upper level column density becomes,
\begin{align}
    N_u^\textnormal{thin} = \frac{4\pi S_\nu\Delta v}{A_{ul}\Omega hc}
\end{align}
Here, $S_\nu\Delta v$ denotes the integrated flux density of the emission line. \\
\indent To obtain the total upper level column density, the optically thin approximation was corrected for the optical depth ($\tau$),
\begin{align} \label{eq:NuCorr}
    N_u = N_u^\textnormal{thin}C_\tau = N_u^\textnormal{thin}\frac{\tau}{1-\exp\left(-\tau\right)}.
\end{align}
The optical depth itself was estimated using,
\begin{align}
    \tau = \frac{A_{ul}N_u^\textnormal{thin}c^3}{8\pi\nu^3\Delta v}\left[\exp\left(\frac{hv}{k_BT_\textnormal{rot}}\right)-1\right].
\end{align}
where $T_\textnormal{rot}$ denotes the rotational temperature. For the disk integrated column densities, displayed in Table \ref{table:Molecules}, we have assumed a constant value of $T_\textnormal{rot}$=35 K. \\
\indent The total column can subsequently be determined through the Boltzmann equation, 
\begin{align} \label{eq:Ntot}
    \frac{N_u}{g_u}=\frac{N_\textnormal{tot}}{Q\left(T_\textnormal{rot}\right)}\exp\left(-\frac{E_u}{k_BT_\textnormal{rot}}\right).
\end{align}
$g_u$ and $E_u$ denote, respectively, the upper level degeneracy and energy. $Q\left(T_\textnormal{rot}\right)$ is the partition function at $T_\textnormal{rot}$ which, through interpolation, has been obtained from the \textit{Cologne Database for Molecular Spectroscopy} (\textit{CDMS}; \citet{CDMS01,CDMS05}). 

\subsection{Rotational diagram analysis} \label{sec:RDA}
To better estimate the total column densities, without having to assume a fixed rotational temperature, we have made use of a rotational diagram analysis. The analysis is carried out by transforming Equation \ref{eq:Ntot} into a likelihood function by taking the logarithm,
\begin{align}
    \ln\left(\frac{N_u}{g_u}\right) = \ln\left(N_\textnormal{tot}\right) - \ln\left(Q\left(T_\textnormal{rot}\right)\right) - \frac{E_u}{k_BT_\textnormal{rot}}.
\end{align}
The left-hand side can be rewritten using Equation \ref{eq:NuCorr}, 
\begin{align}
    \ln\left(\frac{N_u}{g_u}\right) = \ln\left(\frac{N_u^\textnormal{thin}}{g_u}\right) + \ln(C_\tau).
\end{align}
\indent In our rotational diagram analysis, we have used the Markov Chain Monte Carlo (MCMC) implementation of \textit{emcee}-package \citep{emcee} to obtain posterior distributions of $N_\textnormal{tot}$ and $T_\textnormal{rot}$.

\subsection{Radial profiles} \label{sec:RadialProfile}
In order to determine the radial location of the emission of certain molecules, we have created azimuthally averaged, deprojected radial profiles and the corresponding errors using \textit{GoFish}. All radial profiles have been created using bin-sizes of half the size of the beam's minor axis.

\section{Results}
\subsection{Observed molecular emission}
The moment-zero maps of the observed molecules, together with the continuum image, are displayed in Figure \ref{fig:Gallery}. Galleries containing all transitions of the \ce{CO}-isotopologues and the other molecules (\ce{HCO^+}, \ce{CS} and \ce{H_2CO}) are displayed in, respectively, Figures \ref{fig:CO-Gallery} and \ref{fig:OM-Gallery}. \\
\indent Most striking are the different morphologies observed for the different molecular species. First, all \ce{CO} isotopologue images appear to be rather symmetric, indicating that the majority of the gas in the disk is distributed symmetrically. The emission shows a few deviations from being fully symmetric, which we relate to continuum oversubtraction (see Section \ref{sec:mmdust}) and shadowing from the inner disk (see also \citealt{BoehlerEA17}). Interestingly, the \ce{^{12}CO} peaks in the center of the disk, whereas the rarer isotopologues all have cavities present in their emission. \\
\indent As opposed to the \ce{CO} isotopologues some of the other molecules present mostly asymmetric structures. For example, the \ce{HCO^+} emission peaks in the center, enclosed by a weaker, non-axisymmetric ring-like structure (see also \citealt{CasassusEA13,CasassusEA15}), while the \ce{HCN} and \ce{CS} show large asymmetries, peaking at the opposite of the disk compared to the dust trap. The \ce{H_2CO} emission appears to be mostly symmetric, however, there appears to be a decrement in the emission at the location of the dust trap.

\subsubsection{Atomic Carbon}
\begin{figure*}[]
    \centering
    \includegraphics[trim=0cm 5.5cm 0cm 5.5cm, clip=true,width=\textwidth]{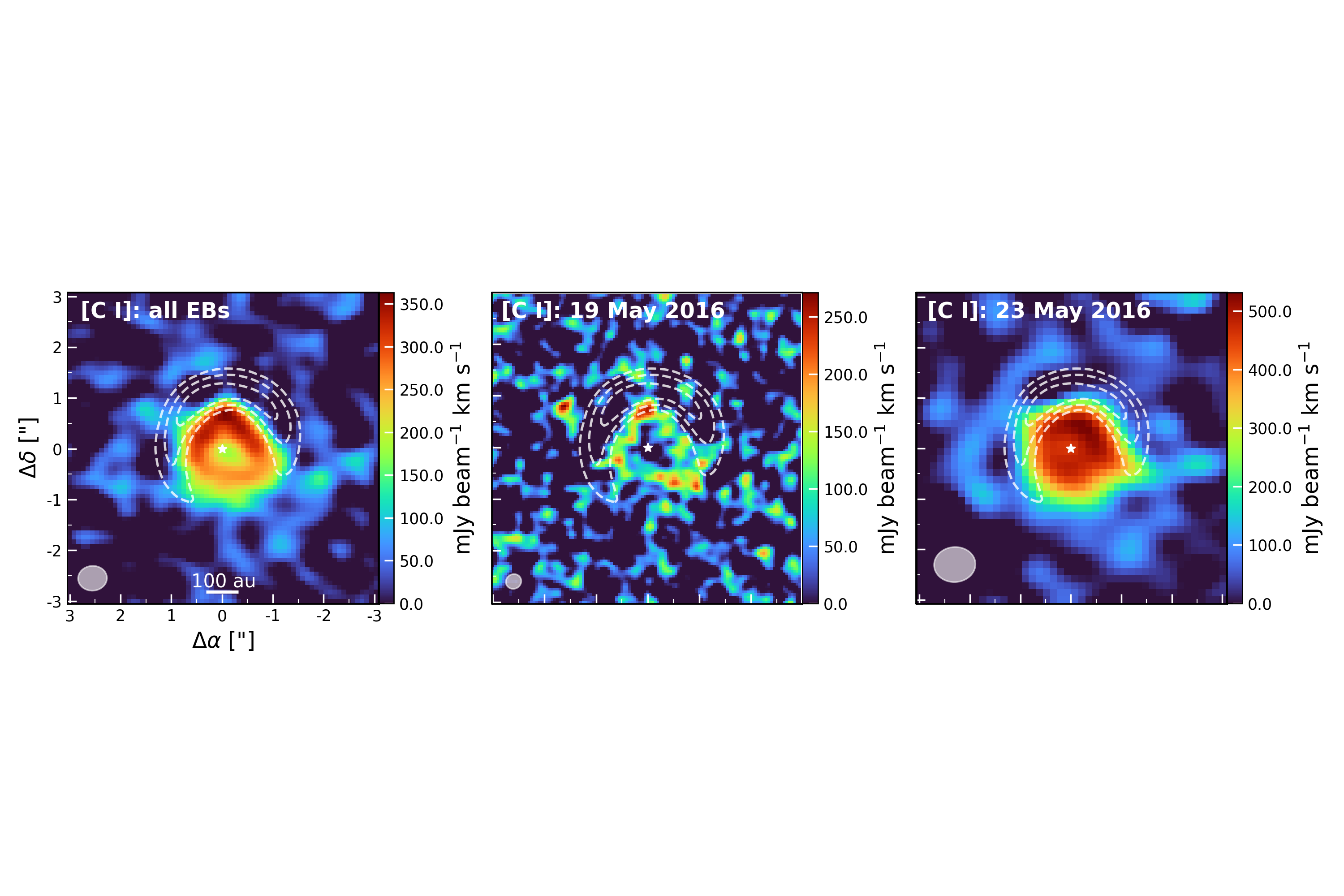}
    \caption{The different moment-zero maps of the [\ce{C} I] $^3$P$_1$-$^3$P$_0$ transition. From left to right are shown all three execution blocks, the higher angular resolution observations from the 19$^\textnormal{th}$ of May 2016 and the other two observations from the 23$^\textnormal{rd}$ of May 2016. The beams are indicated in the lower left and the white star in the center shows the inferred location of the host star and the white, dashed contours indicate the location of the continuum emission.}
    \label{fig:Gallery-CI}
\end{figure*}

Figure \ref{fig:Gallery-CI} shows the different moment-zero maps of the [\ce{C} I] $^3$P$_1$-$^3$P$_0$ transition, as mentioned in Section \ref{sec:2015.1.01137.S}. As apparent from the moment-zero maps, the [\ce{C} I] emission shows, similar to the rarer \ce{CO} isotopologues, a cavity and a ring-like emission structure. As opposed to, for example, the \ce{^{13}CO} and \ce{C^{18}O} emission, the [\ce{C} I] ring appears to be smaller and (almost) fully located inside the dust cavity. The presence of both \ce{^{12}CO} and \ce{[C I]} shows that the gas cavity is smaller than the dust cavity, as has been found for other transitional disks (e.g. \citealt{vdMarelEA16,LeemkerEA22,WoelferEA22}).

\subsubsection{Weak line emission}
We have used the \textit{GoFish} spectral stacking technique for our detections of the \ce{DCO^+} $J$=4-3 and the \ce{H_2CO} $J$=4$_{2,3}$-3$_{2,2}$ and $J$=4$_{2,2}$-3$_{2,1}$ transitions. To also infer the radial location of these transitions, we have acquired integrated spectra between 0.0" and 3.0" using a width of twice the size of the minor axis of the beam and steps of one times the minor axis. The resulting spectra for the \ce{DCO^+} transition are displayed in the top panel of Figure \ref{fig:GFSpectra}, where $>3\sigma$-Keplerian signatures are found between 1.25"-1.75" ($\sim$200-275 au). For the \ce{H_2CO} transitions, displayed in the second and third panels of Figure \ref{fig:GFSpectra}, respectively, $>3\sigma$-detections can be found between 1.5"-2.0" ($\sim$235-315 au) and 1.0"-2.0" ($\sim$150-315 au). \\
\indent For comparison, we have also included the \textit{GoFish} spectra for the \ce{SO} $J$=1$_2$-0$_1$ \ce{CH_3OH} $J$=7$_{1,6}$-7$_{0,7}$ transitions, taken both over the same region as the \ce{H_2CO} $J$=4$_{2,2}$-3$_{2,1}$ transition. These spectra can be found in Figure \ref{fig:GFSpectra-NDs}. For both species no 3$\sigma$-signals are found, providing proof that emission from neither \ce{SO} nor \ce{CH_3OH} is presently detected.

\begin{figure*}[]
    \centering
    \includegraphics[width=\textwidth]{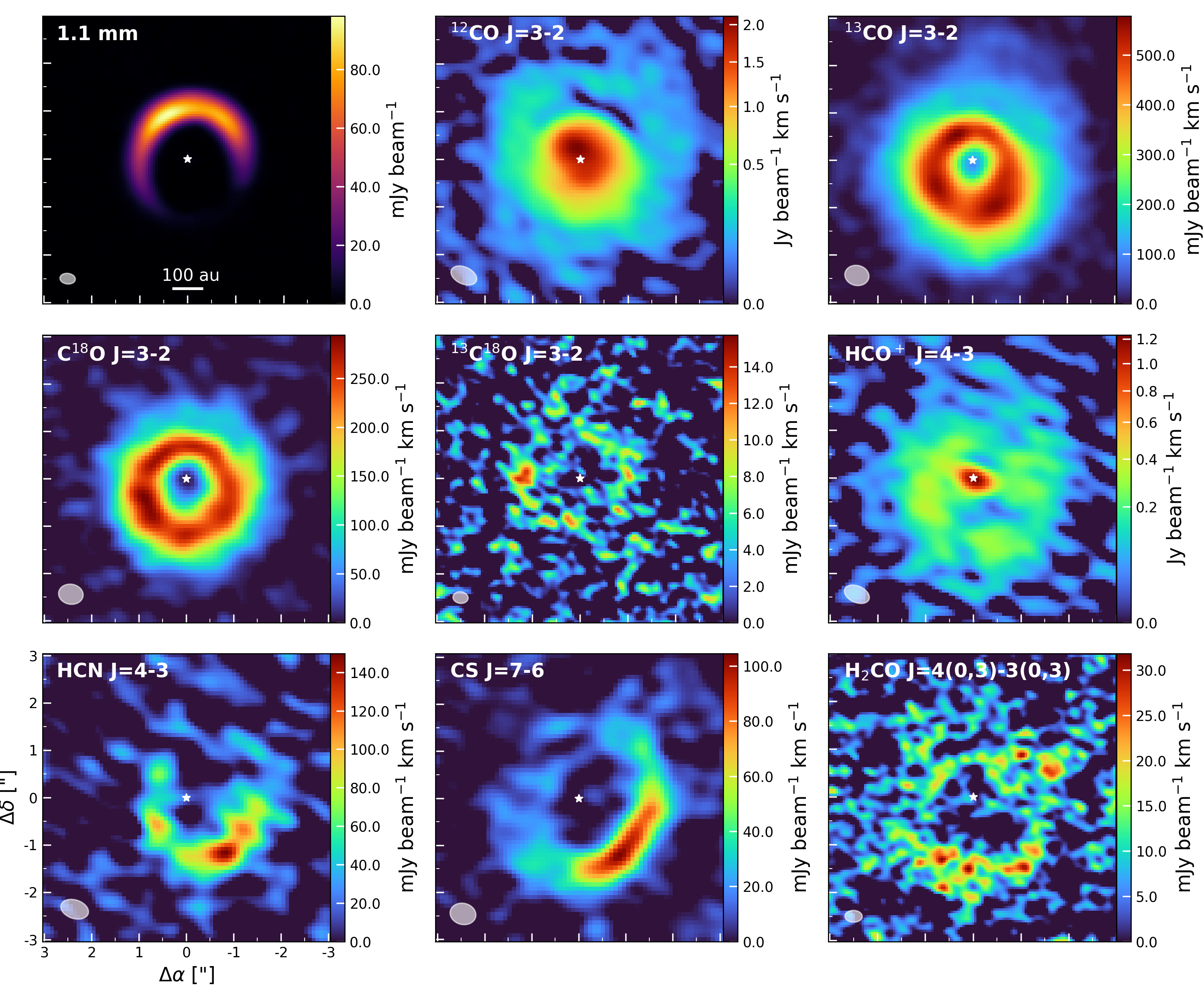}
    \caption{Moment-zero maps of the continuum (upper left) and the majority of the detected molecules. The beams are indicated in the lower left and the yellow star in the center shows the inferred location of the host star based on the position of the inner disk. The \ce{^{12}CO} and \ce{HCO^+} colour-maps are displayed using a power-law scaling with power 0.5.}
    \label{fig:Gallery}
\end{figure*}

\subsubsection{Rotational diagram analysis of \ce{H_2CO}}
\begin{figure}[ht!]
    \centering
    \includegraphics[width=\columnwidth]{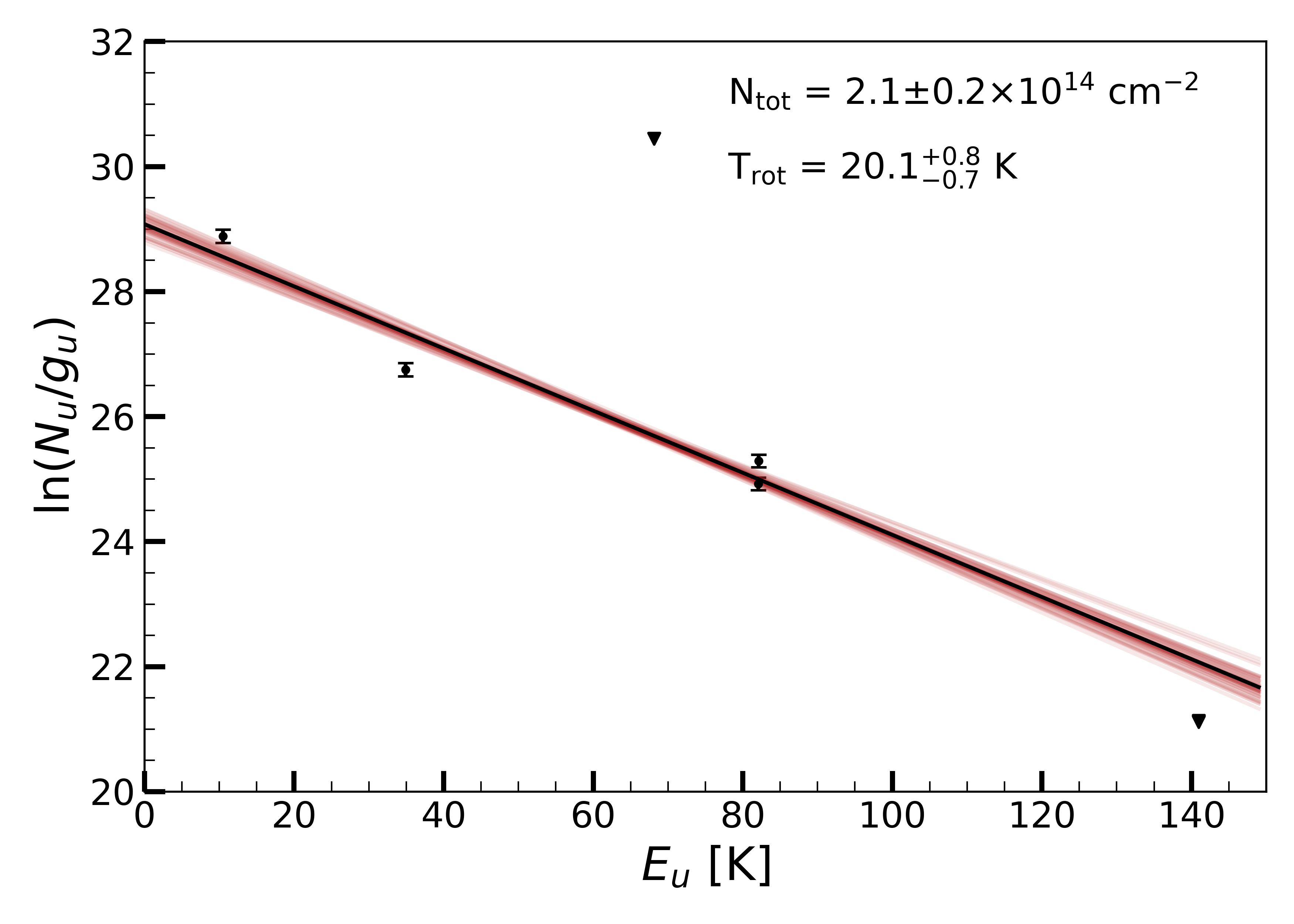}
    \caption{Results of the rotational diagram analysis for \ce{H_2CO}. The triangles display the acquired column densities for the non-detected lines, which have not been taken into account during the analysis. Furthermore, the faint, red lines are solutions randomly sampled from the posterior distributions.}
    \label{fig:H2CO_RDA}
\end{figure}

\indent Using the four detections of \ce{H_2CO} the total column density and rotational temperature were constrained using the rotational diagram analysis. We have used the disk integrated fluxes of the detected molecules listed in Table \ref{table:Molecules} and for the solid angle we have used an annulus extending from $\sim$115 au to $\sim$395 au ($\Omega$=4.7$\times$10$^{-11}$ sr), which encloses the emission of the strongest \ce{H_2CO} transition. The resulting fit can be found in Figure \ref{fig:H2CO_RDA}, yielding a total column density of $N_\textnormal{tot}$=(2.1$\pm$0.2)$\times$10$^{14}$ cm$^{-2}$ and a rotational temperature of $T_\textnormal{rot}$=20.1$^{+0.8}_{-0.7}$ K. The posterior distributions are displayed in Figure \ref{fig:H2CO_RDA_Corner}.

\subsubsection{Comparison with IRS 48}
The observed molecular emission is noticeable different compared to what has been observed in the IRS~48 disk. Most obvious are the different observed molecular species and morphologies. For example, in IRS~48 \ce{SO}, \ce{SO_2}, \ce{H_2CO} and \ce{CH_3OH} are observed, while \ce{CS} is not detected, whereas \ce{HCN}, \ce{CS} and \ce{H_2CO} are observed in HD~142527, with non-detections of \ce{SO}, \ce{SO_2} and \ce{CH_3OH}. The differences in the observed emission morphologies are, however, the most striking: in IRS~48 all molecules are more-or-less co-located with the continuum emission, while the molecules in HD~142527 all peak away from the dust trap, hinting that different chemical processes are responsible for molecular emissions in each disk. The HD~142527 disk also differs from the IRS~48 disk in regarding the overall dust and gas mass, which are discussed in the next subsections.

\subsection{Dust mass} \label{sec:DustMass}
Using the continuum image in band 6 (top left in Figure \ref{fig:Gallery}) we have estimated the dust mass of the mm-sized grains in the HD 142527 disk. The continuum flux ($F_\nu$) can be related to the dust mass ($M_\textnormal{dust}$) under the assumption of optically thin continuum emission \citep{Hildebrand83}:
\begin{align} \label{eq:DustMass}
    M_\textnormal{dust} = \frac{F_\nu d^2}{\kappa_\nu B_\nu(T_\textnormal{dust})}.
\end{align}
Here, $d$ denotes the distance, $\kappa_\nu$ presents the dust opacity and $B_\nu(T_\textnormal{dust})$ is the Planck function, evaluated at a certain dust temperature ($T_\textnormal{dust}$). The continuum flux was extracted from the image using a circular aperture with a radius of 1.625", yielding a flux density of $F_\nu$=1.9$\pm$0.2 Jy at a rest frequency of $\nu$=277.51 GHz. \\
\indent The dust opacity has been taken to be 10 cm$^2$~g$^{-1}$ at a frequency of 1000 GHz \citep{BeckwithEA90}, 
\begin{align}
    \kappa_\nu = 10\left(\frac{\nu}{1000\textnormal{ GHz}}\right)^\beta \textnormal{ cm$^{2}$ g$^{-1}$}.
\end{align}
Following \citet{AnsdellEA16,CazzolettiEA19,StapperEA22} we have used a power-law scaling of $\beta$=1. At the rest frequency of the continuum image, we obtain a dust opacity of 2.8 cm$^2$~g$^{-1}$. \\
\indent The dust temperature has been taken to be equal to the peak value found in the continuum brightness temperature map. The brightness temperatures were obtained by making use of the inverse Planck function,
\begin{align} \label{eq:InversePlanck}
    T = \frac{h\nu}{k_\textnormal{B}}\log\left[\frac{2h\nu^3}{F_\nu c^2}+1\right]^{-1}
\end{align}
The peak brightness temperature was found to be $T_\textnormal{dust}$=28.2 K at $\sim$175 au. \\
\indent Using this method, the minimum dust mass of the mm-sized dust grains was found to be $M_\textnormal{dust}$=(1.5$\pm$0.2)$\times$10$^{-3}$ M$_\odot$. The provided uncertainty here follows from taking a 10\% ALMA flux calibration error on the measured flux. The derived dust mass is a factor $\sim$25 higher compared to that of IRS~48. Using the flux given in Table 3 of \citet{FvdM20} and a dust temperature equal to the brightness temperature of 27 K \citep{vdMarelEA21}, we derive a dust mass for IRS~48 of (6.2$\pm$0.6)$\times$10$^{-5}$ M$_\odot$.

\subsection{Gas mass} \label{sec:GasMass}
To provide an estimate on the gas mass, we have made use of the rare, optically thin \ce{^{13}C^{18}O} $J$=3-2 transition. As the \ce{^{13}C^{18}O} emission only shows an emission ring, we use the \ce{C^{18}O} $J$=3-2 transition to cover the radii ($\lesssim$50 au and $\gtrsim350$ au) where \ce{^{13}C^{18}O} is not detected. To be able to work with the different resolution of the different datasets, we tapered the \ce{^{13}C^{18}O} emission to the beam of the \ce{C^{18}O} transition. We acquired radial profiles extending to $\sim$415 au (beyond which \ce{C^{18}O} is no longer detected) for both transition using \textit{GoFish} as described in Section \ref{sec:RadialProfile}. \\
\indent To estimate the mass, we first converted both radial profiles to radial column density profiles. Using the isotolopogue ratios, \ce{^{12}\ce{C}}/\ce{^{13}}\ce{C}=68$\pm$15 \citep{MilamEA05} and \ce{^{16}\ce{O}}/\ce{^{18}\ce{O}}=557$\pm$30 \citep{Wilson99}, we converted the column densities into \ce{^{12}CO} column densities, assuming optically thin emission. Next, we combined the \ce{^{12}CO} column densities obtained for \ce{C^{18}O} and \ce{^{13}C^{18}O} at the aforementioned radii. Subsequently, we converted the combined \ce{^{12}CO} column densities into \ce{H_2} column densities assuming a ratio of \ce{H_2}/\ce{^{12}CO}=10$^4$ \citep{BWEA17}. \\%\citep{Dickman78,LacyEA17}. \\
\indent Assuming a gas molecular weight of 2.4, considering hydrogen and helium, we obtained radial gas surface densities, which, through an integration, were converted to the gas mass. By summing over all the radial bins we acquired a gas mass estimate of $M_\textnormal{gas}$=(1.6$\pm$0.6)$\times$10$^{-2}$ M$_\odot$. Combining our estimates of the gas and dust mass, we obtain a gas-to-dust ratio of 10.8$\pm$4.2. The estimated gas mass is, just as the found dust mass, a factor of $\sim$100 larger than the gas mass found for IRS~48, using optically thin \ce{C^{17}O} emission ($\sim$1.4$\times$10$^{-4}$, \citealt{BrudererEA14}).\\
\indent To justify the choice of using the \ce{^{13}C^{18}O} emission for the calculation of the gas mass in the emission ring, we investigated the optical depth of both \ce{^{13}C^{18}O} and, subsequently, \ce{C^{18}O}. Using the \ce{^{13}C^{18}O} radial profile (further discussed in Section \ref{sec:CO-snowline}) we inferred a peak optical depth of $\tau_{\textnormal{\ce{^{13}C^{18}O}}}\simeq$0.03 at $\sim$180 au. Using the peak fluxes of the \ce{C^{18}O} $J$=3-2 and the \ce{^{13}C^{18}O} transitions, we infer a flux ratio of $F_\textnormal{\ce{C^{18}O}}/F_\textnormal{\ce{^{13}C^{18}O}}\simeq$23 and an optical depth for \ce{C^{18}O} of $\tau_\textnormal{\ce{C^{18}O}}\simeq$0.7, pointing towards the \ce{C^{18}O} emission being moderately optically thick. However, if we use the \ce{^{12}\ce{C}}/\ce{^{13}}\ce{C} isotopologue ratio, as done above, we infer an optical depth for \ce{C^{18}O} of $\tau_\textnormal{\ce{C^{18}O}}\simeq$2.16. From this analysis, we confirm that the more common (\ce{^{12}CO}, \ce{^{13}CO} and \ce{C^{18}O}) must be optically thick and using their observations to determine the gas mass would only yield a lower limit. 

\subsection{The \ce{CO} snowline} \label{sec:CO-snowline}
\begin{figure}[ht!]
    \centering
    \includegraphics[width=\columnwidth]{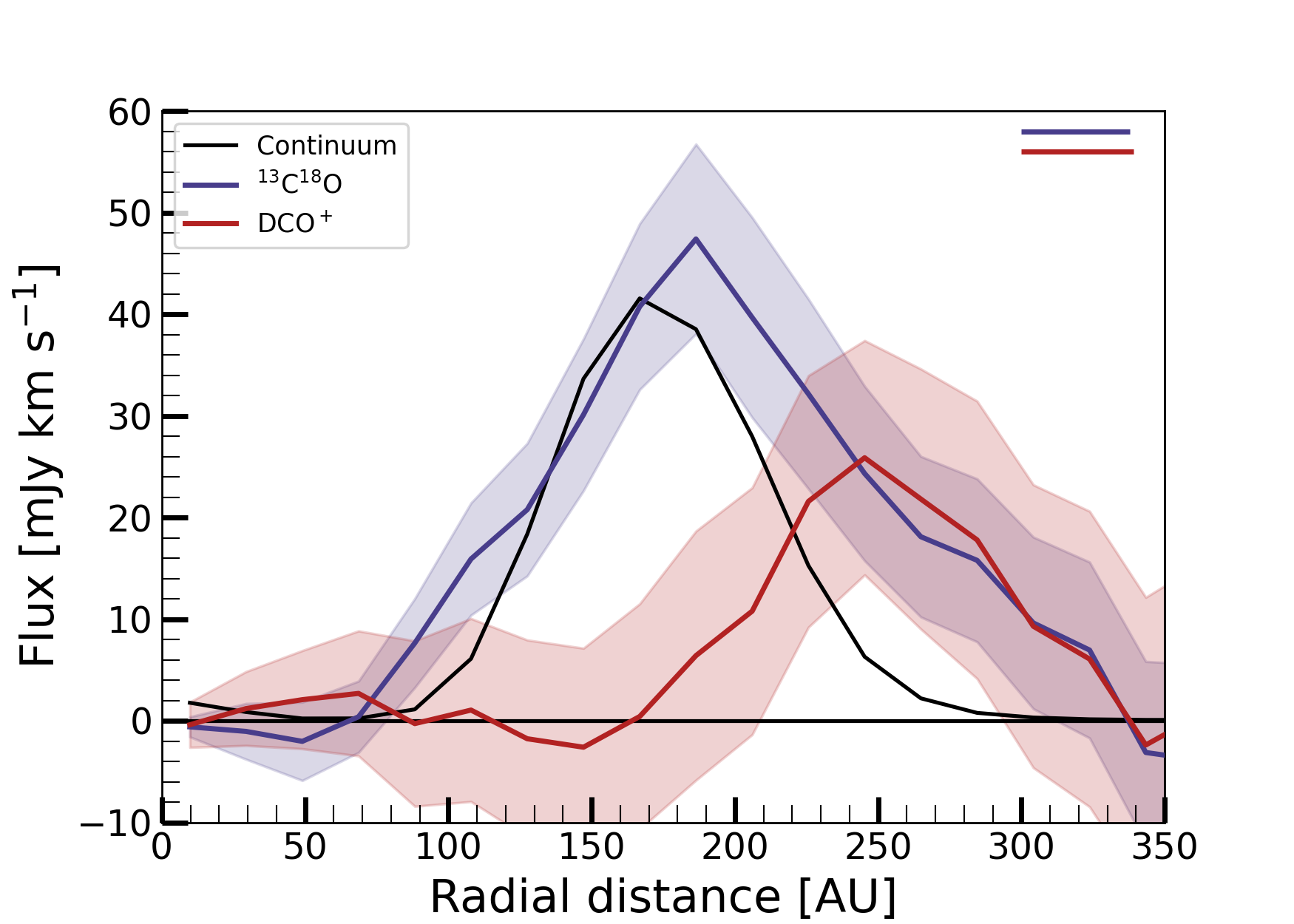}
    \caption{The azimuthally averaged, deprojected radial profiles of \ce{^{13}C^{18}O} (in blue) and \ce{DCO^+} (in red), shown together with the continuum emission (black profile). The shaded areas are the errors on the profiles. The two horizontal bars in the top right indicate the size of the minor axis for \ce{^{13}C^{18}O} (blue) and \ce{DCO^+} (red).}
    \label{fig:CO-Snowline-RP}
\end{figure}

\begin{figure}[h!]
    \centering
    \includegraphics[width=\columnwidth]{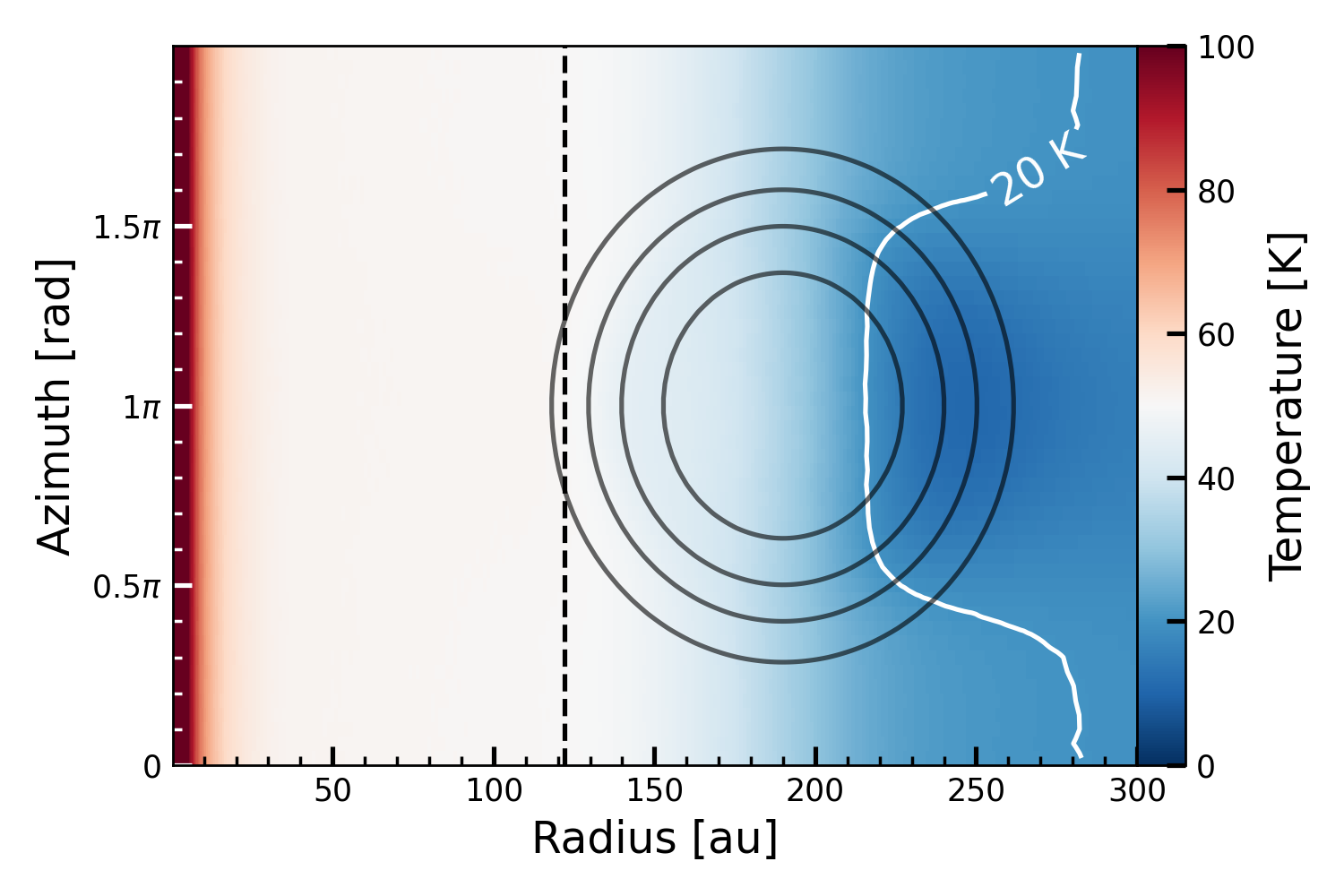}
    \caption{Midplane dust temperature structure of the RADMC-3D model. The black contours indicate the location of the asymmetric dust trap, while the white contour shows where the temperature becomes 20 K. The black dashed line indicates where a temperature of 20 K would be reached assuming a power-law for the midplane temperature \citep{vdMarelEA21b}.}
    \label{fig:RADMC3D-Temp}
\end{figure}

The \ce{CO} snowline is of great importance for the formation of \ce{H_2CO} and \ce{CH_3OH}, as both molecules share an ice-phase formation path: the hydrogenation of CO (e.g., \citealt{WK02,FuchsEa09}). The location of the snowline can be approximated using various molecules, like \ce{N_2H^+} and \ce{DCO^+} (e.g., \citealt{MathewsEA13,QiEA13,vtHoffEA17,CarneyEA18,QiEA19}). Here, we use a combination of the optically thin emission of \ce{^{13}C^{18}O} and the weak detection of \ce{DCO^+}. The optically thin \ce{^{13}C^{18}O} emission allows us to trace emission closer to the midplane and, hence, provides us information where \ce{CO} might be frozen out on the midplane. \ce{DCO^+} on the other hand provides an approximation of the \ce{CO} snowline, as its formation depends on the presence of \ce{CO} and its abundance enhances around the CO snowline. \\
\indent \ce{DCO^+} forms through the ion-molecule reaction \citep{Wootten87}, \ce{H_2D^+}+\ce{CO}$\rightarrow$\ce{DCO^+}+\ce{H_2}. The formation of the parent molecule \ce{H_2D^+} (\ce{HD}+\ce{H_3^+}$\leftrightarrow$\ce{H_2D^+}+\ce{H_2}+$\Delta$E; \citealt{RM00}) requires cold temperatures, which should ensure the depletion of \ce{CO} through freeze-out. However, as \ce{CO} is also a parent molecule of \ce{DCO^+}, \ce{CO} must not be frozen-out completely and a balance between all the aforementioned reactions is required for the formation of \ce{DCO^+}. At higher temperatures ($>$30 K), \ce{DCO^+} dominantly forms through reactions of \ce{CH_2D^+} or \ce{CH_4D^+} \citep{MillarEA89,RoueffEA13,CarneyEA18}. However, due to the large radial distance of the observed \ce{DCO^+} emission and the low gas temperature at this distance ($T_\textnormal{\ce{H_2CO}}\simeq$20 K), we expect the \ce{DCO^+} to have formed through the colder \ce{H_2D^+} route in HD~142527.  \\
\indent To trace the \ce{CO} snowline we have created radial profiles of \ce{^{13}C^{18}O} and \ce{DCO^+} using \textit{GoFish} (see Section \ref{sec:RadialProfile}). The acquired radial profiles, shown together with the continuum radial profile, are displayed in Figure \ref{fig:CO-Snowline-RP}. From Figure \ref{fig:CO-Snowline-RP} it is clear that the \ce{^{13}C^{18}O} is originating from about the same region as the continuum emission. As \ce{^{13}C^{18}O} is the most optically thin \ce{CO} isotopologue observed here, it is the isotopologue that traces closest to the midplane. The \ce{^{13}C^{18}O} thus already hints that \ce{CO} is not frozen out at the location of the dust trap. Furthermore, Figure \ref{fig:CO-Snowline-RP} shows that \ce{DCO^+} peaks beyond the dust trap, allowing us to confirm that the \ce{CO} snowline must lie beyond the dust trap, likely located at a radius of $\sim$300 au. We note that multiple \ce{DCO^+} rings have been detected in some disks \citep{ObergEA15,HuangEA17,SalinasEA17}, where the second ring is inferred to be beyond the midplane \ce{CO} snowline. These outer \ce{DCO^+} rings are expected to be the result of either reactions involving non-thermally desorped \ce{CO} or a thermal inversion in the midplane, following radial, grain growth and settling \citep{Cleeves16,FacchiniEA17}. If either of these scenarios would also hold true for the HD~142527 disk, the inferred location of the \ce{CO} snowline should be treated as an outer radius. The relationship between \ce{DCO+} and the CO snowline is therefore non trivial and we require observations of the preferred CO snowline tracer \ce{N_2H^+} \citep{vtHoffEA17} to confirm our hypothesis for the HD~142527 disk. \\
\indent Furthermore, the conclusion of the snowline lying at the location on/beyond the dust trap is supported by the brightness temperature of \ce{^{12}CO} transitions at the cavity. As the brightness temperature reaches values of $\sim$40 K at the cavity (compared to the peak continuum brightness temperature of $T_\textnormal{dust}\sim$30 K), the gas temperature at this location, albeit higher up in the disk's atmosphere, will have a similar value, considering the \ce{^{12}CO} transitions are optically thick. The derived brightness temperature agrees with those found by \cite{GargEA21} ($\sim$40-50 K) using high-resolution \ce{CO}-isotopologue observations. \\ 
\indent We have used \textit{RADMC-3D}\footnote{\textit{RADMC-3D}: \url{https://www.ita.uni-heidelberg.de/~dullemond/software/radmc-3d/}} \citep{RADMC3D} to create a dust radiative-transfer model (see Section \ref{sec:RADMC3D} for a full description of our model), which allowes us to investigate the location of the \ce{CO} snowline in more detail. As can be seen in Figure \ref{fig:RADMC3D-Temp}, the \ce{CO} snowline is located just beyond the dust trap, as predicted by our analysis of the \ce{^{13}C^{18}O} and \ce{DCO^+} emission. The \ce{CO} snowline is located at smaller radii at the location of the dust trap ($\sim$220 au) compared to the other side of the disk ($\sim$280 au), indicating that the large dust concentration lowers the temperature. Additionally, Figure \ref{fig:RADMC3D-TempSides} displays the modelled midplane temperature of the dust at the peak location of the dust trap and at the other side of the disk, allowing for a better comparison of the temperature differences. \\
\indent There is, however, one caveat tied to the observational analysis of the \ce{CO} snowline. Due to the low SNR of the \ce{^{13}C^{18}O} and \ce{DCO^+} image cubes, we cannot constrain whether the emission of both molecules is symmetric. If the emission is asymmetric, it is possible that \ce{CO} snowline can be found at smaller radial locations at the location of the dust trap (as suggested by our \textit{RADMC-3D} model), due to the presence of the large concentration of mm-sized dust particles, compared to the other side of the disk. Deeper observations of \ce{^{13}C^{18}O} and \ce{DCO^+} are required to investigate the azimuthal distribution and to better constrain the location of the \ce{CO} snowline. In addition, observations of \ce{N_2H^+}, the other \ce{CO} snowline tracer, could further help constraining the location of the \ce{CO} snowline. \\
\indent Detections of the rarer \ce{CO}-isotopologues (\ce{^{13}C^{18}O} and \ce{^{13}C^{17}O}) in the disks of MWC~480 \citep{LoomisEA20} and HD~163296 \citep{ABoothEA19} have been linked to enhanced \ce{CO}/\ce{H_2} abundances within the \ce{CO} snowline due to pebble drift \citep{KrijtEA18,KrijtEA20,ZhangEA19,ZhangEA20b,ZhangEA21}. Interestingly, the HD~142527 disk contains a large dust cavity and the observed \ce{^{13}C^{18}O} emission forms a ring rather than the compact emission observed for both other disks. However, as we expect the \ce{CO} snowline to be located beyond the dust trap, we might still observe enhanced \ce{CO}/\ce{H_2} abundances as observed for the other disks, despite the different emission morphologies.

\section{Discussion}
\subsection{Observed asymmetries}
The observed molecular asymmetries, most notable in the \ce{HCN} and \ce{CS} transitions, are striking. As mentioned before, \cite{vdPlasEA14} provided two possible explanations for the asymmetries: continuum over-subtraction or freeze-out. Here, we explore these possibilities further.

\subsubsection{Influence of the mm-sized dust} \label{sec:mmdust}
\begin{figure*}
    \centering
    \includegraphics[width=0.8\textwidth]{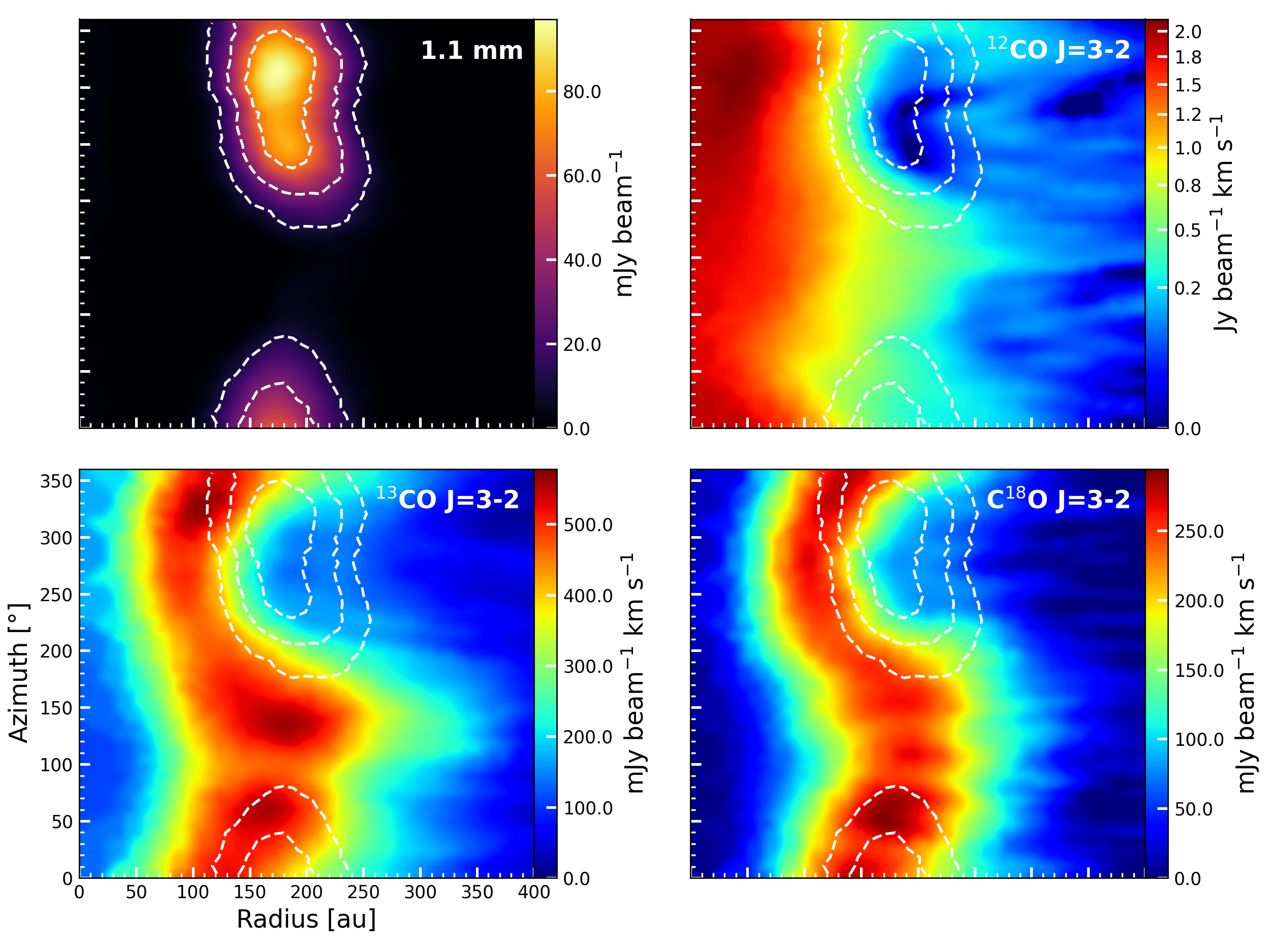}
    \caption{Deprojected moment-zero maps of the continuum and the major \ce{CO}-isotopologues. As in Figure \ref{fig:Gallery}, the \ce{^{12}CO} emission is presented using a power-law scaling with a power of 0.5. The white, dashed contour-lines indicate the location of the continuum emission.}
    \label{fig:DepGallery}
\end{figure*}

Figure \ref{fig:DepGallery} presents the deprojected moment-zero maps of the continuum emission and the main \ce{CO}-isotopologues (\ce{^{12}CO}, \ce{^{13}CO} and \ce{C^{18}O}), to highlight their azimuthal distributions. None of the isotopologues show emission at the peak location of the dust trap. As discussed in Section \ref{sec:CO-snowline}, the \ce{CO} snowline lies beyond the dust trap, indicating that the lack of emission in the other \ce{CO}-isotopologues must be the result of a continuum over-subtraction. Continuum over-subtraction can be the result of the overestimation of the dust emission that should be removed, as the absorption of dust emission at molecular line frequencies is not taken into account \citep{BoehlerEA17,WeaverEA18,BoehlerEA21}. An additional effect follows from the absorption of emission from the back-side of the disk by dust particles located in the midplane, which can theoretically result in a depletion of the emission by up to 50\% \citep{IsellaEA18,RabEA20,BoehlerEA21,RosottiEA21}. \\
\indent For the other molecules (\ce{HCN}, \ce{CS} and \ce{H_2CO}), continuum over-subtraction can also play a role in the lack of observed emission. However, as these molecules all have higher freeze-out temperatures compared to \ce{CO} and compared with the estimated peak brightness temperature of the continuum $T_\textnormal{dust}\simeq$30 K, freeze-out cannot yet be ruled out.  

\subsubsection{Influence of misaligned, warped inner disk}
\begin{figure*}[t]
    \centering
    \includegraphics[trim=0cm 0.5cm 0cm 0.5cm, width=0.8\textwidth]{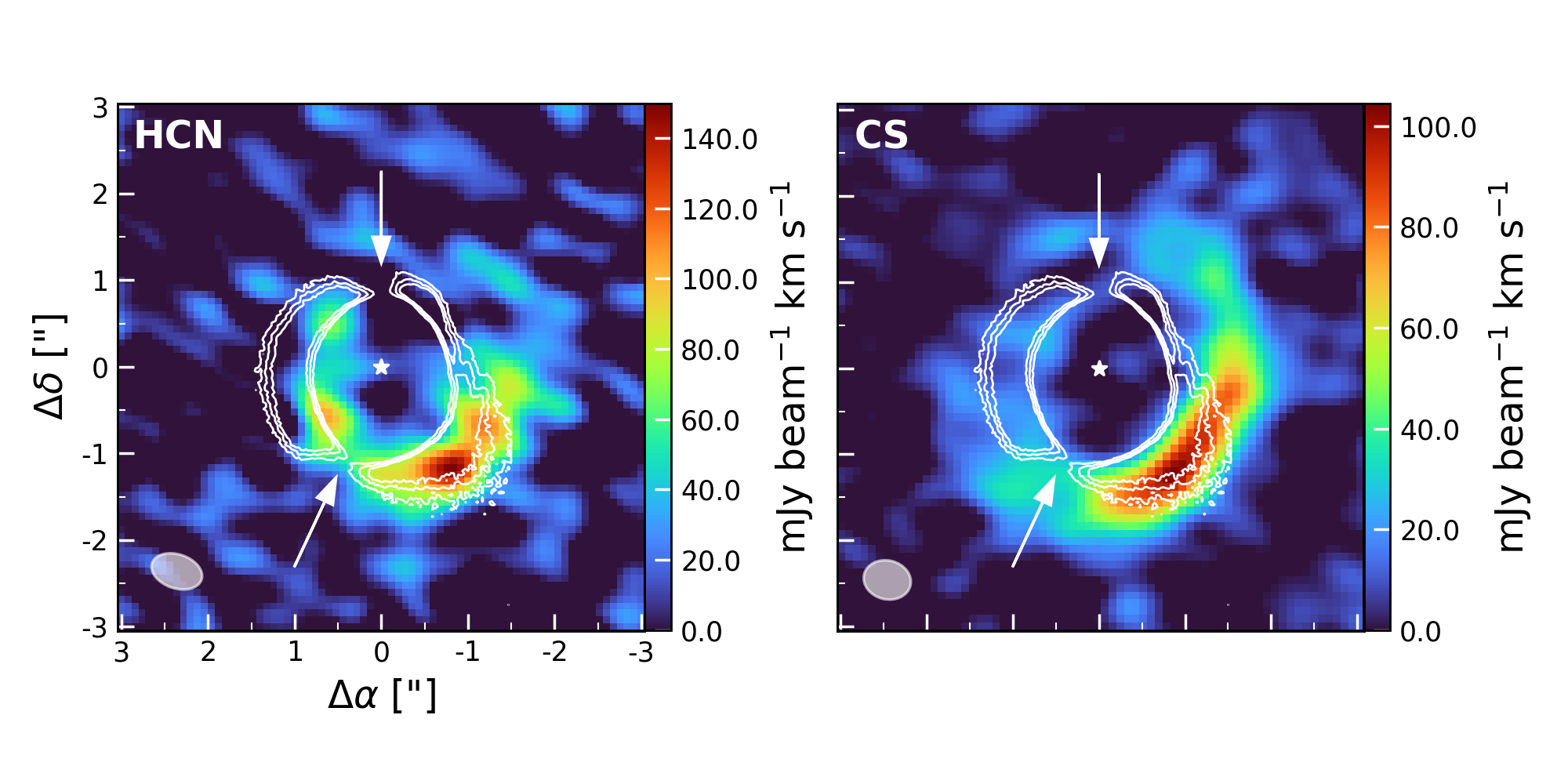}
    \caption{The \ce{HCN} $J$=4-3 and \ce{CS} $J$=7-6 moment-zero maps. The white contours show the scattered light emission from the $\mu$m-sized dust grains, obtained with the NACO instrument in the Ks-band, and the white arrows indicate the location of the shadows.}
    \label{fig:ScatteredLight}
\end{figure*}

\begin{figure*}[]
    \centering
    \includegraphics[width=0.8\textwidth]{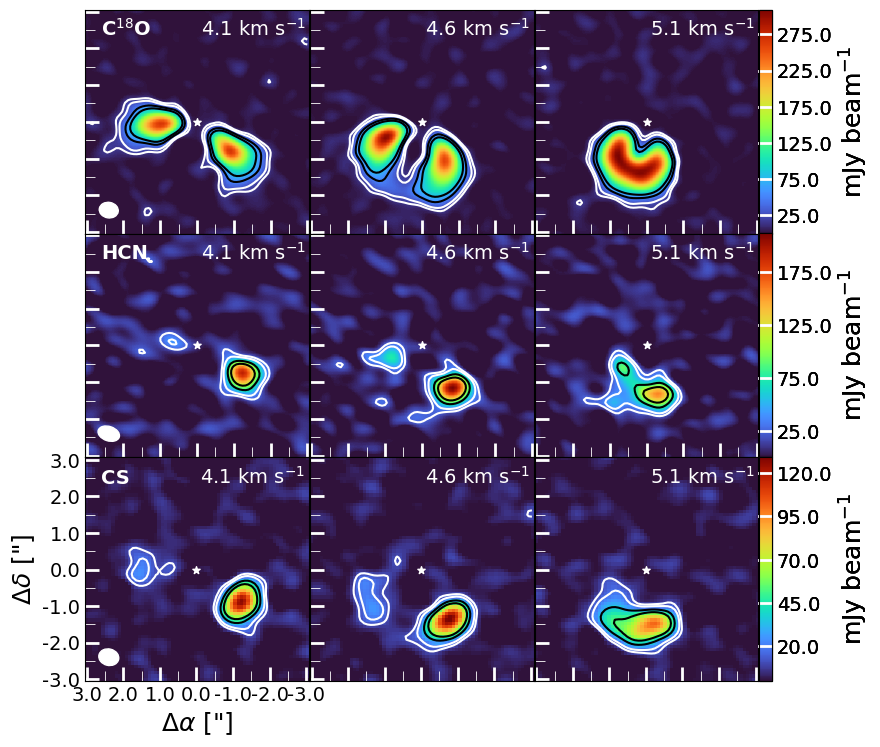}
    \caption{Zoom-in on the 4.1-5.1 km s$^{-1}$ region of the channel maps of \ce{C^{18}O} $J$=3-2 (top), \ce{HCN} $J$=4-3 (middle) and \ce{CS} $J$=7-6 transitions. The \ce{C^{18}O} channel maps show symmetric emission, while the channel maps of \ce{HCN} and \ce{CS} are both clearly asymmetric. The resolving beam is displayed in the lower left and the white star in the centre indicates the inferred location of the host star. The white contours indicate the $3\sigma$- and $5\sigma$-RMS levels, while the black contours show the $10\sigma$- and $15\sigma$-RMS levels.}
    \label{fig:ZoomInChanMaps}
\end{figure*}

For \ce{HCN} and \ce{CS} we would like to propose another scenario to explain the observed asymmetries. As can be seen in their channel maps (Figure \ref{fig:ChanMaps}), both molecules show emission enclosing the dust trap, which means that freeze-out at the dust trap alone can also not explain the observed asymmetry. One possibility left to explore is the influence of the misaligned, warped inner disk. As shown by \citet{YoungEA21}, a misaligned, warped inner disk can cast shadows on certain parts of the disk and, subsequently, lower the temperature.\\ 
\indent Figure \ref{fig:ScatteredLight} shows the \ce{HCN} and \ce{CS} moment-zero maps with the scattered light emission (\citealt{AvenhausEA14}, S. de Regt \textit{priv. comm.}) overplotted. The southern shadow is visible in the emission of both \ce{HCN} and \ce{CS}, most notably in the \ce{HCN} emission. The northern shadow, visible in the scattered light emission, also appears to be present in both channel maps. For both molecules, there is a clear turning-point between the bright west-side of the disk and less bright east-side of the disk, which appears to occur at the location of the northern shadow. \\
\indent Here, we would like to propose that the observed asymmetric emission of \ce{HCN} and \ce{CS} is the result of the eastern-side of the disk being shadowed by the inner disk, yielding a temperature difference across the disk. This is also further supported by the notion that the channel maps display asymmetries in the 4.1-5.1 km s$^{-1}$ range (see Figure \ref{fig:ZoomInChanMaps}), while the \ce{C^{18}O} channel map is fully symmetric. From chemical models, we expect \ce{HCN} and \ce{CS} to come from the same layer above the midplane (see, for example, \citealt{AgundezEA18}), explaining why both molecules would similarly be affected by the misaligned inner disk. The effects of the misalignment are also expected to be visible in the \ce{C^{18}O} emission \citep{WalshEA17,YoungEA21}, as the emission is optically thick and hence tracing the gas temperature. Future work is required to investigate the proper orientation of the inner disk with respect to the other disk and to confirm whether this proposition holds true. 

\subsection{The elemental \ce{C}/\ce{O}-ratio} \label{sec:CO-ratio}
The ratio between the column densities of \ce{CS} and \ce{SO} can be used as a proxy for the volatile \ce{C}/\ce{O}-ratio (e.g. \citealt{LeGalEA19,LeGalEA21}). For example, \citet{LeGalEA21} derived lower limits on the \ce{CS}/\ce{SO}-ratio between $\sim$4 and 12 for the disks in the MAPS program. By taking the ratio of the column density between the \ce{CS} $J$=7-6 transition and the $3\sigma$ upper limit for the \ce{SO} $J$=1$_2$-0$_1$, displayed in Table \ref{table:Molecules}, we obtain \ce{CS}/\ce{SO}$>$1. As we have observations for both \ce{CS} and \ce{HCN}, and no observations of other oxygen-bearing species, such as \ce{SO}, \ce{SO_2} or \ce{NO}, this agrees with an expected gas-phase [\ce{C}]/[\ce{O}]$\gtrsim$1. To confirm and further explore the \ce{C}/\ce{O}-ratio, deeper SO observations of brighter lines (e.g., the $J$=5$_{3,3}$-4$_{2,2}$ transition observed in IRS~48, \citealt{ABoothEA21} or the $J$=7$_7$-6$_6$ transition observed in HD~100546 \citealt{ABoothEA22}) are required.

\subsection{The formation of \ce{H_2CO}} \label{sec:H2COForm}
The ratio between the derived total column density of \ce{H_2CO} and the lowest upper limit of \ce{CH_3OH} (using the $J$=7$_{1,6}$-7$_{0,7}$ transition) provides insight of the formation of \ce{H_2CO}. Taking the ratio, we find \ce{H_2CO}/\ce{CH_3OH}$\gtrsim$114$\pm14$ As expected from the non-detections of \ce{CH_3OH}, \ce{H_2CO} is much more abundant. Taking into account that the molecules share a grain-surface formation route - the hydrogenation of \ce{CO} - it is very unlikely that the observed \ce{H_2CO} is only thermally desorbed from the ices. This is further supported by the found location of the \ce{CO} snowline and the low ($\sim$30 K) brightness temperature for the continuum emission. As the \ce{CO} snowline is located beyond the dust trap, we do not expect \ce{CO} to be frozen out at the grains in the dust trap. This means that \ce{H_2CO} (and \ce{CH_3OH}) cannot actively form on the grains through \ce{CO}-hydrogenation, but could have done so in a previous, colder phase, as was found for IRS~48 \citep{vdMarelEA21} and HD~100546 \citep{ABoothEA21b}. Furthermore, as we assume the \ce{^{12}CO} emission to be optically thick, the kinematic temperature can be approximated to be equal to the brightness temperature, $T_\textnormal{kin}\sim$40 K (see Section \ref{sec:CO-snowline}, \citealt{GargEA21}). As this is lower than the sublimation temperature of \ce{H_2CO} ($\sim$66 K \citealt{PenteadoEA17}), thermal desorption of \ce{H_2CO} can be ruled out.\\
\indent Subsequently, there are two options left to explain the observed \ce{H_2CO} emission: (1) through cold gas-phase, neutral-neutral reactions, mainly \ce{CH_3}+\ce{O}$\rightarrow$\ce{H_2CO}+\ce{H} and \ce{CH_2}+\ce{OH}$\rightarrow$\ce{H_2CO}+\ce{H} (e.g., \citealt{LoomisEA15,TvSEA21}), or (2) non-thermal desorption processes. The found rotational temperature for \ce{H_2CO} of $\sim$20 K hints that the observed emission, assuming that the densities are high enough for LTE, likely originates from the molecular layer (20-50 K), but close to the midplane (10-30 K) (e.g., \citealt{WalshEA10,PeguesEA20}). As the majority of the \ce{H_2CO} emission is found on the opposite side of the disk, away from the mm-sized dust trap, we expect the observed gaseous \ce{H_2CO} emission to mainly have formed through cold gas-phase chemistry. At the location of the dust trap, however, a contribution from non-thermal desorption processes cannot be ruled out. The above analysis agrees with that for TW Hya by \cite{TvSEA21}, where gas-phase formation can explain the observed \ce{H_2CO}, but ice-phase formation and contributions in the cold outer disk cannot be ruled out. In addition, \citet{PeguesEA20} found \ce{H_2CO} emission to originate from both inside and outside the \ce{CO} snowline for a small variety of disks, which suggests that \ce{H_2CO} forms through both the gas- and ice-phase formation pathways. \\
\indent Both proposed options for the observed \ce{H_2CO} emission, gas-phase formation and non-thermal desorption processes are consistent with the lack of \ce{CH_3OH}. First, \ce{CH3OH} is known to not form efficiently in the gas-phase and, second, \ce{CH_3OH} fragments during non-thermal desorption processes \citep{BertinEA16,CruzDiazEA16,WalshEA18,SantosEA23}.

\subsection{Comparisons with IRS~48}
As mentioned before, the differences between the disks of IRS~48 (stellar type A0, L=17.8$L_\odot$, \citealt{FvdM20}) and HD~142527 (stellar type F6, L=9.9$L_\odot$ \citealt{FvdM20}) are remarkable. Not only are different molecules observed in the disks, but the emission morphologies are also strikingly different. In addition, the estimations for both the gas and dust mass are much higher for HD~142527 compared to IRS~48. In the following subsections, we discuss the implications of these differences.

\begin{table}[ht!]
\caption{Column densities (in cm$^{-2}$) and molecular ratios in the IRS~48 and HD~142527 disks.}           
\label{table:AbundComps}     
    \begin{tabular}{c c c}
    \hline\hline  
    Molecules & IRS~48 & HD~142527 \\
    \hline
    \ce{CS} & $^{(a)}\leq$(5.9$\pm$0.6)$\times$10$^{13}$ & 1.6$\pm$0.2)$\times10^{13}$ \\
    \ce{SO} & $^{(a)}\geq$(4.7$\pm$0.5)$\times$10$^{15}$ & $<$(1.9$\pm$0.2)$\times10^{13}$ \\
    \ce{H_2CO} & (7.7$\pm$0.5)$\times$10$^{13}$ & (1.6$\pm$0.1)$\times$10$^{14}$ \\
    \ce{CH_3OH} & (4.9$\pm$0.2)$\times$10$^{14}$ & $<$(1.4$\pm$0.1)$\times10^{12}$ \\
    \hline
    \ce{CS}/\ce{SO} & $\leq$0.012 & $\geq$1.0 \\
    \ce{H_2CO}/\ce{CH_3OH} & $\sim$0.2 & $\gtrsim$114 \\
    \hline
    \end{tabular}
\tablefoot{The column densities have been calculated using different beams. For IRS~48 beam sizes of 0.20"$\times$0.16" have been used for both \ce{CS} and \ce{SO}, while the \ce{H_2CO} and \ce{CH_3OH} column densities were acquired using a rotational diagram analysis ($\Omega$=1.4$\times$10$^{-11}$ sr, \citealt{vdMarelEA21}). For the HD~142527 we have used sizes of 0.55"$\times$0.45" (\ce{CS}), 0.62"$\times$0.47" (\ce{SO}) and 0.31"$\times$0.24" (\ce{CH_3OH}), and we acquired the \ce{H_2CO} column density through a rotational diagram analysis ($\Omega$=4.1$\times$10$^{-11}$ sr, see Section \ref{sec:RDA}).\\
$^{(a)}$: the \ce{CS} and \ce{SO} column densities for IRS~48 have been taken at a rotational temperature of $T_\textnormal{rot}$=50 K \citep{ABoothEA21}.}
\end{table}

\subsubsection{\ce{CS} \& \ce{SO}}
\begin{figure}[]
    \centering
    \includegraphics[width=\columnwidth]{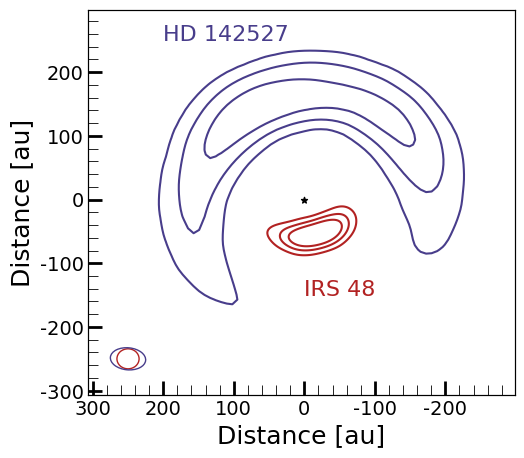}
    \caption{Contour maps of the continuum images for IRS~48 (band 7, in red) and HD~142527 (band 6, in blue). For IRS~48 the contours indicate the 25$\sigma$, 75$\sigma$ and 150$\sigma$ levels, while for HD~142527 the 50$\sigma$, 150$\sigma$ and 300$\sigma$ levels are shown. The IRS 48 continuum emission peaks at $\sim$175 au, whereas the HD 142527 continuum emission peaks at $\sim$60 au. The location of the host star is displayed by the black star and the resolving beams for both images are shown in the lower left in the respective colours. }
    \label{fig:SVDTs}
\end{figure}

\indent Using the (non-)detections of \ce{CS} and \ce{SO} the elemental C/O-ratios in both disks can be compared. For IRS~48, \citet{ABoothEA21} found a \ce{CS}/\ce{SO} ratio of $\leq$0.012 (see Table \ref{table:AbundComps}), while for HD~142527 we derived, as discussed in Section \ref{sec:CO-ratio}, a ratio \ce{CS}/\ce{SO}>1. The different ratios can be explained by considering the disks to be either 'warm' or 'cold' \citep{vdMarelEA21b}. The inner cavity walls of the warm disks, such as IRS~48, are irradiated and, subsequently, oxygen-bearing species can be returned to the gas-phase through sublimation of ices. For disks such as HD~142527, which have dust traps at much larger radii, the temperature is too low for the sublimation of these kinds of ices, yielding higher \ce{C}/\ce{O}-ratios. The location of the dust trap (displayed for both dust traps in Figure \ref{fig:SVDTs}) plays a major role in determining whether the temperature at the inner wall is high enough for the sublimation of ices. To elaborate, the IRS~48 dust trap is located at $\sim$60 au, whereas the HD~142527 dust trap is located at $\sim$150 au. The location of the dust trap must thus be taken into account when investigating the influence of the dust trap in the observable chemistry of the disk. While for both disks the dust trap is found to be located inside \ce{CO} snowline, sublimation is not observed in HD~142527, suggesting that the temperature might be lower compared to IRS~48. Potentially, the temperature at the cavity wall of IRS~48 is higher compared to that of HD~142527, allowing ices to be sublimation. Another possible explanation is that the HD~142527 disk might be less turbulent compared to the IRS~48 disk. As proposed in \cite{vdMarelEA21}, icy dust grains end up in warmer upper layers of the disk at the cavity wall due to both radial and vertical transportation, allowing for the sublimation of ices. If vertical transportation does not or do to a lesser extent take place in the HD~142527 disk, the grains do not end up in the warmer layers where ice sublimation can take place. \\
\indent Furthermore, the elemental ratios in the inner disks are also impacted in these disks, affecting the chemical reservoirs available for terrestrial planets. If the sublimation front of the dust trap in a warm disk is located inside the \ce{H_2O} snowline, the inner disk has high gas-phase \ce{C}/\ce{H}- and \ce{O}/\ce{H}-ratios. In a cold disk, where the dust trap is located further out, the inner disk will have low gas-phase \ce{C}/\ce{H}- and \ce{O}/\ce{H}-ratios. These ratios can be probed with infrared spectroscopy, either from the ground or from space. However, as IRS~48 and HD~142527 are both too bright for the \textit{James Webb Space Telescope}, observations from, for example, the Herschel-PACS (see, for example, \citealt{FedeleEA13}) and the VLT-CRIRES(+) instruments could be used to probe these ratios. \cite{FedeleEA13} presented detections of [\ce{O} I] transition(s) for both HD~142527 and IRS~48, \ce{H_2O} transitions for HD~142527 and detections of \ce{[C II]} and \ce{CO} for IRS~48, while \ce{CO} has been in HD~142527 detected by \citet{BrownEA13} using CRIRES. In addition, other carbon-bearing species have not (yet) been observed, potentially hinting at a different \ce{C}/\ce{O}-ratio in the inner disk of HD~142527 compared to the outer disk. Ground-based observations with CRIRES(+) could provide detections of carbon-bearing species and further constrain elemental ratios in the inner disk. For example, \citet{AdamsEA19} has detected \ce{H_2O} and \ce{OH} with CRIRES in the Herbig Ae/Be star HD~101412, while \cite{BanzattiEA22} has presented \ce{CO}-observations in a large sample of disks using both iSHELL on the Infrared Telescope Facility and CRIRES. \\
\indent These elemental ratios, as alluded to in the introduction, have implications for the (giant) planets forming in these disks. If a giant planet is accreting its atmosphere in the cavity of a disk such as IRS~48, the atmosphere will have an elemental ratio of \ce{C}/\ce{O}$<$1. On the other hand, a giant planet accreting its atmospheres in a disk like HD~142527 will have an atmospheric elemental ratio of \ce{C}/\ce{O}$>$1. 

\subsubsection{\ce{H_2CO} \& \ce{CH_3OH}}
The different formation routes for \ce{H_2CO} in both disks is also reflected in their \ce{H_2CO}/\ce{CH_3OH}-ratios (see Table \ref{table:AbundComps}). For HD~142527 we established in Section \ref{sec:H2COForm} that the observed \ce{H_2CO} must be dominantly formed in the gas. Using the ratio of the total derived column density for \ce{H_2CO} (see Section \ref{sec:RDA}) and the calculated $3\sigma$-upper limit for the \ce{CH_3OH} $J$=7$_{1,6}$-6$_{0,7}$ transition, we derive \ce{H_2CO}/\ce{CH_3OH}$\gtrsim$114 (see Table \ref{table:AbundComps}). For IRS~48, \citet{vdMarelEA21} found a ratio of \ce{H_2CO}/\ce{CH_3OH}$\sim$0.2. The higher \ce{H_2CO}/\ce{CH_3OH}-ratio thus points to a cold gas-phase formation of \ce{H_2CO}, whereas a low ratio points towards the grain-surface route, through which both molecules form efficiently. Bulk of the \ce{H_2CO} and \ce{CH_3OH} might, however, still be locked up in ices and possibly have a different ratio.

\subsubsection{Effect of the different gas mass}
The difference in the gas mass might also play a role in producing the large differences between observed column densities and ratios (see Table \ref{table:AbundComps}). The gas mass of the IRS~48 disk is a factor of $\sim$100 lower compared to that of HD~142527, while the observed abundances of both \ce{SO} and \ce{CH_3OH} in IRS~48 are a factor of $\sim$100 larger than the estimated lower limits for HD~142527, yielding for both molecules a total difference of a factor $\sim$10000 regarding gas mass and observed abundance. The low gas mass in IRS~48 might result in low column densities of species formed in the gas phase (e.g., \ce{CN} and \ce{C_2H}, Leemker et al. in press), such that only those sublimating from the ices can easily be observed. The current observations of HD~142527 have not reached high enough sensitivity to detect these gas-phase species like \ce{SO} that could be located at its dust trap.

\subsection{Comparisons with HD~100546}
Besides IRS~48, \ce{SO}, \ce{H_2CO} and \ce{CH_3OH} have also been observed in the disk of HD~100546 \citep{ABoothEA21b,ABoothEA22}, a Herbig Be \citep{VioqueEA18} disk with two symmetric dust rings. The inner dust ring of HD~100546 extends between $\sim$20 and 40 au, while the second one centred at $\sim$200 au \citep{FedeleEA21}. The emission of \ce{SO}, \ce{H_2CO} and \ce{CH_3OH} emission has a central, compact emission component and an outer ring, which is well detected for both \ce{SO} and \ce{H_2CO}. For the inner component, the emission is thought to be the result of thermal sublimation at the inner dust cavity edge, as for IRS~48, but for the outer ring, the emission might be the result of other desorption processes. As the radial column densities of \ce{SO} and \ce{CH_3OH} at the location of the outer dust ring are of the same order of magnitude as our 3$\sigma$-upper limits, the outer ring of HD~100546 might close represent the asymmetric dust trap of HD~142527. The outer ring of HD~100546 contains, however, significantly less dust mass compared the dust trap of HD~142527. Another difference is that we do not (yet) reach the sensitivity in the HD~142527 observations to detect the products of non-thermal desorption processes. Deeper observations focusing on \ce{SO} and \ce{CH_3OH} are necessary to investigate the possibility of observing products of non-thermal desorption processes in the HD~142527 disk.

\section{Summary}
In this work we have investigated the chemistry of the HD~142527 disk using ALMA archival data. We summarise our main conclusions here:
\begin{itemize}
    \item We report the first ALMA detections of \ce{[C}~\ce{I]}, \ce{^{13}C^{18}O}, \ce{DCO^+} and \ce{H_2CO} transitions in the HD~142527 planet-forming disk. Additionally, we report observations of the HCO$^+$ $J$=8-7 and CS $J$=10-9 transitions in this disk.
    \item The observed molecular asymmetries are thought to be partially caused by continuum over-subtraction. Additionally, we propose that shadowing of the misaligned, warped inner disk on the eastern side of the disk causes the asymmetries observed in the \ce{HCN} and \ce{CS} emission.
    \item The dust and gas mass have been recalculated, yielding values of $M_\textnormal{dust}$=1.5$\times$10$^{-3}$ M$_\odot$ and $M_\textnormal{gas}$=1.6$\times$10$^{-2}$ M$_\odot$, respectively. These estimates yield a gas-to-dust ratio of $\sim$11. 
    \item Based on our observations of \ce{^{13}C^{18}O} and \ce{DCO^+}, we expect the \ce{CO} snowline to be located beyond the dust trap at a location of $\sim$300 au, which is supported by our RADMC-3D model.
    \item Using our detection of the \ce{CS} $J$=7-6 transition and the $3\sigma$ upper limit on the \ce{SO} $J$=1$_2$-0$_1$ transition, we estimate \ce{C}/\ce{O}$>$1. The larger ratio is likely due to oxygen-bearing species being frozen-out on the dust trap and is higher than that found for IRS~48, where oxygen-bearing species are returned to the gas-phase through sublimation. 
    \item Finally, the observed \ce{H_2CO} is expected to come from the molecular layer above the midplane. Furthermore, we propose that the observed \ce{H_2CO} is mainly formed in the gas-phase, but contributions from ice-phase formation pathways and non-thermal desorption processes cannot be ruled out. By taking the ratio of the found \ce{H_2CO} column density and the 3$\sigma$-upper limit for \ce{CH_3OH}, we find a \ce{H_2CO}/\ce{CH_3OH}-ratio of $\sim$114. The ratio is significantly higher than that found for IRS~48 ($\sim$0.2). In IRS~48, both \ce{H_2CO} and \ce{CH_3OH} are observed through the sublimation of ices, yielding the lower ratio and further confirming that the observed \ce{H_2CO} must have formed primarily through gas-phase reactions. 
\end{itemize}
The observed molecular content of the HD~142527 disk is clearly different from that of IRS~48 and HD~100546. In particular, in HD~142527 all of the molecules, aside from the \ce{CO} isotopologues, peak in the region of the disk where the mm-sized dust is absent, which is the opposite of what has been observed in IRS~48. Deeper observations targeting \ce{SO} and \ce{CH_3OH} are needed detect or obtain meaningful upper-limits on their relative abundances in order to understand the dominant chemical process(es) occurring in the HD~142527 disk are compared to that in IRS~48. The chemical differences between IRS~48 and HD~142527 might arise due to the location of the dust trap with respect to the host star, shadowing due to a misaligned inner disk and/or a different degree of turbulence in the azimuthal dust traps.

%% === Acknowledgements
\begin{acknowledgements}
    We acknowledge assistance from Allegro, the European ALMA Regional Centre node in the Netherlands. \\
    Astrochemistry in Leiden is supported by funding from the European Research Council (ERC) under the European Union’s Horizon 2020 research and innovation programme (grant agreement No. 101019751 MOLDISK) and by the Netherlands Research School for Astronomy (NOVA). \\
    This work makes use of the following ALMA data: ADS/JAO.ALMA\#2011.0.00318.S, ADS/JAO.ALMA\#2011.0.00465.S, ADS/JAO.ALMA\#2012.1.00613.S, ADS/JAO.ALMA\#2013.1.00305.S, ADS/JAO.ALMA\#2015.1.00614.S, ADS/JAO.ALMA\#2015.1.00805.S, ADS/JAO.ALMA\#2015.1.01137.S. ALMA is a partnership of ESO (representing its member states), NSF (USA) and NINS (Japan), together with NRC (Canada), MOST and ASIAA (Taiwan), and KASI (Republic of Korea), in cooperation with the Republic of Chile. The Joint ALMA Observatory is operated by ESO, AUI/NRAO and NAOJ. \\
    This work has used the following additional software packages that have not been referred to in the main text: Astropy, IPython, Jupyter, Matplotlib, NumPy, pandas and SciPy \citep{Astropy,IPython,Jupyter,Matplotlib,NumPy,Pandas,Pandas2,SciPy}.
\end{acknowledgements}

%% === References
\bibliographystyle{aa}
\bibliography{References}

\begin{thebibliography}{130}
\expandafter\ifx\csname natexlab\endcsname\relax\def\natexlab#1{#1}\fi

\bibitem[{{Adams} {et~al.}(2019){Adams}, {{\'A}d{\'a}mkovics}, {Carr},
  {Najita}, \& {Brittain}}]{AdamsEA19}
{Adams}, S.~C., {{\'A}d{\'a}mkovics}, M., {Carr}, J.~S., {Najita}, J.~R., \&
  {Brittain}, S.~D. 2019, \apj, 871, 173

\bibitem[{{Ag{\'u}ndez} {et~al.}(2018){Ag{\'u}ndez}, {Roueff}, {Le Petit}, \&
  {Le Bourlot}}]{AgundezEA18}
{Ag{\'u}ndez}, M., {Roueff}, E., {Le Petit}, F., \& {Le Bourlot}, J. 2018,
  \aap, 616, A19

\bibitem[{{Andrews} {et~al.}(2018){Andrews}, {Huang}, {P{\'e}rez}, {Isella},
  {Dullemond}, {Kurtovic}, {Guzm{\'a}n}, {Carpenter}, {Wilner}, {Zhang}, {Zhu},
  {Birnstiel}, {Bai}, {Benisty}, {Hughes}, {{\"O}berg}, \& {Ricci}}]{DSHARPI}
{Andrews}, S.~M., {Huang}, J., {P{\'e}rez}, L.~M., {et~al.} 2018, \apjl, 869,
  L41

\bibitem[{{Ansdell} {et~al.}(2016){Ansdell}, {Williams}, {van der Marel},
  {Carpenter}, {Guidi}, {Hogerheijde}, {Mathews}, {Manara}, {Miotello},
  {Natta}, {Oliveira}, {Tazzari}, {Testi}, {van Dishoeck}, \& {van
  Terwisga}}]{AnsdellEA16}
{Ansdell}, M., {Williams}, J.~P., {van der Marel}, N., {et~al.} 2016, \apj,
  828, 46

\bibitem[{{Astropy Collaboration} {et~al.}(2022){Astropy Collaboration},
  {Price-Whelan}, {Lim}, {Earl}, {Starkman}, {Bradley}, {Shupe}, {Patil},
  {Corrales}, {Brasseur}, {N{\"o}the}, {Donath}, {Tollerud}, {Morris},
  {Ginsburg}, {Vaher}, {Weaver}, {Tocknell}, {Jamieson}, {van Kerkwijk},
  {Robitaille}, {Merry}, {Bachetti}, {G{\"u}nther}, {Aldcroft},
  {Alvarado-Montes}, {Archibald}, {B{\'o}di}, {Bapat}, {Barentsen},
  {Baz{\'a}n}, {Biswas}, {Boquien}, {Burke}, {Cara}, {Cara}, {Conroy},
  {Conseil}, {Craig}, {Cross}, {Cruz}, {D'Eugenio}, {Dencheva}, {Devillepoix},
  {Dietrich}, {Eigenbrot}, {Erben}, {Ferreira}, {Foreman-Mackey}, {Fox},
  {Freij}, {Garg}, {Geda}, {Glattly}, {Gondhalekar}, {Gordon}, {Grant},
  {Greenfield}, {Groener}, {Guest}, {Gurovich}, {Handberg}, {Hart},
  {Hatfield-Dodds}, {Homeier}, {Hosseinzadeh}, {Jenness}, {Jones}, {Joseph},
  {Kalmbach}, {Karamehmetoglu}, {Ka{\l}uszy{\'n}ski}, {Kelley}, {Kern},
  {Kerzendorf}, {Koch}, {Kulumani}, {Lee}, {Ly}, {Ma}, {MacBride}, {Maljaars},
  {Muna}, {Murphy}, {Norman}, {O'Steen}, {Oman}, {Pacifici}, {Pascual},
  {Pascual-Granado}, {Patil}, {Perren}, {Pickering}, {Rastogi}, {Roulston},
  {Ryan}, {Rykoff}, {Sabater}, {Sakurikar}, {Salgado}, {Sanghi}, {Saunders},
  {Savchenko}, {Schwardt}, {Seifert-Eckert}, {Shih}, {Jain}, {Shukla}, {Sick},
  {Simpson}, {Singanamalla}, {Singer}, {Singhal}, {Sinha}, {Sip{\H{o}}cz},
  {Spitler}, {Stansby}, {Streicher}, {{\v{S}}umak}, {Swinbank}, {Taranu},
  {Tewary}, {Tremblay}, {de Val-Borro}, {Van Kooten}, {Vasovi{\'c}}, {Verma},
  {de Miranda Cardoso}, {Williams}, {Wilson}, {Winkel}, {Wood-Vasey}, {Xue},
  {Yoachim}, {Zhang}, {Zonca}, \& {Astropy Project Contributors}}]{Astropy}
{Astropy Collaboration}, {Price-Whelan}, A.~M., {Lim}, P.~L., {et~al.} 2022,
  \apj, 935, 167

\bibitem[{{Avenhaus} {et~al.}(2017){Avenhaus}, {Quanz}, {Schmid}, {Dominik},
  {Stolker}, {Ginski}, {de Boer}, {Szul{\'a}gyi}, {Garufi}, {Zurlo},
  {Hagelberg}, {Benisty}, {Henning}, {M{\'e}nard}, {Meyer}, {Baruffolo},
  {Bazzon}, {Beuzit}, {Costille}, {Dohlen}, {Girard}, {Gisler}, {Kasper},
  {Mouillet}, {Pragt}, {Roelfsema}, {Salasnich}, \& {Sauvage}}]{AvenhausEA17}
{Avenhaus}, H., {Quanz}, S.~P., {Schmid}, H.~M., {et~al.} 2017, \aj, 154, 33

\bibitem[{{Avenhaus} {et~al.}(2014){Avenhaus}, {Quanz}, {Schmid}, {Meyer},
  {Garufi}, {Wolf}, \& {Dominik}}]{AvenhausEA14}
{Avenhaus}, H., {Quanz}, S.~P., {Schmid}, H.~M., {et~al.} 2014, \apj, 781, 87

\bibitem[{{Balmer} {et~al.}(2022){Balmer}, {Follette}, {Close}, {Males}, {De
  Rosa}, {Adams Redai}, {Watson}, {Weinberger}, {Morzinski}, {Morales},
  {Ward-Duong}, \& {Pueyo}}]{BalmerEA22}
{Balmer}, W.~O., {Follette}, K.~B., {Close}, L.~M., {et~al.} 2022, \aj, 164, 29

\bibitem[{{Banzatti} {et~al.}(2022){Banzatti}, {Abernathy}, {Brittain},
  {Bosman}, {Pontoppidan}, {Boogert}, {Jensen}, {Carr}, {Najita}, {Grant},
  {Sigler}, {Sanchez}, {Kern}, \& {Rayner}}]{BanzattiEA22}
{Banzatti}, A., {Abernathy}, K.~M., {Brittain}, S., {et~al.} 2022, \aj, 163,
  174

\bibitem[{{Banzatti} {et~al.}(2020){Banzatti}, {Pascucci}, {Bosman}, {Pinilla},
  {Salyk}, {Herczeg}, {Pontoppidan}, {Vazquez}, {Watkins}, {Krijt}, {Hendler},
  \& {Long}}]{BanzattiEA20}
{Banzatti}, A., {Pascucci}, I., {Bosman}, A.~D., {et~al.} 2020, \apj, 903, 124

\bibitem[{{Beckwith} {et~al.}(1990){Beckwith}, {Sargent}, {Chini}, \&
  {Guesten}}]{BeckwithEA90}
{Beckwith}, S. V.~W., {Sargent}, A.~I., {Chini}, R.~S., \& {Guesten}, R. 1990,
  \aj, 99, 924

\bibitem[{{Bergin} \& {Williams}(2017)}]{BWEA17}
{Bergin}, E.~A. \& {Williams}, J.~P. 2017, in Astrophysics and Space Science
  Library, Vol. 445, Formation, Evolution, and Dynamics of Young Solar Systems,
  ed. M.~{Pessah} \& O.~{Gressel}, 1

\bibitem[{{Bertin} {et~al.}(2016){Bertin}, {Romanzin}, {Doronin}, {Philippe},
  {Jeseck}, {Ligterink}, {Linnartz}, {Michaut}, \& {Fillion}}]{BertinEA16}
{Bertin}, M., {Romanzin}, C., {Doronin}, M., {et~al.} 2016, \apjl, 817, L12

\bibitem[{{Biller} {et~al.}(2012){Biller}, {Lacour}, {Juh{\'a}sz}, {Benisty},
  {Chauvin}, {Olofsson}, {Pott}, {M{\"u}ller}, {Sicilia-Aguilar}, {Bonnefoy},
  {Tuthill}, {Thebault}, {Henning}, \& {Crida}}]{BillerEA12}
{Biller}, B., {Lacour}, S., {Juh{\'a}sz}, A., {et~al.} 2012, \apjl, 753, L38

\bibitem[{{Boehler} {et~al.}(2021){Boehler}, {M{\'e}nard}, {Robert}, {Isella},
  {Pinte}, {Gonzalez}, {van der Plas}, {Weaver}, {Teague}, {Garg}, \&
  {M{\'e}heut}}]{BoehlerEA21}
{Boehler}, Y., {M{\'e}nard}, F., {Robert}, C.~M.~T., {et~al.} 2021, \aap, 650,
  A59

\bibitem[{{Boehler} {et~al.}(2017){Boehler}, {Weaver}, {Isella}, {Ricci},
  {Grady}, {Carpenter}, \& {Perez}}]{BoehlerEA17}
{Boehler}, Y., {Weaver}, E., {Isella}, A., {et~al.} 2017, \apj, 840, 60

\bibitem[{{Bohn} {et~al.}(2022){Bohn}, {Benisty}, {Perraut}, {van der Marel},
  {W{\"o}lfer}, {van Dishoeck}, {Facchini}, {Manara}, {Teague}, {Francis},
  {Berger}, {Garcia-Lopez}, {Ginski}, {Henning}, {Kenworthy}, {Kraus},
  {M{\'e}nard}, {M{\'e}rand}, \& {P{\'e}rez}}]{BohnEA22}
{Bohn}, A.~J., {Benisty}, M., {Perraut}, K., {et~al.} 2022, \aap, 658, A183

\bibitem[{{Booth} {et~al.}(2022){Booth}, {Ilee}, {Walsh}, {Kama}, {Keyte}, {van
  Dishoeck}, \& {Nomura}}]{ABoothEA22}
{Booth}, A.~S., {Ilee}, J.~D., {Walsh}, C., {et~al.} 2022, arXiv e-prints,
  arXiv:2210.14820

\bibitem[{{Booth} {et~al.}(2021{\natexlab{a}}){Booth}, {van der Marel},
  {Leemker}, {van Dishoeck}, \& {Ohashi}}]{ABoothEA21}
{Booth}, A.~S., {van der Marel}, N., {Leemker}, M., {van Dishoeck}, E.~F., \&
  {Ohashi}, S. 2021{\natexlab{a}}, \aap, 651, L6

\bibitem[{{Booth} {et~al.}(2019){Booth}, {Walsh}, {Ilee}, {Notsu}, {Qi},
  {Nomura}, \& {Akiyama}}]{ABoothEA19}
{Booth}, A.~S., {Walsh}, C., {Ilee}, J.~D., {et~al.} 2019, \apjl, 882, L31

\bibitem[{{Booth} {et~al.}(2021{\natexlab{b}}){Booth}, {Walsh}, {Terwisscha van
  Scheltinga}, {van Dishoeck}, {Ilee}, {Hogerheijde}, {Kama}, \&
  {Nomura}}]{ABoothEA21b}
{Booth}, A.~S., {Walsh}, C., {Terwisscha van Scheltinga}, J., {et~al.}
  2021{\natexlab{b}}, Nature Astronomy, 5, 684

\bibitem[{{Booth} {et~al.}(2017){Booth}, {Clarke}, {Madhusudhan}, \&
  {Ilee}}]{RBoothEA17}
{Booth}, R.~A., {Clarke}, C.~J., {Madhusudhan}, N., \& {Ilee}, J.~D. 2017,
  \mnras, 469, 3994

\bibitem[{{Bosman} \& {Banzatti}(2019)}]{BB19}
{Bosman}, A.~D. \& {Banzatti}, A. 2019, \aap, 632, L10

\bibitem[{{Bouwman} {et~al.}(2004){Bouwman}, {Dominik}, {Dullemond}, {Henning},
  {Leinert}, {Min}, {Thi}, {Tielens}, {Vandenbussche}, {Waelkens}, {Waters},
  {de Koter}, {van Boekel}, \& {van den Ancker}}]{BouwmanEA04}
{Bouwman}, J., {Dominik}, C., {Dullemond}, C.~P., {et~al.} 2004, {The
  mineralogy of proto-planetary disks surrounding Herbig Ae/Be stars}, Spitzer
  Proposal ID 3470

\bibitem[{{Brown} {et~al.}(2013){Brown}, {Pontoppidan}, {van Dishoeck},
  {Herczeg}, {Blake}, \& {Smette}}]{BrownEA13}
{Brown}, J.~M., {Pontoppidan}, K.~M., {van Dishoeck}, E.~F., {et~al.} 2013,
  \apj, 770, 94

\bibitem[{{Bruderer} {et~al.}(2014){Bruderer}, {van der Marel}, {van Dishoeck},
  \& {van Kempen}}]{BrudererEA14}
{Bruderer}, S., {van der Marel}, N., {van Dishoeck}, E.~F., \& {van Kempen},
  T.~A. 2014, \aap, 562, A26

\bibitem[{{Brunken} {et~al.}(2022){Brunken}, {Booth}, {Leemker}, {Nazari}, {van
  der Marel}, \& {van Dishoeck}}]{BrunkenEA22}
{Brunken}, N. G.~C., {Booth}, A.~S., {Leemker}, M., {et~al.} 2022, \aap, 659,
  A29

\bibitem[{{Calahan} {et~al.}(2021){Calahan}, {Bergin}, {Zhang}, {Teague},
  {Cleeves}, {Bergner}, {Blake}, {Cazzoletti}, {Guzm{\'a}n}, {Hogerheijde},
  {Huang}, {Kama}, {Loomis}, {{\"O}berg}, {Qi}, {van Dishoeck}, {Terwisscha van
  Scheltinga}, {Walsh}, \& {Wilner}}]{CalahanEA21TWH}
{Calahan}, J.~K., {Bergin}, E., {Zhang}, K., {et~al.} 2021, \apj, 908, 8

\bibitem[{{Canovas} {et~al.}(2013){Canovas}, {M{\'e}nard}, {Hales},
  {Jord{\'a}n}, {Schreiber}, {Casassus}, {Gledhill}, \& {Pinte}}]{CanovasEA13}
{Canovas}, H., {M{\'e}nard}, F., {Hales}, A., {et~al.} 2013, \aap, 556, A123

\bibitem[{{Carney} {et~al.}(2018){Carney}, {Fedele}, {Hogerheijde}, {Favre},
  {Walsh}, {Bruderer}, {Miotello}, {Murillo}, {Klaassen}, {Henning}, \& {van
  Dishoeck}}]{CarneyEA18}
{Carney}, M.~T., {Fedele}, D., {Hogerheijde}, M.~R., {et~al.} 2018, \aap, 614,
  A106

\bibitem[{{Carney} {et~al.}(2019){Carney}, {Hogerheijde}, {Guzm{\'a}n},
  {Walsh}, {{\"O}berg}, {Fayolle}, {Cleeves}, {Carpenter}, \&
  {Qi}}]{CarneyEA19}
{Carney}, M.~T., {Hogerheijde}, M.~R., {Guzm{\'a}n}, V.~V., {et~al.} 2019,
  \aap, 623, A124

\bibitem[{{Casassus} {et~al.}(2015){Casassus}, {Marino}, {P{\'e}rez}, {Roman},
  {Dunhill}, {Armitage}, {Cuadra}, {Wootten}, {van der Plas}, {Cieza}, {Moral},
  {Christiaens}, \& {Montesinos}}]{CasassusEA15}
{Casassus}, S., {Marino}, S., {P{\'e}rez}, S., {et~al.} 2015, \apj, 811, 92

\bibitem[{{Casassus} {et~al.}(2013){Casassus}, {van der Plas}, {Perez}, {Dent},
  {Fomalont}, {Hagelberg}, {Hales}, {Jord{\'a}n}, {Mawet}, {M{\'e}nard},
  {Wootten}, {Wilner}, {Hughes}, {Schreiber}, {Girard}, {Ercolano}, {Canovas},
  {Rom{\'a}n}, \& {Salinas}}]{CasassusEA13}
{Casassus}, S., {van der Plas}, G.~M., {Perez}, S., {et~al.} 2013, \nat, 493,
  191

\bibitem[{{Cazzoletti} {et~al.}(2019){Cazzoletti}, {Manara}, {Liu}, {van
  Dishoeck}, {Facchini}, {Alcal{\`a}}, {Ansdell}, {Testi}, {Williams},
  {Carrasco-Gonz{\'a}lez}, {Dong}, {Forbrich}, {Fukagawa}, {Galv{\'a}n-Madrid},
  {Hirano}, {Hogerheijde}, {Hasegawa}, {Muto}, {Pinilla}, {Takami}, {Tamura},
  {Tazzari}, \& {Wisniewski}}]{CazzolettiEA19}
{Cazzoletti}, P., {Manara}, C.~F., {Liu}, H.~B., {et~al.} 2019, \aap, 626, A11

\bibitem[{{Chiang} \& {Goldreich}(1997)}]{CG97}
{Chiang}, E.~I. \& {Goldreich}, P. 1997, \apj, 490, 368

\bibitem[{{Cleeves}(2016)}]{Cleeves16}
{Cleeves}, L.~I. 2016, \apjl, 816, L21

\bibitem[{{Cleeves} {et~al.}(2021){Cleeves}, {Loomis}, {Teague}, {Bergin},
  {Wilner}, {Bergner}, {Blake}, {Calahan}, {Cazzoletti}, {van Dishoeck},
  {Guzm{\'a}n}, {Hogerheijde}, {Huang}, {Kama}, {{\"O}berg}, {Qi}, {Terwisscha
  van Scheltinga}, \& {Walsh}}]{CleevesEA21TWH}
{Cleeves}, L.~I., {Loomis}, R.~A., {Teague}, R., {et~al.} 2021, \apj, 911, 29

\bibitem[{{Cruz-Diaz} {et~al.}(2016){Cruz-Diaz}, {Mart{\'\i}n-Dom{\'e}nech},
  {Mu{\~n}oz Caro}, \& {Chen}}]{CruzDiazEA16}
{Cruz-Diaz}, G.~A., {Mart{\'\i}n-Dom{\'e}nech}, R., {Mu{\~n}oz Caro}, G.~M., \&
  {Chen}, Y.~J. 2016, \aap, 592, A68

\bibitem[{{Dominik} {et~al.}(2021){Dominik}, {Min}, \& {Tazaki}}]{OpTool}
{Dominik}, C., {Min}, M., \& {Tazaki}, R. 2021, {OpTool: Command-line driven
  tool for creating complex dust opacities}, Astrophysics Source Code Library,
  record ascl:2104.010

\bibitem[{{Dullemond} {et~al.}(2001){Dullemond}, {Dominik}, \&
  {Natta}}]{DullemondEA01}
{Dullemond}, C.~P., {Dominik}, C., \& {Natta}, A. 2001, \apj, 560, 957

\bibitem[{{Dullemond} {et~al.}(2012){Dullemond}, {Juhasz}, {Pohl}, {Sereshti},
  {Shetty}, {Peters}, {Commercon}, \& {Flock}}]{RADMC3D}
{Dullemond}, C.~P., {Juhasz}, A., {Pohl}, A., {et~al.} 2012, {RADMC-3D: A
  multi-purpose radiative transfer tool}, Astrophysics Source Code Library,
  record ascl:1202.015

\bibitem[{{Eistrup}(2022)}]{Eistrup22}
{Eistrup}, C. 2022, arXiv e-prints, arXiv:2210.16921

\bibitem[{{Eistrup} {et~al.}(2018){Eistrup}, {Walsh}, \& {van
  Dishoeck}}]{EistrupEA18}
{Eistrup}, C., {Walsh}, C., \& {van Dishoeck}, E.~F. 2018, \aap, 613, A14

\bibitem[{{Facchini} {et~al.}(2017){Facchini}, {Birnstiel}, {Bruderer}, \& {van
  Dishoeck}}]{FacchiniEA17}
{Facchini}, S., {Birnstiel}, T., {Bruderer}, S., \& {van Dishoeck}, E.~F. 2017,
  \aap, 605, A16

\bibitem[{{Facchini} {et~al.}(2021){Facchini}, {Teague}, {Bae}, {Benisty},
  {Keppler}, \& {Isella}}]{FacchiniEA21}
{Facchini}, S., {Teague}, R., {Bae}, J., {et~al.} 2021, \aj, 162, 99

\bibitem[{{Fairlamb} {et~al.}(2015){Fairlamb}, {Oudmaijer}, {Mendigut{\'\i}a},
  {Ilee}, \& {van den Ancker}}]{FairlambEA15}
{Fairlamb}, J.~R., {Oudmaijer}, R.~D., {Mendigut{\'\i}a}, I., {Ilee}, J.~D., \&
  {van den Ancker}, M.~E. 2015, \mnras, 453, 976

\bibitem[{{Fedele} {et~al.}(2013){Fedele}, {Bruderer}, {van Dishoeck}, {Carr},
  {Herczeg}, {Salyk}, {Evans}, {Bouwman}, {Meeus}, {Henning}, {Green},
  {Najita}, \& {G{\"u}del}}]{FedeleEA13}
{Fedele}, D., {Bruderer}, S., {van Dishoeck}, E.~F., {et~al.} 2013, \aap, 559,
  A77

\bibitem[{{Fedele} {et~al.}(2021){Fedele}, {Toci}, {Maud}, \&
  {Lodato}}]{FedeleEA21}
{Fedele}, D., {Toci}, C., {Maud}, L., \& {Lodato}, G. 2021, \aap, 651, A90

\bibitem[{{Foreman-Mackey} {et~al.}(2013){Foreman-Mackey}, {Hogg}, {Lang}, \&
  {Goodman}}]{emcee}
{Foreman-Mackey}, D., {Hogg}, D.~W., {Lang}, D., \& {Goodman}, J. 2013, \pasp,
  125, 306

\bibitem[{{Francis} \& {van der Marel}(2020)}]{FvdM20}
{Francis}, L. \& {van der Marel}, N. 2020, \apj, 892, 111

\bibitem[{{Fuchs} {et~al.}(2009){Fuchs}, {Cuppen}, {Ioppolo}, {Romanzin},
  {Bisschop}, {Andersson}, {van Dishoeck}, \& {Linnartz}}]{FuchsEa09}
{Fuchs}, G.~W., {Cuppen}, H.~M., {Ioppolo}, S., {et~al.} 2009, \aap, 505, 629

\bibitem[{{Fukagawa} {et~al.}(2006){Fukagawa}, {Tamura}, {Itoh}, {Kudo},
  {Imaeda}, {Oasa}, {Hayashi}, \& {Hayashi}}]{FukagawaEA06}
{Fukagawa}, M., {Tamura}, M., {Itoh}, Y., {et~al.} 2006, \apjl, 636, L153

\bibitem[{{Gaia Collaboration} {et~al.}(2018){Gaia Collaboration}, {Brown},
  {Vallenari}, {Prusti}, {de Bruijne}, {Babusiaux}, {Bailer-Jones}, {Biermann},
  {Evans}, {Eyer}, {Jansen}, {Jordi}, {Klioner}, {Lammers}, {Lindegren},
  {Luri}, {Mignard}, {Panem}, {Pourbaix}, {Randich}, {Sartoretti}, {Siddiqui},
  {Soubiran}, {van Leeuwen}, {Walton}, {Arenou}, {Bastian}, {Cropper},
  {Drimmel}, {Katz}, {Lattanzi}, {Bakker}, {Cacciari}, {Casta{\~n}eda},
  {Chaoul}, {Cheek}, {De Angeli}, {Fabricius}, {Guerra}, {Holl}, {Masana},
  {Messineo}, {Mowlavi}, {Nienartowicz}, {Panuzzo}, {Portell}, {Riello},
  {Seabroke}, {Tanga}, {Th{\'e}venin}, {Gracia-Abril}, {Comoretto},
  {Garcia-Reinaldos}, {Teyssier}, {Altmann}, {Andrae}, {Audard},
  {Bellas-Velidis}, {Benson}, {Berthier}, {Blomme}, {Burgess}, {Busso},
  {Carry}, {Cellino}, {Clementini}, {Clotet}, {Creevey}, {Davidson}, {De
  Ridder}, {Delchambre}, {Dell'Oro}, {Ducourant},
  {Fern{\'a}ndez-Hern{\'a}ndez}, {Fouesneau}, {Fr{\'e}mat}, {Galluccio},
  {Garc{\'\i}a-Torres}, {Gonz{\'a}lez-N{\'u}{\~n}ez}, {Gonz{\'a}lez-Vidal},
  {Gosset}, {Guy}, {Halbwachs}, {Hambly}, {Harrison}, {Hern{\'a}ndez},
  {Hestroffer}, {Hodgkin}, {Hutton}, {Jasniewicz}, {Jean-Antoine-Piccolo},
  {Jordan}, {Korn}, {Krone-Martins}, {Lanzafame}, {Lebzelter}, {L{\"o}ffler},
  {Manteiga}, {Marrese}, {Mart{\'\i}n-Fleitas}, {Moitinho}, {Mora}, {Muinonen},
  {Osinde}, {Pancino}, {Pauwels}, {Petit}, {Recio-Blanco}, {Richards},
  {Rimoldini}, {Robin}, {Sarro}, {Siopis}, {Smith}, {Sozzetti}, {S{\"u}veges},
  {Torra}, {van Reeven}, {Abbas}, {Abreu Aramburu}, {Accart}, {Aerts},
  {Altavilla}, {{\'A}lvarez}, {Alvarez}, {Alves}, {Anderson}, {Andrei},
  {Anglada Varela}, {Antiche}, {Antoja}, {Arcay}, {Astraatmadja}, {Bach},
  {Baker}, {Balaguer-N{\'u}{\~n}ez}, {Balm}, {Barache}, {Barata}, {Barbato},
  {Barblan}, {Barklem}, {Barrado}, {Barros}, {Barstow}, {Bartholom{\'e}
  Mu{\~n}oz}, {Bassilana}, {Becciani}, {Bellazzini}, {Berihuete}, {Bertone},
  {Bianchi}, {Bienaym{\'e}}, {Blanco-Cuaresma}, {Boch}, {Boeche}, {Bombrun},
  {Borrachero}, {Bossini}, {Bouquillon}, {Bourda}, {Bragaglia}, {Bramante},
  {Breddels}, {Bressan}, {Brouillet}, {Br{\"u}semeister}, {Brugaletta},
  {Bucciarelli}, {Burlacu}, {Busonero}, {Butkevich}, {Buzzi}, {Caffau},
  {Cancelliere}, {Cannizzaro}, {Cantat-Gaudin}, {Carballo}, {Carlucci},
  {Carrasco}, {Casamiquela}, {Castellani}, {Castro-Ginard}, {Charlot},
  {Chemin}, {Chiavassa}, {Cocozza}, {Costigan}, {Cowell}, {Crifo}, {Crosta},
  {Crowley}, {Cuypers}, {Dafonte}, {Damerdji}, {Dapergolas}, {David}, {David},
  {de Laverny}, {De Luise}, {De March}, {de Martino}, {de Souza}, {de Torres},
  {Debosscher}, {del Pozo}, {Delbo}, {Delgado}, {Delgado}, {Di Matteo},
  {Diakite}, {Diener}, {Distefano}, {Dolding}, {Drazinos}, {Dur{\'a}n},
  {Edvardsson}, {Enke}, {Eriksson}, {Esquej}, {Eynard Bontemps}, {Fabre},
  {Fabrizio}, {Faigler}, {Falc{\~a}o}, {Farr{\`a}s Casas}, {Federici},
  {Fedorets}, {Fernique}, {Figueras}, {Filippi}, {Findeisen}, {Fonti},
  {Fraile}, {Fraser}, {Fr{\'e}zouls}, {Gai}, {Galleti}, {Garabato},
  {Garc{\'\i}a-Sedano}, {Garofalo}, {Garralda}, {Gavel}, {Gavras}, {Gerssen},
  {Geyer}, {Giacobbe}, {Gilmore}, {Girona}, {Giuffrida}, {Glass}, {Gomes},
  {Granvik}, {Gueguen}, {Guerrier}, {Guiraud}, {Guti{\'e}rrez-S{\'a}nchez},
  {Haigron}, {Hatzidimitriou}, {Hauser}, {Haywood}, {Heiter}, {Helmi}, {Heu},
  {Hilger}, {Hobbs}, {Hofmann}, {Holland}, {Huckle}, {Hypki}, {Icardi},
  {Jan{\ss}en}, {Jevardat de Fombelle}, {Jonker}, {Juh{\'a}sz}, {Julbe},
  {Karampelas}, {Kewley}, {Klar}, {Kochoska}, {Kohley}, {Kolenberg},
  {Kontizas}, {Kontizas}, {Koposov}, {Kordopatis}, {Kostrzewa-Rutkowska},
  {Koubsky}, {Lambert}, {Lanza}, {Lasne}, {Lavigne}, {Le Fustec}, {Le
  Poncin-Lafitte}, {Lebreton}, {Leccia}, {Leclerc}, {Lecoeur-Taibi},
  {Lenhardt}, {Leroux}, {Liao}, {Licata}, {Lindstr{\o}m}, {Lister}, {Livanou},
  {Lobel}, {L{\'o}pez}, {Managau}, {Mann}, {Mantelet}, {Marchal}, {Marchant},
  {Marconi}, {Marinoni}, {Marschalk{\'o}}, {Marshall}, {Martino}, {Marton},
  {Mary}, {Massari}, {Matijevi{\v{c}}}, {Mazeh}, {McMillan}, {Messina},
  {Michalik}, {Millar}, {Molina}, {Molinaro}, {Moln{\'a}r}, {Montegriffo},
  {Mor}, {Morbidelli}, {Morel}, {Morris}, {Mulone}, {Muraveva}, {Musella},
  {Nelemans}, {Nicastro}, {Noval}, {O'Mullane}, {Ord{\'e}novic},
  {Ord{\'o}{\~n}ez-Blanco}, {Osborne}, {Pagani}, {Pagano}, {Pailler},
  {Palacin}, {Palaversa}, {Panahi}, {Pawlak}, {Piersimoni}, {Pineau}, {Plachy},
  {Plum}, {Poggio}, {Poujoulet}, {Pr{\v{s}}a}, {Pulone}, {Racero}, {Ragaini},
  {Rambaux}, {Ramos-Lerate}, {Regibo}, {Reyl{\'e}}, {Riclet}, {Ripepi}, {Riva},
  {Rivard}, {Rixon}, {Roegiers}, {Roelens}, {Romero-G{\'o}mez}, {Rowell},
  {Royer}, {Ruiz-Dern}, {Sadowski}, {Sagrist{\`a} Sell{\'e}s}, {Sahlmann},
  {Salgado}, {Salguero}, {Sanna}, {Santana-Ros}, {Sarasso}, {Savietto},
  {Schultheis}, {Sciacca}, {Segol}, {Segovia}, {S{\'e}gransan}, {Shih},
  {Siltala}, {Silva}, {Smart}, {Smith}, {Solano}, {Solitro}, {Sordo}, {Soria
  Nieto}, {Souchay}, {Spagna}, {Spoto}, {Stampa}, {Steele},
  {Steidelm{\"u}ller}, {Stephenson}, {Stoev}, {Suess}, {Surdej}, {Szabados},
  {Szegedi-Elek}, {Tapiador}, {Taris}, {Tauran}, {Taylor}, {Teixeira},
  {Terrett}, {Teyssandier}, {Thuillot}, {Titarenko}, {Torra Clotet}, {Turon},
  {Ulla}, {Utrilla}, {Uzzi}, {Vaillant}, {Valentini}, {Valette}, {van Elteren},
  {Van Hemelryck}, {van Leeuwen}, {Vaschetto}, {Vecchiato}, {Veljanoski},
  {Viala}, {Vicente}, {Vogt}, {von Essen}, {Voss}, {Votruba}, {Voutsinas},
  {Walmsley}, {Weiler}, {Wertz}, {Wevers}, {Wyrzykowski}, {Yoldas},
  {{\v{Z}}erjal}, {Ziaeepour}, {Zorec}, {Zschocke}, {Zucker}, {Zurbach}, \&
  {Zwitter}}]{GC18}
{Gaia Collaboration}, {Brown}, A.~G.~A., {Vallenari}, A., {et~al.} 2018, \aap,
  616, A1

\bibitem[{{Garg} {et~al.}(2021){Garg}, {Pinte}, {Christiaens}, {Price},
  {Lazendic}, {Boehler}, {Casassus}, {Marino}, {Perez}, \& {Zuleta}}]{GargEA21}
{Garg}, H., {Pinte}, C., {Christiaens}, V., {et~al.} 2021, \mnras, 504, 782

\bibitem[{{Goldsmith} \& {Langer}(1999)}]{GL99}
{Goldsmith}, P.~F. \& {Langer}, W.~D. 1999, \apj, 517, 209

\bibitem[{Harris {et~al.}(2020)Harris, Millman, van~der Walt, Gommers,
  Virtanen, Cournapeau, Wieser, Taylor, Berg, Smith, Kern, Picus, Hoyer, van
  Kerkwijk, Brett, Haldane, del R{\'{i}}o, Wiebe, Peterson,
  G{\'{e}}rard-Marchant, Sheppard, Reddy, Weckesser, Abbasi, Gohlke, \&
  Oliphant}]{NumPy}
Harris, C.~R., Millman, K.~J., van~der Walt, S.~J., {et~al.} 2020, Nature, 585,
  357

\bibitem[{{Hildebrand}(1983)}]{Hildebrand83}
{Hildebrand}, R.~H. 1983, \qjras, 24, 267

\bibitem[{{Huang} {et~al.}(2017){Huang}, {{\"O}berg}, {Qi}, {Aikawa},
  {Andrews}, {Furuya}, {Guzm{\'a}n}, {Loomis}, {van Dishoeck}, \&
  {Wilner}}]{HuangEA17}
{Huang}, J., {{\"O}berg}, K.~I., {Qi}, C., {et~al.} 2017, \apj, 835, 231

\bibitem[{Hunter(2007)}]{Matplotlib}
Hunter, J.~D. 2007, Computing in Science \& Engineering, 9, 90

\bibitem[{{Isella} {et~al.}(2018){Isella}, {Huang}, {Andrews}, {Dullemond},
  {Birnstiel}, {Zhang}, {Zhu}, {Guzm{\'a}n}, {P{\'e}rez}, {Bai}, {Benisty},
  {Carpenter}, {Ricci}, \& {Wilner}}]{IsellaEA18}
{Isella}, A., {Huang}, J., {Andrews}, S.~M., {et~al.} 2018, \apjl, 869, L49

\bibitem[{Kluyver {et~al.}(2016)Kluyver, Ragan-Kelley, P{\'e}rez, Granger,
  Bussonnier, Frederic, Kelley, Hamrick, Grout, Corlay, Ivanov, Avila, Abdalla,
  \& Willing}]{Jupyter}
Kluyver, T., Ragan-Kelley, B., P{\'e}rez, F., {et~al.} 2016, in Positioning and
  Power in Academic Publishing: Players, Agents and Agendas, ed. F.~Loizides \&
  B.~Schmidt, IOS Press, 87 -- 90

\bibitem[{{Krijt} {et~al.}(2020){Krijt}, {Bosman}, {Zhang}, {Schwarz},
  {Ciesla}, \& {Bergin}}]{KrijtEA20}
{Krijt}, S., {Bosman}, A.~D., {Zhang}, K., {et~al.} 2020, \apj, 899, 134

\bibitem[{{Krijt} {et~al.}(2018){Krijt}, {Schwarz}, {Bergin}, \&
  {Ciesla}}]{KrijtEA18}
{Krijt}, S., {Schwarz}, K.~R., {Bergin}, E.~A., \& {Ciesla}, F.~J. 2018, \apj,
  864, 78

\bibitem[{{Lacour} {et~al.}(2016){Lacour}, {Biller}, {Cheetham}, {Greenbaum},
  {Pearce}, {Marino}, {Tuthill}, {Pueyo}, {Mamajek}, {Girard},
  {Sivaramakrishnan}, {Bonnefoy}, {Baraffe}, {Chauvin}, {Olofsson}, {Juhasz},
  {Benisty}, {Pott}, {Sicilia-Aguilar}, {Henning}, {Cardwell}, {Goodsell},
  {Graham}, {Hibon}, {Ingraham}, {Konopacky}, {Macintosh}, {Oppenheimer},
  {Perrin}, {Rantakyr{\"o}}, {Sadakuni}, \& {Thomas}}]{LacourEA16}
{Lacour}, S., {Biller}, B., {Cheetham}, A., {et~al.} 2016, \aap, 590, A90

\bibitem[{{Le Gal} {et~al.}(2019){Le Gal}, {Brady}, {{\"O}berg}, {Roueff}, \&
  {Le Petit}}]{LeGalEA19}
{Le Gal}, R., {Brady}, M.~T., {{\"O}berg}, K.~I., {Roueff}, E., \& {Le Petit},
  F. 2019, \apj, 886, 86

\bibitem[{{Le Gal} {et~al.}(2021){Le Gal}, {{\"O}berg}, {Teague}, {Loomis},
  {Law}, {Walsh}, {Bergin}, {M{\'e}nard}, {Wilner}, {Andrews}, {Aikawa},
  {Booth}, {Cataldi}, {Bergner}, {Bosman}, {Cleeves}, {Czekala}, {Furuya},
  {Guzm{\'a}n}, {Huang}, {Ilee}, {Nomura}, {Qi}, {Schwarz}, {Tsukagoshi},
  {Yamato}, \& {Zhang}}]{LeGalEA21}
{Le Gal}, R., {{\"O}berg}, K.~I., {Teague}, R., {et~al.} 2021, \apjs, 257, 12

\bibitem[{{Leemker} {et~al.}(2022){Leemker}, {Booth}, {van Dishoeck},
  {P{\'e}rez-S{\'a}nchez}, {Szul{\'a}gyi}, {Bosman}, {Bruderer}, {Facchini},
  {Hogerheijde}, {Paneque-Carre{\~n}o}, \& {Sturm}}]{LeemkerEA22}
{Leemker}, M., {Booth}, A.~S., {van Dishoeck}, E.~F., {et~al.} 2022, \aap, 663,
  A23

\bibitem[{{Leemker} {et~al.}(2023){Leemker}, {Booth}, {van Dishoeck}, {van der
  Marel}, {Tabone}, {Ligterink}, {Brunken}, \& {Hogerheijde}}]{LeemkerEA23}
{Leemker}, M., {Booth}, A.~S., {van Dishoeck}, E.~F., {et~al.} 2023, arXiv
  e-prints, arXiv:2303.00768

\bibitem[{{Long} {et~al.}(2018){Long}, {Pinilla}, {Herczeg}, {Harsono},
  {Dipierro}, {Pascucci}, {Hendler}, {Tazzari}, {Ragusa}, {Salyk}, {Edwards},
  {Lodato}, {van de Plas}, {Johnstone}, {Liu}, {Boehler}, {Cabrit}, {Manara},
  {Menard}, {Mulders}, {Nisini}, {Fischer}, {Rigliaco}, {Banzatti}, {Avenhaus},
  \& {Gully-Santiago}}]{LongEA18}
{Long}, F., {Pinilla}, P., {Herczeg}, G.~J., {et~al.} 2018, \apj, 869, 17

\bibitem[{{Loomis} {et~al.}(2015){Loomis}, {Cleeves}, {{\"O}berg}, {Guzman}, \&
  {Andrews}}]{LoomisEA15}
{Loomis}, R.~A., {Cleeves}, L.~I., {{\"O}berg}, K.~I., {Guzman}, V.~V., \&
  {Andrews}, S.~M. 2015, \apjl, 809, L25

\bibitem[{{Loomis} {et~al.}(2020){Loomis}, {{\"O}berg}, {Andrews}, {Bergin},
  {Bergner}, {Blake}, {Cleeves}, {Czekala}, {Huang}, {Le Gal}, {M{\'e}nard},
  {Pegues}, {Qi}, {Walsh}, {Williams}, \& {Wilner}}]{LoomisEA20}
{Loomis}, R.~A., {{\"O}berg}, K.~I., {Andrews}, S.~M., {et~al.} 2020, \apj,
  893, 101

\bibitem[{{Marino} {et~al.}(2015){Marino}, {Perez}, \& {Casassus}}]{MarinoEA15}
{Marino}, S., {Perez}, S., \& {Casassus}, S. 2015, \apjl, 798, L44

\bibitem[{{Mathews} {et~al.}(2013){Mathews}, {Klaassen}, {Juh{\'a}sz},
  {Harsono}, {Chapillon}, {van Dishoeck}, {Espada}, {de Gregorio-Monsalvo},
  {Hales}, {Hogerheijde}, {Mottram}, {Rawlings}, {Takahashi}, \&
  {Testi}}]{MathewsEA13}
{Mathews}, G.~S., {Klaassen}, P.~D., {Juh{\'a}sz}, A., {et~al.} 2013, \aap,
  557, A132

\bibitem[{{McClure} {et~al.}(2020){McClure}, {Dominik}, \&
  {Kama}}]{McClureEA20}
{McClure}, M.~K., {Dominik}, C., \& {Kama}, M. 2020, \aap, 642, L15

\bibitem[{{McMullin} {et~al.}(2007){McMullin}, {Waters}, {Schiebel}, {Young},
  \& {Golap}}]{CASA}
{McMullin}, J.~P., {Waters}, B., {Schiebel}, D., {Young}, W., \& {Golap}, K.
  2007, in Astronomical Society of the Pacific Conference Series, Vol. 376,
  Astronomical Data Analysis Software and Systems XVI, ed. R.~A. {Shaw},
  F.~{Hill}, \& D.~J. {Bell}, 127

\bibitem[{{Milam} {et~al.}(2005){Milam}, {Savage}, {Brewster}, {Ziurys}, \&
  {Wyckoff}}]{MilamEA05}
{Milam}, S.~N., {Savage}, C., {Brewster}, M.~A., {Ziurys}, L.~M., \& {Wyckoff},
  S. 2005, \apj, 634, 1126

\bibitem[{{Miley} {et~al.}(2021){Miley}, {Pani{\'c}}, {Booth}, {Ilee}, {Ida},
  \& {Kunitomo}}]{MileyEA21}
{Miley}, J.~M., {Pani{\'c}}, O., {Booth}, R.~A., {et~al.} 2021, \mnras, 500,
  4658

\bibitem[{{Millar} {et~al.}(1989){Millar}, {Bennett}, \& {Herbst}}]{MillarEA89}
{Millar}, T.~J., {Bennett}, A., \& {Herbst}, E. 1989, \apj, 340, 906

\bibitem[{{Miyake} \& {Nakagawa}(1993)}]{MN93}
{Miyake}, K. \& {Nakagawa}, Y. 1993, \icarus, 106, 20

\bibitem[{{M{\"u}ller} {et~al.}(2005){M{\"u}ller}, {Schl{\"o}der}, {Stutzki},
  \& {Winnewisser}}]{CDMS05}
{M{\"u}ller}, H. S.~P., {Schl{\"o}der}, F., {Stutzki}, J., \& {Winnewisser}, G.
  2005, Journal of Molecular Structure, 742, 215

\bibitem[{{M{\"u}ller} {et~al.}(2001){M{\"u}ller}, {Thorwirth}, {Roth}, \&
  {Winnewisser}}]{CDMS01}
{M{\"u}ller}, H.~S.~P., {Thorwirth}, S., {Roth}, D.~A., \& {Winnewisser}, G.
  2001, \aap, 370, L49

\bibitem[{{{\"O}berg} \& {Bergin}(2021)}]{OB21}
{{\"O}berg}, K.~I. \& {Bergin}, E.~A. 2021, \physrep, 893, 1

\bibitem[{{{\"O}berg} {et~al.}(2021{\natexlab{a}}){{\"O}berg}, {Cleeves},
  {Bergner}, {Cavanaro}, {Teague}, {Huang}, {Loomis}, {Bergin}, {Blake},
  {Calahan}, {Cazzoletti}, {Guzm{\'a}n}, {Hogerheijde}, {Kama}, {Terwisscha van
  Scheltinga}, {Qi}, {van Dishoeck}, {Walsh}, \& {Wilner}}]{ObergEA21TWH}
{{\"O}berg}, K.~I., {Cleeves}, L.~I., {Bergner}, J.~B., {et~al.}
  2021{\natexlab{a}}, \aj, 161, 38

\bibitem[{{{\"O}berg} {et~al.}(2015){{\"O}berg}, {Furuya}, {Loomis}, {Aikawa},
  {Andrews}, {Qi}, {van Dishoeck}, \& {Wilner}}]{ObergEA15}
{{\"O}berg}, K.~I., {Furuya}, K., {Loomis}, R., {et~al.} 2015, \apj, 810, 112

\bibitem[{{{\"O}berg} {et~al.}(2021{\natexlab{b}}){{\"O}berg}, {Guzm{\'a}n},
  {Walsh}, {Aikawa}, {Bergin}, {Law}, {Loomis}, {Alarc{\'o}n}, {Andrews},
  {Bae}, {Bergner}, {Boehler}, {Booth}, {Bosman}, {Calahan}, {Cataldi},
  {Cleeves}, {Czekala}, {Furuya}, {Huang}, {Ilee}, {Kurtovic}, {Le Gal}, {Liu},
  {Long}, {M{\'e}nard}, {Nomura}, {P{\'e}rez}, {Qi}, {Schwarz}, {Sierra},
  {Teague}, {Tsukagoshi}, {Yamato}, {van't Hoff}, {Waggoner}, {Wilner}, \&
  {Zhang}}]{MAPSI}
{{\"O}berg}, K.~I., {Guzm{\'a}n}, V.~V., {Walsh}, C., {et~al.}
  2021{\natexlab{b}}, \apjs, 257, 1

\bibitem[{{{\"O}berg} {et~al.}(2011){{\"O}berg}, {Murray-Clay}, \&
  {Bergin}}]{ObergEA11}
{{\"O}berg}, K.~I., {Murray-Clay}, R., \& {Bergin}, E.~A. 2011, \apjl, 743, L16

\bibitem[{{Pacetti} {et~al.}(2022){Pacetti}, {Turrini}, {Schisano}, {Molinari},
  {Fonte}, {Politi}, {Hennebelle}, {Klessen}, {Testi}, \&
  {Lebreuilly}}]{PacettiEA22}
{Pacetti}, E., {Turrini}, D., {Schisano}, E., {et~al.} 2022, \apj, 937, 36

\bibitem[{pandas~development team(2020)}]{Pandas}
pandas~development team, T. 2020, pandas-dev/pandas: Pandas

\bibitem[{{Pegues} {et~al.}(2020){Pegues}, {{\"O}berg}, {Bergner}, {Loomis},
  {Qi}, {Le Gal}, {Cleeves}, {Guzm{\'a}n}, {Huang}, {J{\o}rgensen}, {Andrews},
  {Blake}, {Carpenter}, {Schwarz}, {Williams}, \& {Wilner}}]{PeguesEA20}
{Pegues}, J., {{\"O}berg}, K.~I., {Bergner}, J.~B., {et~al.} 2020, \apj, 890,
  142

\bibitem[{{Penteado} {et~al.}(2017){Penteado}, {Walsh}, \&
  {Cuppen}}]{PenteadoEA17}
{Penteado}, E.~M., {Walsh}, C., \& {Cuppen}, H.~M. 2017, \apj, 844, 71

\bibitem[{P\'erez \& Granger(2007)}]{IPython}
P\'erez, F. \& Granger, B.~E. 2007, Computing in Science and Engineering, 9, 21

\bibitem[{{Qi} {et~al.}(2019){Qi}, {{\"O}berg}, {Espaillat}, {Robinson},
  {Andrews}, {Wilner}, {Blake}, {Bergin}, \& {Cleeves}}]{QiEA19}
{Qi}, C., {{\"O}berg}, K.~I., {Espaillat}, C.~C., {et~al.} 2019, \apj, 882, 160

\bibitem[{{Qi} {et~al.}(2013{\natexlab{a}}){Qi}, {{\"O}berg}, \&
  {Wilner}}]{QiEA13a}
{Qi}, C., {{\"O}berg}, K.~I., \& {Wilner}, D.~J. 2013{\natexlab{a}}, \apj, 765,
  34

\bibitem[{{Qi} {et~al.}(2013{\natexlab{b}}){Qi}, {{\"O}berg}, {Wilner},
  {D'Alessio}, {Bergin}, {Andrews}, {Blake}, {Hogerheijde}, \& {van
  Dishoeck}}]{QiEA13}
{Qi}, C., {{\"O}berg}, K.~I., {Wilner}, D.~J., {et~al.} 2013{\natexlab{b}},
  Science, 341, 630

\bibitem[{{Rab} {et~al.}(2020){Rab}, {Kamp}, {Dominik}, {Ginski}, {Muro-Arena},
  {Thi}, {Waters}, \& {Woitke}}]{RabEA20}
{Rab}, C., {Kamp}, I., {Dominik}, C., {et~al.} 2020, \aap, 642, A165

\bibitem[{{Roberts} \& {Millar}(2000)}]{RM00}
{Roberts}, H. \& {Millar}, T.~J. 2000, \aap, 361, 388

\bibitem[{{Rosotti} {et~al.}(2021){Rosotti}, {Ilee}, {Facchini}, {Tazzari},
  {Booth}, {Clarke}, \& {Kama}}]{RosottiEA21}
{Rosotti}, G.~P., {Ilee}, J.~D., {Facchini}, S., {et~al.} 2021, \mnras, 501,
  3427

\bibitem[{{Roueff} {et~al.}(2013){Roueff}, {Gerin}, {Lis}, {Wootten},
  {Marcelino}, {Cernicharo}, \& {Tercero}}]{RoueffEA13}
{Roueff}, E., {Gerin}, M., {Lis}, D.~C., {et~al.} 2013, Journal of Physical
  Chemistry A, 117, 9959

\bibitem[{{Salinas} {et~al.}(2017){Salinas}, {Hogerheijde}, {Mathews},
  {{\"O}berg}, {Qi}, {Williams}, \& {Wilner}}]{SalinasEA17}
{Salinas}, V.~N., {Hogerheijde}, M.~R., {Mathews}, G.~S., {et~al.} 2017, \aap,
  606, A125

\bibitem[{{Santos} {et~al.}(2023){Santos}, {Chuang}, {Schrauwen}, {Traspas
  Mui{\~n}a}, {Zhang}, {Cuppen}, {Redlich}, {Linnartz}, \&
  {Ioppolo}}]{SantosEA23}
{Santos}, J.~C., {Chuang}, K.~J., {Schrauwen}, J.~G.~M., {et~al.} 2023, arXiv
  e-prints, arXiv:2302.11591

\bibitem[{{Stapper} {et~al.}(2022){Stapper}, {Hogerheijde}, {van Dishoeck}, \&
  {Mentel}}]{StapperEA22}
{Stapper}, L.~M., {Hogerheijde}, M.~R., {van Dishoeck}, E.~F., \& {Mentel}, R.
  2022, \aap, 658, A112

\bibitem[{{Sturm} {et~al.}(2022){Sturm}, {Booth}, {McClure}, {Leemker}, \& {van
  Dishoeck}}]{SturmEA22}
{Sturm}, J.~A., {Booth}, A.~S., {McClure}, M.~K., {Leemker}, M., \& {van
  Dishoeck}, E.~F. 2022, arXiv e-prints, arXiv:2209.09286

\bibitem[{Teague(2019)}]{GoFish}
Teague, R. 2019, The Journal of Open Source Software, 4, 1632

\bibitem[{{Terwisscha van Scheltinga} {et~al.}(2021){Terwisscha van
  Scheltinga}, {Hogerheijde}, {Cleeves}, {Loomis}, {Walsh}, {{\"O}berg},
  {Bergin}, {Bergner}, {Blake}, {Calahan}, {Cazzoletti}, {van Dishoeck},
  {Guzm{\'a}n}, {Huang}, {Kama}, {Qi}, {Teague}, \& {Wilner}}]{TvSEA21}
{Terwisscha van Scheltinga}, J., {Hogerheijde}, M.~R., {Cleeves}, L.~I.,
  {et~al.} 2021, \apj, 906, 111

\bibitem[{{van der Marel}(2022)}]{vdMarelEA22}
{van der Marel}, N. 2022, arXiv e-prints, arXiv:2210.05539

\bibitem[{{van der Marel} {et~al.}(2021{\natexlab{a}}){van der Marel},
  {Birnstiel}, {Garufi}, {Ragusa}, {Christiaens}, {Price}, {Sallum}, {Muley},
  {Francis}, \& {Dong}}]{vdMarelEA21c}
{van der Marel}, N., {Birnstiel}, T., {Garufi}, A., {et~al.}
  2021{\natexlab{a}}, \aj, 161, 33

\bibitem[{{van der Marel} {et~al.}(2021{\natexlab{b}}){van der Marel}, {Booth},
  {Leemker}, {van Dishoeck}, \& {Ohashi}}]{vdMarelEA21}
{van der Marel}, N., {Booth}, A.~S., {Leemker}, M., {van Dishoeck}, E.~F., \&
  {Ohashi}, S. 2021{\natexlab{b}}, \aap, 651, L5

\bibitem[{{van der Marel} {et~al.}(2021{\natexlab{c}}){van der Marel},
  {Bosman}, {Krijt}, {Mulders}, \& {Bergner}}]{vdMarelEA21b}
{van der Marel}, N., {Bosman}, A.~D., {Krijt}, S., {Mulders}, G.~D., \&
  {Bergner}, J.~B. 2021{\natexlab{c}}, \aap, 653, L9

\bibitem[{{van der Marel} {et~al.}(2016){van der Marel}, {van Dishoeck},
  {Bruderer}, {Andrews}, {Pontoppidan}, {Herczeg}, {van Kempen}, \&
  {Miotello}}]{vdMarelEA16}
{van der Marel}, N., {van Dishoeck}, E.~F., {Bruderer}, S., {et~al.} 2016,
  \aap, 585, A58

\bibitem[{{van der Marel} {et~al.}(2013){van der Marel}, {van Dishoeck},
  {Bruderer}, {Birnstiel}, {Pinilla}, {Dullemond}, {van Kempen}, {Schmalzl},
  {Brown}, {Herczeg}, {Mathews}, \& {Geers}}]{vdMarelEA13}
{van der Marel}, N., {van Dishoeck}, E.~F., {Bruderer}, S., {et~al.} 2013,
  Science, 340, 1199

\bibitem[{{van der Plas} {et~al.}(2014){van der Plas}, {Casassus},
  {M{\'e}nard}, {Perez}, {Thi}, {Pinte}, \& {Christiaens}}]{vdPlasEA14}
{van der Plas}, G., {Casassus}, S., {M{\'e}nard}, F., {et~al.} 2014, \apjl,
  792, L25

\bibitem[{{van 't Hoff} {et~al.}(2017){van 't Hoff}, {Walsh}, {Kama},
  {Facchini}, \& {van Dishoeck}}]{vtHoffEA17}
{van 't Hoff}, M.~L.~R., {Walsh}, C., {Kama}, M., {Facchini}, S., \& {van
  Dishoeck}, E.~F. 2017, \aap, 599, A101

\bibitem[{{Verhoeff} {et~al.}(2011){Verhoeff}, {Min}, {Pantin}, {Waters},
  {Tielens}, {Honda}, {Fujiwara}, {Bouwman}, {van Boekel}, {Dougherty}, {de
  Koter}, {Dominik}, \& {Mulders}}]{VerhoeffEA11}
{Verhoeff}, A.~P., {Min}, M., {Pantin}, E., {et~al.} 2011, \aap, 528, A91

\bibitem[{{Vioque} {et~al.}(2018){Vioque}, {Oudmaijer}, {Baines},
  {Mendigut{\'\i}a}, \& {P{\'e}rez-Mart{\'\i}nez}}]{VioqueEA18}
{Vioque}, M., {Oudmaijer}, R.~D., {Baines}, D., {Mendigut{\'\i}a}, I., \&
  {P{\'e}rez-Mart{\'\i}nez}, R. 2018, \aap, 620, A128

\bibitem[{Virtanen {et~al.}(2020)Virtanen, Gommers, Oliphant, Haberland, Reddy,
  Cournapeau, Burovski, Peterson, Weckesser, Bright, {van der Walt}, Brett,
  Wilson, Millman, Mayorov, Nelson, Jones, Kern, Larson, Carey, Polat, Feng,
  Moore, {VanderPlas}, Laxalde, Perktold, Cimrman, Henriksen, Quintero, Harris,
  Archibald, Ribeiro, Pedregosa, {van Mulbregt}, \& {SciPy 1.0
  Contributors}}]{SciPy}
Virtanen, P., Gommers, R., Oliphant, T.~E., {et~al.} 2020, Nature Methods, 17,
  261

\bibitem[{{Walsh} {et~al.}(2017){Walsh}, {Daley}, {Facchini}, \&
  {Juh{\'a}sz}}]{WalshEA17}
{Walsh}, C., {Daley}, C., {Facchini}, S., \& {Juh{\'a}sz}, A. 2017, \aap, 607,
  A114

\bibitem[{{Walsh} {et~al.}(2010){Walsh}, {Millar}, \& {Nomura}}]{WalshEA10}
{Walsh}, C., {Millar}, T.~J., \& {Nomura}, H. 2010, \apj, 722, 1607

\bibitem[{{Walsh} {et~al.}(2018){Walsh}, {Vissapragada}, \&
  {McGee}}]{WalshEA18}
{Walsh}, C., {Vissapragada}, S., \& {McGee}, H. 2018, IAU Symposium, 332, 395

\bibitem[{{Watanabe} \& {Kouchi}(2002)}]{WK02}
{Watanabe}, N. \& {Kouchi}, A. 2002, \apjl, 571, L173

\bibitem[{{Weaver} {et~al.}(2018){Weaver}, {Isella}, \& {Boehler}}]{WeaverEA18}
{Weaver}, E., {Isella}, A., \& {Boehler}, Y. 2018, \apj, 853, 113

\bibitem[{{W}es {M}c{K}inney(2010)}]{Pandas2}
{W}es {M}c{K}inney. 2010, in {P}roceedings of the 9th {P}ython in {S}cience
  {C}onference, ed. {S}t\'efan van~der {W}alt \& {J}arrod {M}illman, 56 -- 61

\bibitem[{{Wilson}(1999)}]{Wilson99}
{Wilson}, T.~L. 1999, Reports on Progress in Physics, 62, 143

\bibitem[{{Woitke} {et~al.}(2016){Woitke}, {Min}, {Pinte}, {Thi}, {Kamp},
  {Rab}, {Anthonioz}, {Antonellini}, {Baldovin-Saavedra}, {Carmona}, {Dominik},
  {Dionatos}, {Greaves}, {G{\"u}del}, {Ilee}, {Liebhart}, {M{\'e}nard},
  {Rigon}, {Waters}, {Aresu}, {Meijerink}, \& {Spaans}}]{WoitkeEA16}
{Woitke}, P., {Min}, M., {Pinte}, C., {et~al.} 2016, \aap, 586, A103

\bibitem[{{W{\"o}lfer} {et~al.}(2022){W{\"o}lfer}, {Facchini}, {van der Marel},
  {van Dishoeck}, {Benisty}, {Bohn}, {Francis}, {Izquierdo}, \&
  {Teague}}]{WoelferEA22}
{W{\"o}lfer}, L., {Facchini}, S., {van der Marel}, N., {et~al.} 2022, arXiv
  e-prints, arXiv:2208.09494

\bibitem[{{Wootten}(1987)}]{Wootten87}
{Wootten}, A. 1987, in Astrochemistry, ed. M.~S. {Vardya} \& S.~P. {Tarafdar},
  Vol. 120, 311--319

\bibitem[{{Yen} {et~al.}(2016){Yen}, {Koch}, {Liu}, {Puspitaningrum}, {Hirano},
  {Lee}, \& {Takakuwa}}]{YenEA16}
{Yen}, H.-W., {Koch}, P.~M., {Liu}, H.~B., {et~al.} 2016, \apj, 832, 204

\bibitem[{{Young} {et~al.}(2021){Young}, {Alexander}, {Walsh}, {Nealon},
  {Booth}, \& {Pinte}}]{YoungEA21}
{Young}, A.~K., {Alexander}, R., {Walsh}, C., {et~al.} 2021, \mnras, 505, 4821

\bibitem[{{Zhang} {et~al.}(2019){Zhang}, {Bergin}, {Schwarz}, {Krijt}, \&
  {Ciesla}}]{ZhangEA19}
{Zhang}, K., {Bergin}, E.~A., {Schwarz}, K., {Krijt}, S., \& {Ciesla}, F. 2019,
  \apj, 883, 98

\bibitem[{{Zhang} {et~al.}(2021){Zhang}, {Booth}, {Law}, {Bosman}, {Schwarz},
  {Bergin}, {{\"O}berg}, {Andrews}, {Guzm{\'a}n}, {Walsh}, {Qi}, {van't Hoff},
  {Long}, {Wilner}, {Huang}, {Czekala}, {Ilee}, {Cataldi}, {Bergner}, {Aikawa},
  {Teague}, {Bae}, {Loomis}, {Calahan}, {Alarc{\'o}n}, {M{\'e}nard}, {Le Gal},
  {Sierra}, {Yamato}, {Nomura}, {Tsukagoshi}, {P{\'e}rez}, {Trapman}, {Liu}, \&
  {Furuya}}]{ZhangEA21}
{Zhang}, K., {Booth}, A.~S., {Law}, C.~J., {et~al.} 2021, \apjs, 257, 5

\bibitem[{{Zhang} {et~al.}(2020){Zhang}, {Bosman}, \& {Bergin}}]{ZhangEA20b}
{Zhang}, K., {Bosman}, A.~D., \& {Bergin}, E.~A. 2020, \apjl, 891, L16

\end{thebibliography}

%% === Appendix
\newpage
\onecolumn
\begin{appendix}

\section{Observed molecules}
\begin{table*}[ht!]
\caption{The datasets used throughout this work.}           
\label{table:Datasets}     
    \begin{tabular}{c c c c c c c c}
    \hline\hline       
    Project Code & PI & Band & Baselines & Int. Time & Ang. res & MRS$^{(a)}$ & Main molecules covered \\
    & & & [m] & [hours] & ["] & ["] & \\
    \hline
    2011.0.00318.S & Fukagawa, M. & 7 & 21 - 453 & 3.695 & 0.420 & 4.587 & \textbf{\ce{^{13}CO}}, \textbf{\ce{C^18O}}, \textbf{\ce{CS}}, \ce{SO}, \ce{^{34}SO_2}, \ce{CH_3OH} \\
    2011.0.00465.S & Cassasus, S. & 6 & 21 - 402 & 1.403 & 0.200 & 6.003 & \textbf{\ce{^{12}CO}}, \textbf{\ce{^{13}CO}}, \textbf{\ce{C^18O}} \\
      2011.0.00465.S & Cassasus, S.& 7 & 21 - 402 & 0.882 & 0.439 & 3.547 & \textbf{\ce{^{12}CO}}, \textbf{\ce{HCO^+}}, \textbf{\ce{HCN}}, \ce{SO_2} \\
    2012.1.00613.S & Fukagawa, M. & 7 & 15 - 1574 & 1.856 & 0.147 & 2.073 & \textbf{\ce{^{13}C^{18}O}}, \ce{CH_3OH} \\
    2013.1.00305.S & Cassasus, S. & 6 & 15 - 1574 & 1.025 & 0.167 & 2.139 & \textbf{\ce{DCO^+}}, \textbf{\ce{H_2CO}}, \ce{H_2CO} \\
    2015.1.00614.S & Cassasus, S. & 9 & 15 - 460 & 0.395 & 0.229 & 2.113 & \textbf{\ce{^{12}CO}}, \textbf{\ce{HCO^+}}, \ce{HCN}, \ce{H_2CO}, \ce{CH_3OH} \\
    2015.1.00805.S & Cassasus, S. & 4 & 15 - 1462 & 0.151 & 0.357 & 7.991 & \textbf{\ce{H_2CO}} \\
    2015.1.01137.S & Tsukagoski, T. & 8 & 15 - 640 & 0.402 & 0.247 & 6.151 & \textbf{\ce{C}}, \textbf{\ce{CS}} \\
    \hline
    \end{tabular}
\tablefoot{The final column shows the detected (in bold) and non-detected molecules for each dataset. \\
$^{(a)}$ MRS = maximum recoverable scale}
\end{table*}

\clearpage
\section{Additional moment-zero maps}
\begin{figure}[ht!]
    \centering
    \includegraphics[width=\textwidth]{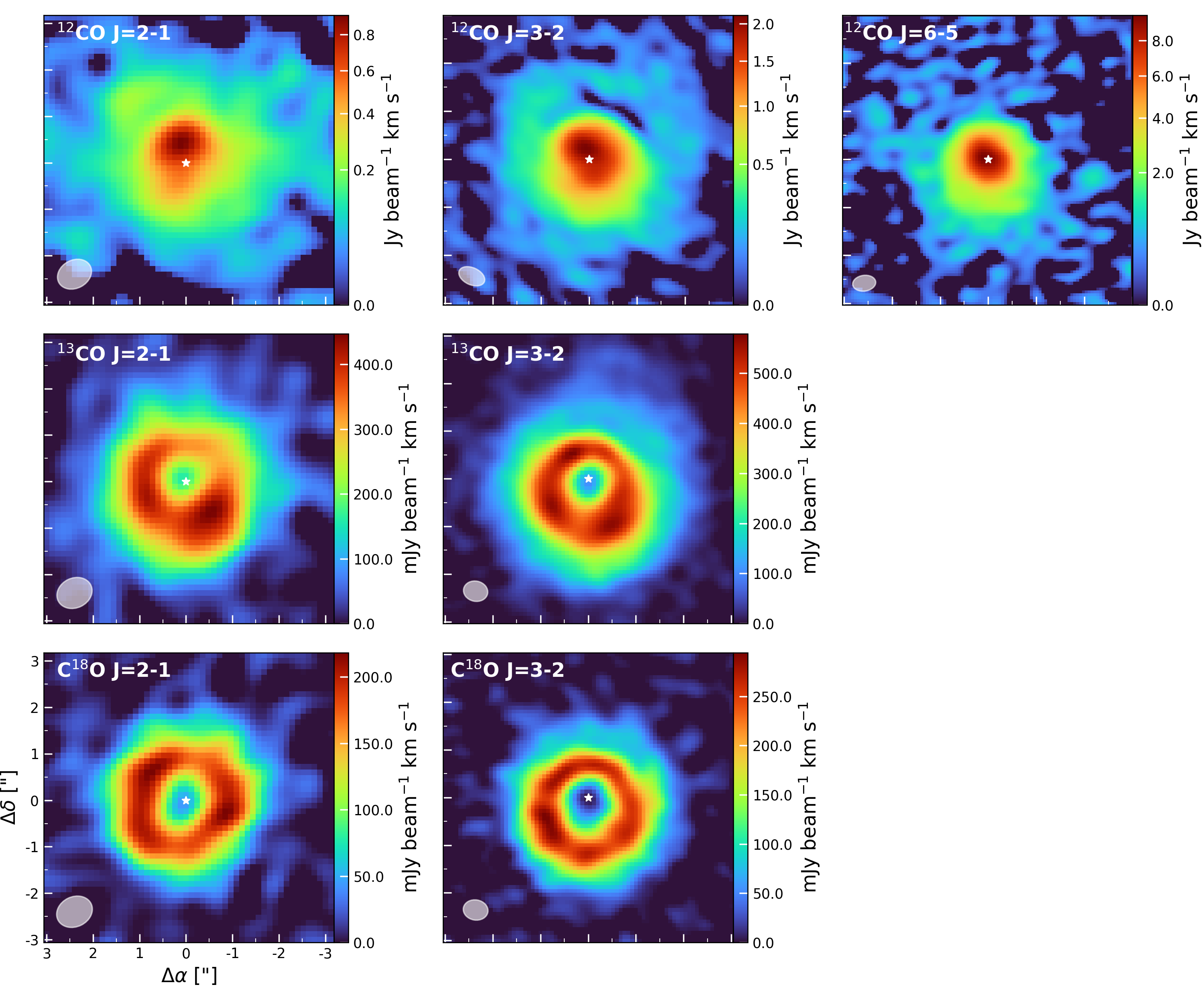}
    \caption{moment-zero maps of the observed CO isotopologue transitions. From top to bottom are shown the \ce{^{12}CO}, \ce{^{13}CO} and \ce{C^{18}O} isotopologues, while from left to right are shown the $J$=2-1, $J$=3-2 and $J$=6-5 transitions. The beams are displayed in the lower left, while the approximate location of the host star is inferred by the yellow star. The \ce{^{12}CO} colour-maps are displayed using a power-law scaling with an exponent of 0.5.}
    \label{fig:CO-Gallery}
\end{figure}

\begin{figure}[ht!]
    \centering
    \includegraphics[width=0.7\textwidth]{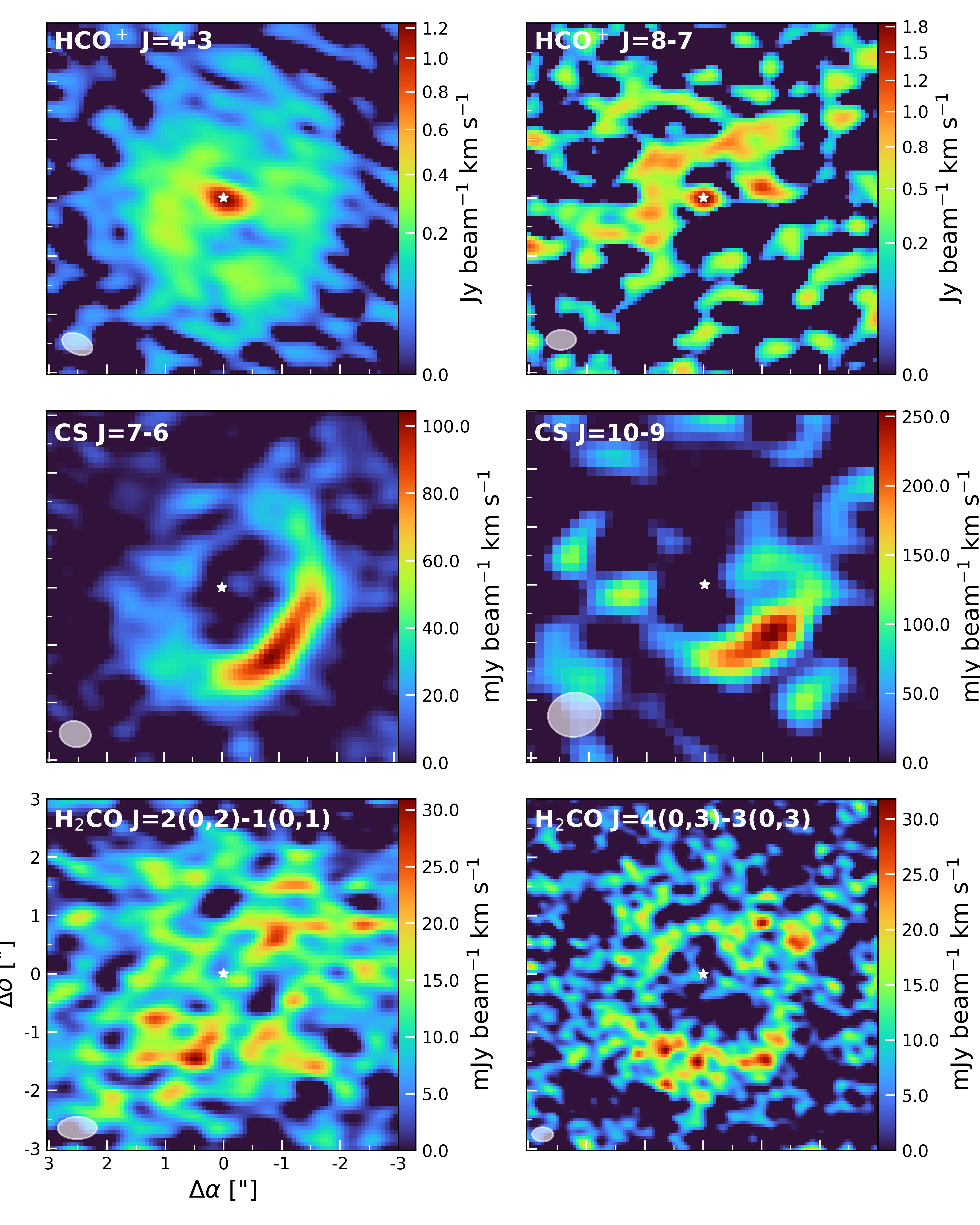}
    \caption{moment-zero maps of the observed \ce{HCO^+} (top row), \ce{CS} (middle row) and \ce{H_2CO} (bottom row) transitions. The beams are displayed in the lower left and the yellow star in the center indicates the inferred location of the host star. The \ce{HCO^+} colour-maps are displayed using a power-law scaling with an exponent of 0.5.}
    \label{fig:OM-Gallery}
\end{figure}

\clearpage
\section{GoFish Spectra}
\begin{figure}[ht!]
    \centering
    \includegraphics[trim=0cm 0.25cm 0cm 0cm, clip=true,width=\textwidth]{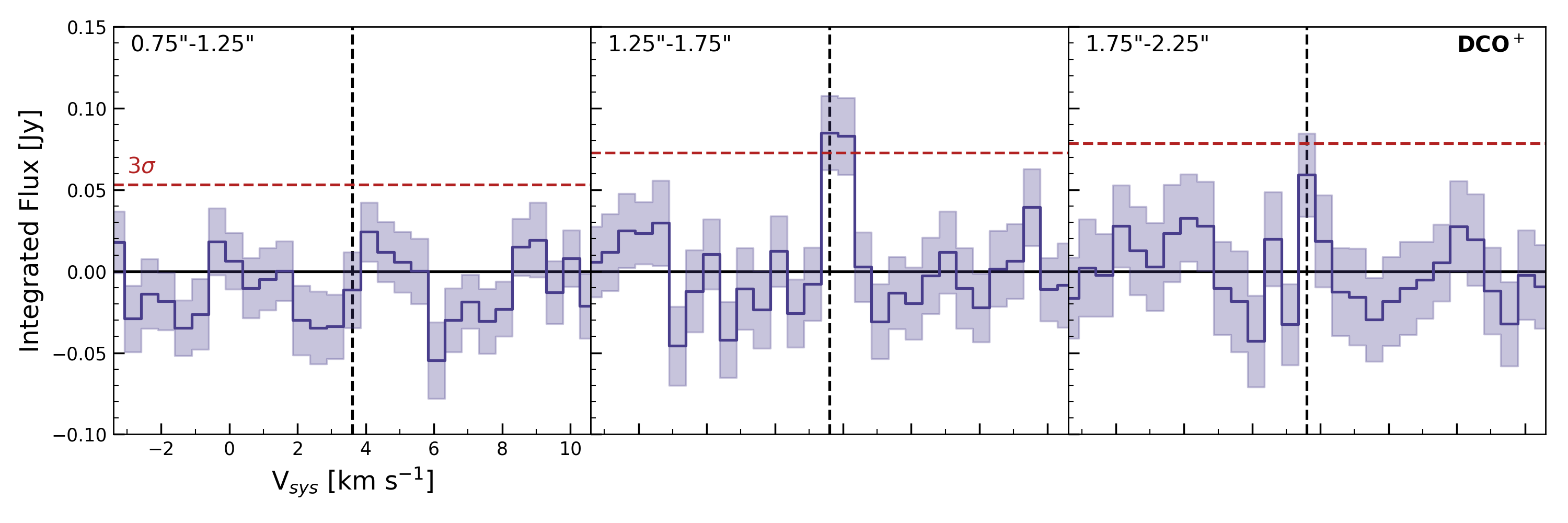}
    \includegraphics[trim=0cm 0.25cm 0cm 0cm, clip=true,width=\textwidth]{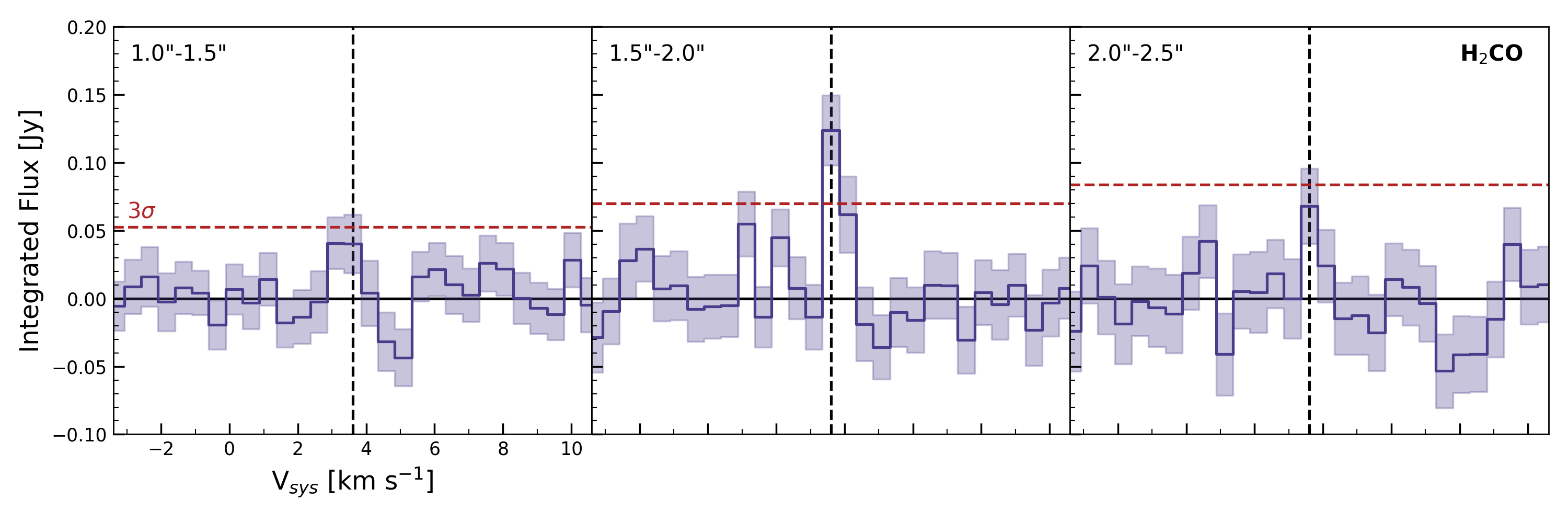}
    \includegraphics[trim=0cm 0.25cm 0cm 0cm, clip=true,width=\textwidth]{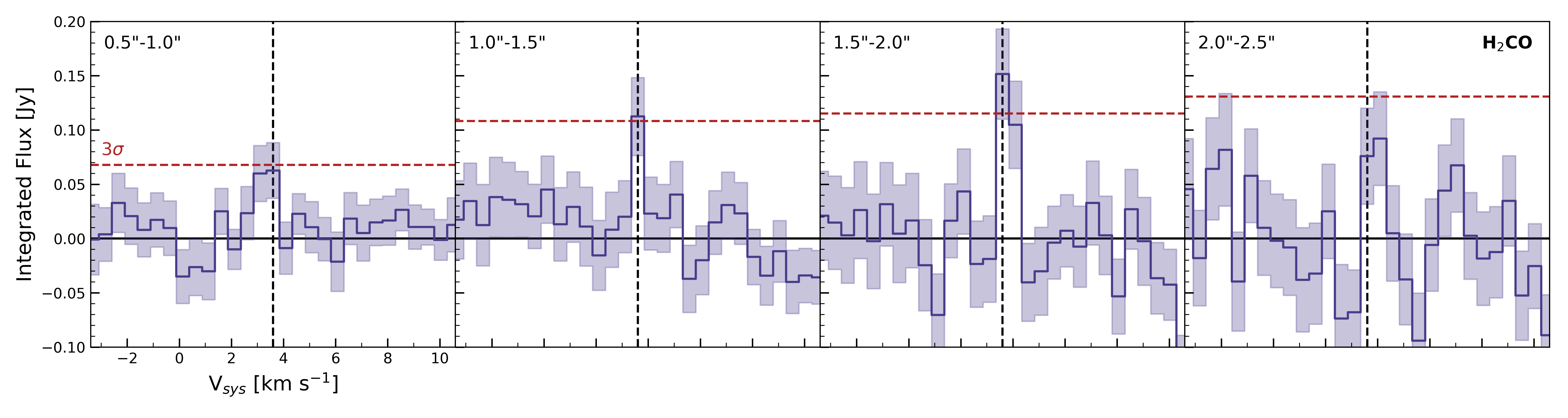}
    \caption{Spectra acquired from different radial regimes with \textit{GoFish}. The radial regimes for each spectrum are indicated in the top left. From top to bottom are shown the \ce{DCO^+} $J$=4-3, \ce{H_2CO} $J$=4$_{2,3}$-3$_{2,2}$ and \ce{H_2CO} $J$=4$_{2,2}$-3$_{2,1}$. The horizontal, red dashed line in each spectrum indicates the $3\sigma$-RMS level, while the vertical, black dashed line indicates the systemic velocity of 3.6 km s$^{-1}$.}
    \label{fig:GFSpectra}
\end{figure}

\begin{figure}[ht!]
    \centering
    \includegraphics[trim=0cm 0.25cm 0cm 0cm, clip=true,width=\textwidth]{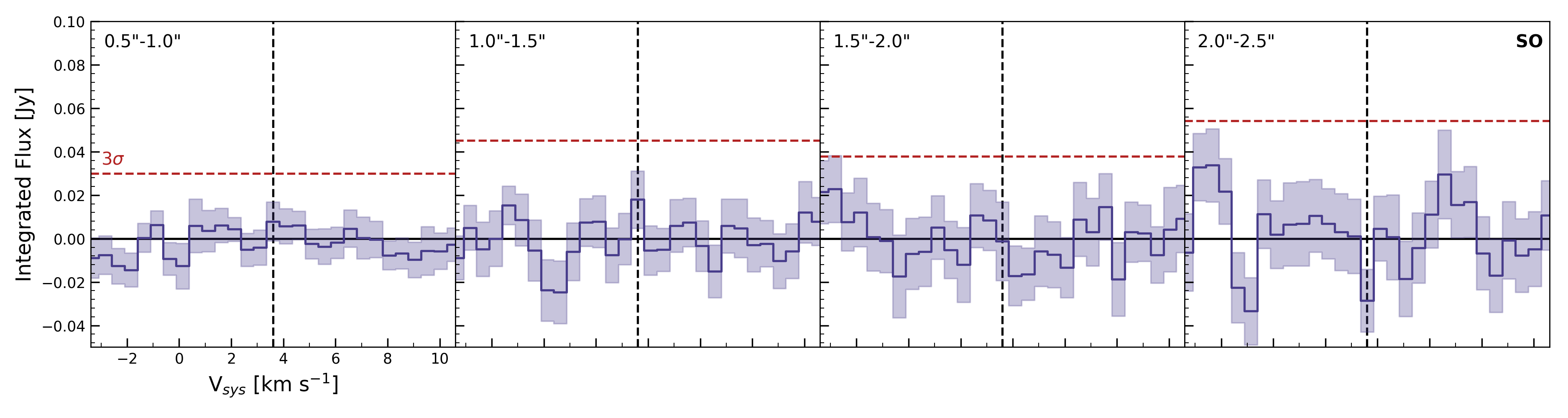}
    \includegraphics[trim=0cm 0.25cm 0cm 0cm, clip=true,width=\textwidth]{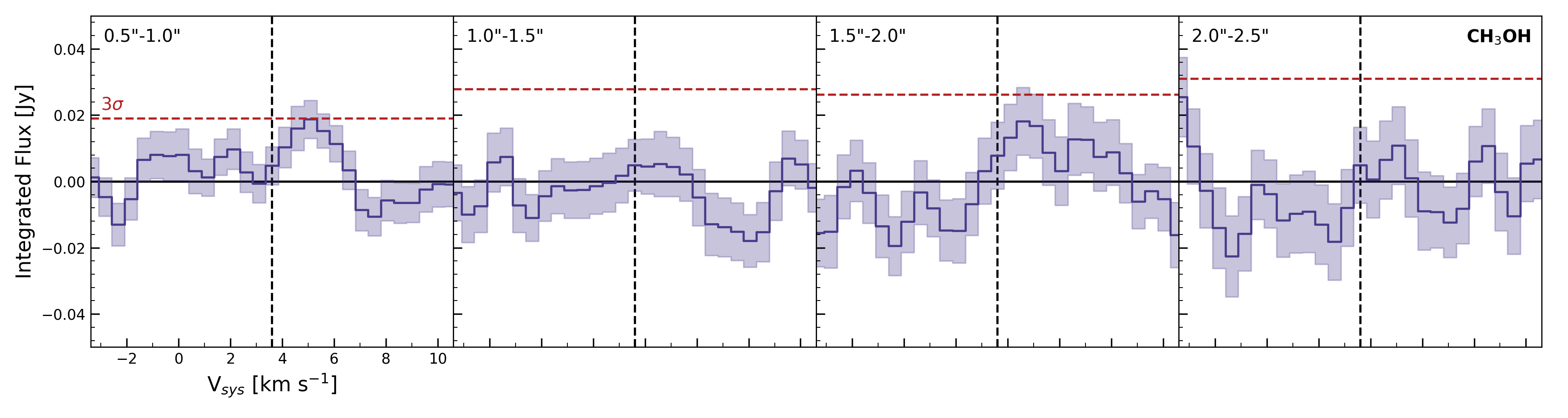}
    \caption{The same as Figure \ref{fig:GFSpectra}, but for the non-detections of the \ce{SO} $J$=1$_2$-0$_1$ (top panel) and the \ce{CH_3OH} $J$=7$_{1,6}$-7$_{0,7}$ (bottom panel) transitions.}
    \label{fig:GFSpectra-NDs}
\end{figure}

\clearpage
\section{Rotational diagram analysis: Posterior distributions}
\begin{figure}[ht!]
    \centering
    \includegraphics[width=0.5\textwidth]{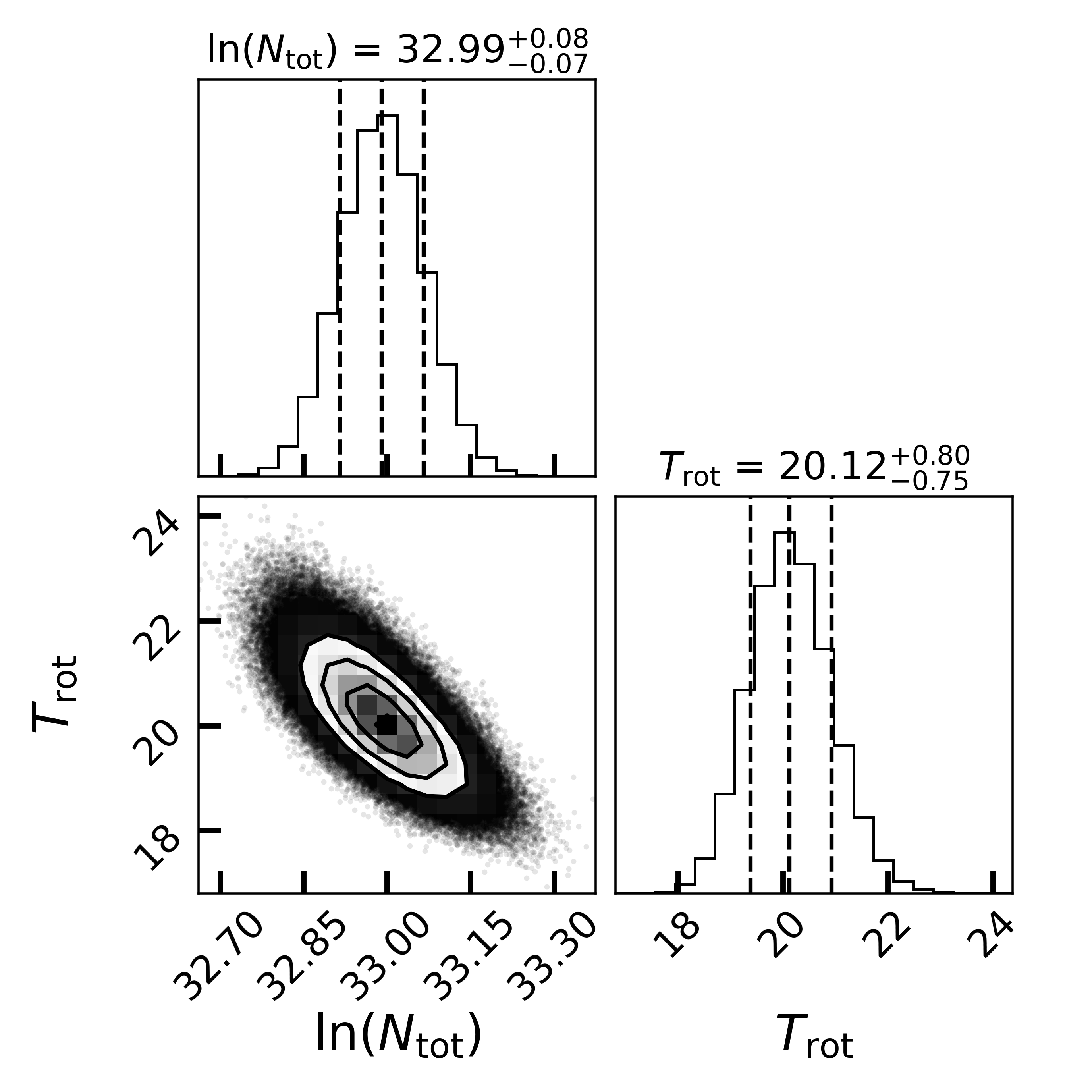}
    \caption{Posterior distributions of the rotational diagram analysis for \ce{H_2CO}.}
    \label{fig:H2CO_RDA_Corner}
\end{figure}

\clearpage
\twocolumn
\section{RADMC-3D Model} \label{sec:RADMC3D}
\begin{figure*}[h!]
    \centering
    \includegraphics[width=\textwidth]{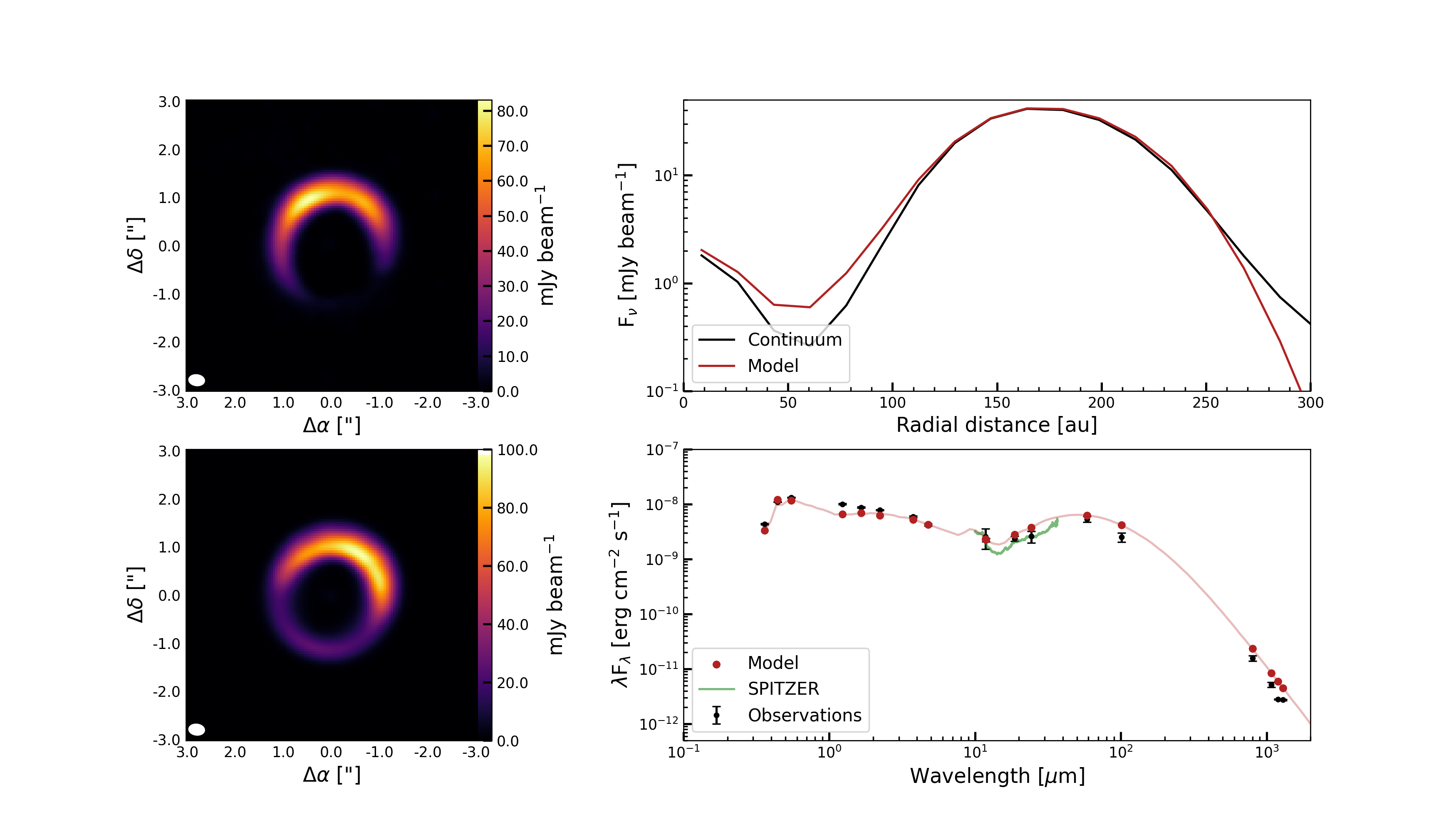}
    \caption{The resulting RADMC-3D fit to  the radial profile (top right) and SED (bottom right). The continuum image is shown on the top left, while the model image is shown on the bottom left. For both the radial profile and the SED, the observations are shown in black, while the model is shown in red. The green line within the SED displays high-resolution Spitzer observations \citep{BouwmanEA04}.}
    \label{fig:RADMC3D-Fit}
\end{figure*}

The \textit{RADMC-3D} model has been visually fitted to both the dereddened SED (obtained from \citealt{VerhoeffEA11}), using $A_V$=0.3 mag \citep{FvdM20} and the ALMA continuum radial profile. For the model we have used the standard \textit{DIANA} opacities \citep{WoitkeEA16}, implemented using \textit{OpTool} \citep{OpTool}. We have used two different dust size distributions (both with 100 sample sizes), one for the small grains with sizes from 0.005 $\mu$m to 1 $\mu$m and one for the large grains with sizes from 0.005 $\mu$m to 1000 $\mu$m. These size distributions have been used for both the inner and the outer disk. \\
\indent We have modelled the inner disk by assuming an exponentially-tapered surface density,
\begin{align}
    \sigma(R) = \sigma_c\left(\frac{R}{R_\textnormal{c,ID}}\right)^\gamma\exp\left[-\left(\frac{R}{R_\textnormal{c,ID}}\right)^{2-\gamma}\right],
\end{align}
where $R_\textnormal{c,ID}$=40 au and $\gamma$=1 have been used. In addition, the inner disks starts at the dust sublimation radius, estimated to be located at $R_\textnormal{sub}$=$0.07\left(L/L_\odot\right)^{1/2}\simeq$0.22 au, assuming a sublimation radius of 1500 K \citep{DullemondEA01}. The outer disk has been modelled by a Gaussian in both the radial and azimuthal directions to create the asymmetric ring. The median for the azimuthal Gaussian has been set to $\pi$ and we have used a standard deviation of 0.18$\pi$, yielding the arbitrary orientation, but similar shape, of the model asymmetry displayed in the bottom left of Figure \ref{fig:RADMC3D-Fit}. Additionally, the cavity, which is fully depleted from both grain distributions, extends from 40 to 100 au. \\
\indent Both inner and outer disks have been given different vertical structures, described with the equation,
\begin{align}
    h(R) = h_\textnormal{c}\left(\frac{R}{R_\textnormal{c}}\right)^\psi.
\end{align}
For the inner disk we have used $h_\textnormal{c}$=0.2, $R_\textnormal{c}$=$R_\textnormal{c,ID}$ and $\psi$=0.05, while for the outer disk we have used $h_\textnormal{c}$=0.35, $R_\textnormal{c}$=150 and $\psi$=0.2. Furthermore, we have taken for the inner and outer disk two separate fractions of large grains, 50\% and 85\%, respectively. For both the inner and outer disk we have set the large grains settling factor to $\chi$=0.1. \\
\indent Furthermore, in order to properly fit the radial profile we had to increase the dust mass to $M_\textnormal{dust}$=7.5$\times$10$^{-3}$ $M_\odot$. This is about 5 times larger than our estimated in Section \ref{sec:DustMass}. The higher dust mass is likely partially linked to the used opacities, but also provides further hints that the continuum emission is optically thick and our dust mass estimate is a lower limit. \\
\indent The resulting fit is displayed in Figure \ref{fig:RADMC3D-Fit}, displaying the continuum and model images on the left, and the radial profile and SED on the right. Overall the model fits reasonable well to both the radial profile and the SED, only slightly overproducing the inner disk. The fit is, however, good enough for the aims of the model, to provide a second check of the approximate location of the \ce{CO} snowline. \\
\indent Additionally, to compare with the estimated location of the snowline in \citet{vdMarelEA22}, the white dashed line displays the 20 K line when assuming a power-law midplane temperature structure \citep{CG97,DullemondEA01},
\begin{align}
    T(R) = \left(\frac{\phi L}{8\pi\sigma_\textnormal{B}R^2}\right)^{1/4}.
\end{align}
Here, following \citet{vdMarelEA21b}, we have taken $\phi=0.02$. As can be seen the white dashed line is well within the dust trap, showing that a simple power-law for the midplane temperature does not yield the right location of the \ce{CO} snowline for the HD~142527 disk. This is very likely due to the large cavity, indicating that temperature structure and, hence, the snowline locations of transition disks with large cavities cannot be determined by a power-law approximation of the midplane temperature.

\begin{figure}[ht!]
    \centering
    \includegraphics[width=\columnwidth]{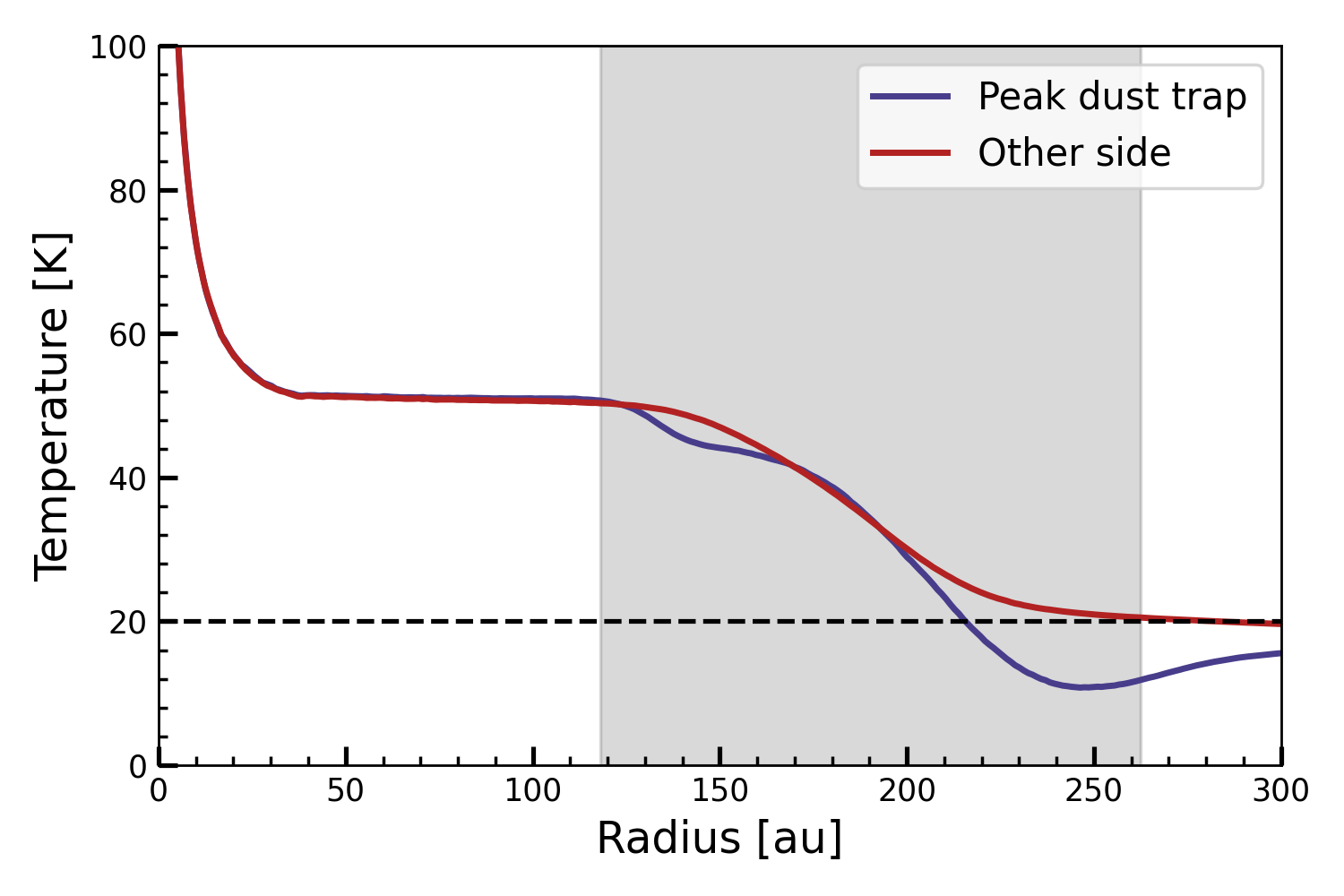}
    \caption{The dust midplane temperature at the peak location of the model dust trap (blue line) and the other side of the disk (red line). The horizontal, dashed black line indicates a temperature of 20 K. The grey shaded area displays the outermost contour levels of the modelled dust trap in Figure \ref{fig:RADMC3D-Temp}.}
    \label{fig:RADMC3D-TempSides}
\end{figure}

\onecolumn
\clearpage
\section{Full channel maps}
\begin{figure}[ht!]
    \centering
    \includegraphics[trim=2cm .5cm 0cm 2cm, clip=true,width=\textwidth]{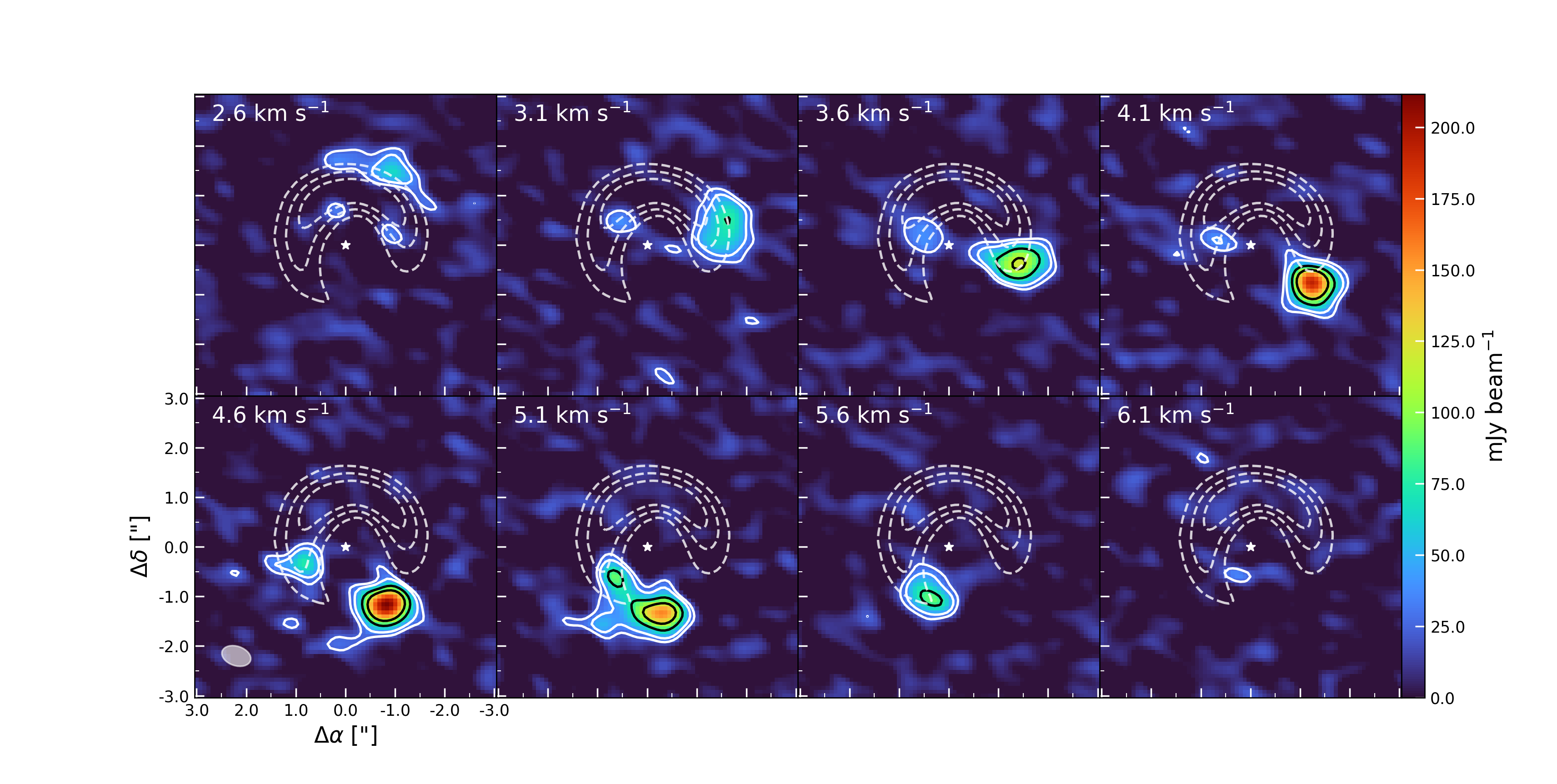}
    \includegraphics[trim=2cm .5cm 0cm 2cm, clip=true,width=\textwidth]{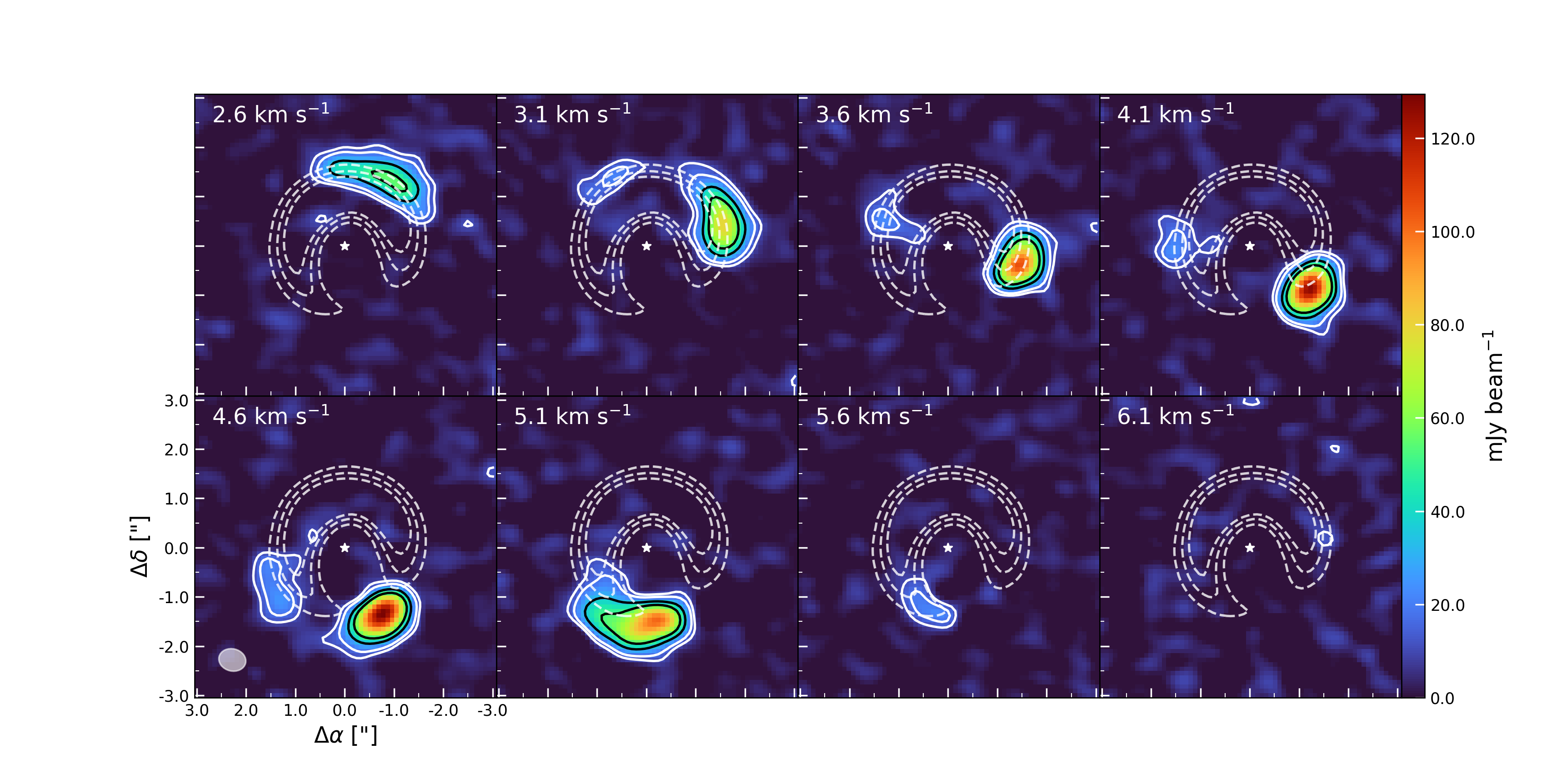}
    \caption{Channels maps of the \ce{HCN} $J$=4-3 (top) and \ce{CS} $J$=7-6 (bottom) transitions. The resolving beam is displayed in the lower left and the white star in the center indicates the approximate location of the host star. The white contours indicate the $3\sigma$- and $5\sigma$-RMS levels, while the black contours show the $10\sigma$- and $15\sigma$-RMS levels. The location of the continuum emission is in each map indicated by the white, dashed contours.}
    \label{fig:ChanMaps}
\end{figure}

\end{appendix}

\end{document}